\documentclass[prd,aps,10pt,twocolumn,nofootinbib,superscriptaddress,preprintnumbers,floatfix]{revtex4}

\usepackage[hyperfootnotes=false,bookmarks=false]{hyperref}
\usepackage{graphicx}
\usepackage{amsmath}

\newcommand{\eq}[1]{Eq.~\eqref{eq:#1}}
\newcommand{\eqs}[2]{Eqs.~\eqref{eq:#1} and \eqref{eq:#2}}
\renewcommand{\sec}[1]{Sec.~\ref{sec:#1}}

\newcommand{\subsec}[1]{Sec.~\ref{subsec:#1}}
\newcommand{\subsecs}[2]{Secs.~\ref{subsec:#1} and \ref{subsec:#2}}
\newcommand{\fig}[1]{Fig.~\ref{fig:#1}}

\newcommand{\figsthree}[3]{Figs.~\ref{fig:#1}, \ref{fig:#2}, and \ref{fig:#3}}
\newcommand{\app}[1]{App.~\ref{app:#1}}

\newcommand{\tab}[1]{Table~\ref{table:#1}}

\newcommand{\abs}[1]{\lvert#1\rvert}
\newcommand{\ord}[1]{\mathcal{O}(#1)}

\newcommand{\be}{\begin{equation}}
\newcommand{\ee}{\end{equation}}

\newcommand{\df}{\mathrm{d}}
\newcommand{\img}{\mathrm{i}}

\newcommand{\e}{\epsilon}

\newcommand{\cM}{\mathcal{M}}

\newcommand{\cI}{\mathcal{I}}

\newcommand{\GeV}{\,\mathrm{GeV}}

\newcommand{\nn}{\nonumber}

\newcommand{\cut}{\mathrm{cut}}
\newcommand{\jet}{\mathrm{jet}}
\newcommand{\as}{\alpha_s}

\newcommand{\pTcut}{p_T^\mathrm{cut}}
\newcommand{\Ecm}{E_\mathrm{cm}}
\newcommand{\Gcusp}{\Gamma_{\rm cusp}^g}

\newcommand{\sing}{\mathrm{s}}
\newcommand{\nons}{\mathrm{ns}}
\newcommand{\Rnons}{{R\mathrm{sub}}}
\newcommand{\muns}{\mu_\mathrm{ns}}
\newcommand{\tot}{\mathrm{tot}}

\newcommand{\kt}{{\rm k}_{\rm T}}
\newcommand{\LO}{\mathrm{LO}}
\newcommand{\NLO}{\mathrm{NLO}}
\newcommand{\NNLO}{\mathrm{NNLO}}
\newcommand{\NNLL}{\mathrm{NNLL}}
\newcommand{\FO}{\mathrm{FO}}
\newcommand{\resum}{\mathrm{resum}}

\begin{document}


\preprint{\vbox{
\hbox{DESY 13-122}
\hbox{MIT--CTP 4479}
}}

\title{\boldmath Jet $p_T$ Resummation in Higgs Production at NNLL$'+$NNLO}

\author{Iain W.~Stewart}
\affiliation{Center for Theoretical Physics, Massachusetts Institute of Technology, Cambridge, MA 02139, USA\vspace{0.5ex}}

\author{Frank J.~Tackmann}
\affiliation{Theory Group, Deutsches Elektronen-Synchrotron (DESY), D-22607 Hamburg, Germany\vspace{0.5ex}}

\author{Jonathan R.~Walsh}
\affiliation{Ernest Orlando Lawrence Berkeley National Laboratory, University of California, Berkeley, CA 94720\vspace{0.5ex}}

\author{Saba Zuberi\vspace{0.2cm}}
\affiliation{Ernest Orlando Lawrence Berkeley National Laboratory, University of California, Berkeley, CA 94720\vspace{0.5ex}}

\date{July 5, 2013}

\begin{abstract}

We present predictions for Higgs production via gluon fusion with a $p_T$ veto on jets and with the resummation of jet-veto logarithms at NNLL$'+$NNLO order. These results incorporate explicit ${\cal O}(\alpha_s^2)$ calculations of soft and beam functions, which include the dominant dependence on the jet radius $R$. In particular the NNLL$'$ order accounts for the correct boundary conditions for the N$^3$LL resummation, for which the only unknown ingredients are higher-order anomalous dimensions. We use scale variations in a factorization theorem in both rapidity and virtuality space to estimate the perturbative uncertainties, accounting for both higher fixed-order corrections as well as higher-order towers of jet-$p_T$ logarithms. This formalism also predicts the correlations in the theory uncertainty between the exclusive 0-jet and inclusive $1$-jet bins. At the values of $R$ used experimentally, there are important corrections due to jet algorithm clustering that include logarithms of $R$.  Although we do not sum logarithms of $R$, we do include an explicit contribution in our uncertainty estimate to account for higher-order jet clustering logarithms. Precision predictions for this $H+0$-jet cross section and its theoretical uncertainty are an integral part of Higgs analyses that employ jet binning.

\end{abstract}

\maketitle

\section{Introduction}
\label{sec:intro}

After the discovery of a Higgs boson~\cite{Aad:2012tfa, Chatrchyan:2012ufa}, a central objective of the LHC physics program is to measure the properties of the new particle by exploiting all accessible production and decay channels.
The $gg\to H\to WW$ channel is very sensitive to the Higgs coupling to $W$ gauge bosons. The $gg\to H\to\tau\tau$ channel provides direct sensitivity to the Higgs couplings to fermions and is the only measurable channel that gives direct access to the Higgs couplings in the leptonic sector of the Standard Model. In both these channels the experimental analyses separate the data into jet bins to take advantage of the fact that the signal over background ratio, as well as the dominant background contributions, strongly depend on the number of jets in the final state. Of particular importance is the 0-jet bin, where any hard jets are vetoed, as it contains the largest signal cross section.

Extracting the Higgs couplings from the measured exclusive 0-jet cross section requires precise theoretical predictions.
Any type of jet veto introduces a veto scale, $k^\cut$. For a tight jet veto, $k^\cut \ll m_H$, large Sudakov logarithms of the veto scale, $\as^n \ln^m(k^\cut/m_H)$, appear in the perturbative series and must be resummed to all orders to obtain a meaningful perturbative prediction. For $k^\cut\sim m_H$, fixed-order perturbation theory can safely be applied, and the cross section with arbitrary cuts has been calculated at fixed next-to-next-to-leading order (NNLO)~\cite{Anastasiou:2004xq, Anastasiou:2005qj,Catani:2007vq, Grazzini:2008tf}.  In the transition region between these two limits, both the veto logarithms and nonlogarithmic fixed-order corrections are numerically important, and a complete description including both types of perturbative corrections must be used to obtain the best possible theoretical precision.  For earlier theoretical work on analytic resummation for Higgs jet vetoes see for example~\cite{Berger:2010xi, Stewart:2011cf,  Banfi:2012yh, Becher:2012qa, Tackmann:2012bt, Banfi:2012jm, Liu:2012sz, Liu:2013hba}.

In principle, there are several different ways to implement a veto on additional emissions due to initial-state and final-state radiation in a given process. A ``global jet veto'' corresponds to a restriction applied to the sum of all radiation, for example through a global event shape such as beam thrust~\cite{Stewart:2009yx} [or equivalently $(N=0)$-jettiness~\cite{Stewart:2010tn}] or $E_T = \sum\abs{p_T}$, and allows for precise resummed predictions~\cite{Stewart:2009yx, Stewart:2010tn, Stewart:2010pd, Berger:2010xi,Papaefstathiou:2010bw}.

The current experimental analyses use a jet clustering algorithm (the anti-$\kt$ algorithm~\cite{Cacciari:2008gp} with a jet radius $R = 0.4$ for ATLAS and $R = 0.5$ for CMS) to identify jets. The jet veto is then implemented by requiring $p_T^\jet < \pTcut$ for any jets with $\abs{\eta^\jet} < \eta^\cut$ (while jets at larger pseudorapidities are unrestricted). The typical experimental ranges are $\pTcut \sim 25 - 30 \GeV$ for $\eta^\cut \sim 4.5 - 5$ (with the high value of $\eta^\cut$ having a small effect on the cross section). In contrast to a global jet veto, this procedure corresponds to a ``local jet veto'', since the restriction on final state radiation is applied separately to each individual local cluster of emissions.

For a cut on either $E_T < \pTcut$ or $p_T^\jet < \pTcut$, the jet-veto scale is set by $p_T$ and Sudakov double logarithms of the ratio $\pTcut / m_H$ arise. The leading correction to the $0$-jet cross section for Higgs production via gluon fusion has the form
\begin{equation} \label{eq:sig0doublelog}
\sigma_0(\pTcut) = \sigma_\LO\Bigl(1 - \frac{\as C_A}{\pi}\, 2 \ln^2 \frac{\pTcut}{m_H} + \ldots \Bigr)
\,,\end{equation}
where $\sigma_\LO$ denotes the lowest-order cross section. The hierarchy between $\pTcut$ and $m_H$ implies that resummation of logarithms of $\pTcut/m_H$ should be performed. For the $p_T^\jet \leq \pTcut$ veto, the resummation of $\pTcut$-logarithms up to NNLL has been presented in Refs.~\cite{Banfi:2012yh,Becher:2012qa, Banfi:2012jm}.

In this paper, we calculate the resummed $H+0$-jet cross section from gluon fusion using the framework of soft-collinear effective theory (SCET)~\cite{Bauer:2000ew, Bauer:2000yr, Bauer:2001ct, Bauer:2001yt, Bauer:2002nz}, where the cross section is factorized into calculable pieces and the resummation is performed by renormalization group evolution (RGE) in both virtuality and rapidity space. We determine the cross section at NNLL$_{p_T}'+$NNLO order, where we use the subscript $p_T$ to explicitly denote the fact that the resummation order only counts logarithms of $\pTcut/m_H$ (and not $R^2$).  The primed order counting is described for example in Ref.~\cite{Berger:2010xi}. It includes the NNLL resummation and in addition the full $\ord{\as^2}$ dependence of the functions appearing in the factorization theorem (including in our case the $\ord{\as^2}$ effects from jet clustering). These corrections incorporate the dominant NNLO corrections at small $\pTcut$ into the resummed result. They are formally part of the N$^3$LL resummation for which they provide the correct RGE boundary conditions. The missing ingredients for a complete N$^3$LL resummation are the unknown three-loop non-cusp and four-loop cusp anomalous dimensions. We also include the ``nonsingular'' $\ord{\alpha_s^2}$ corrections that vanish as $\pTcut\to 0$, which are not part of the resummation. Thus our results incorporate the complete NNLO cross section for all values of $\pTcut$, including the total NNLO cross section in the limit of large $\pTcut$. This allows us to also obtain resummed predictions for the exclusive 0-jet event fraction (or efficiency) and the inclusive 1-jet cross section with a cut $p_T^\jet \geq \pTcut$ on the leading jet.

In our analysis, we place a particular emphasis on a careful estimate of the remaining perturbative uncertainties in our predictions. The different contributions to the uncertainty are estimated by appropriate variations of the different scales in virtuality and rapidity space appearing in the factorization theorem. This allows us to distinguish between and account for both resummation and  fixed-order uncertainties. This formalism then automatically determines the correlations in the perturbative uncertainties between the total inclusive, exclusive 0-jet, and inclusive $1$-jet cross sections.

The 0-jet cross section defined by $p_T^\jet \leq \pTcut$ has a complex dependence on the jet algorithm, whose effect is to introduce a nontrivial dependence on the jet radius $R$,
\begin{align} \label{eq:sigmaclus}
\ln\frac{\sigma_0^{(n)}(\pTcut)}{\sigma_\LO}  
&\supset  \Bigl( \frac{\as}{4\pi} \Bigr)^n \ln \frac{m_H}{\pTcut} \, C_n(R)
\,.
\end{align}
For small $R^2$ the numerically most relevant terms contain logarithms of $R^2$ and at $\ord{\as^n}$ are of the form~\cite{Tackmann:2012bt}
\begin{align} \label{eq:CnRdef}
  C_n(R) \sim \ln^{n-1} R^2 + \ln^{n-2} R^2 + \ldots  \ln R^2 + {\cal O}(R^2) \,. 
\end{align}
They also contain subleading power corrections of $\ord{R^2}$. The jet clustering effects start at $\ord{\as^2}$ ($n=2$) relative to $\sigma_\LO$. They were first calculated in Ref.~\cite{Banfi:2012yh} and at present are the only clustering corrections that are known. They turn out to have a sizeable effect on the cross section for jet radii $R = 0.4$ and $0.5$. The large clustering effects for such small values of $R$ imply that the logarithms of $R^2$ should be formally treated as being of similar size as the logarithms of $\pTcut/m_H$ and hence should also be resummed.  In particular, as \eq{sigmaclus} shows, counting $\ln R^2 \sim \ln (\pTcut/m_H)$ implies that there are NLL terms from clustering at each order in $\alpha_s$.  However, the clustering coefficients $C_{n>2}(R)$ are unknown, and in principle a separate fixed-order calculation is required to obtain each one.  This renders the resummation of the clustering logarithms intractable at present.  In our analysis, we incorporate the known $\ord{\as^2}$ clustering effects, calculate the $\ord{\as^2}$ clustering effects that involve $\ln R^2$ without a $\ln(m_H/\pTcut)$, and include an explicit contribution in our uncertainty estimate for unknown higher-order clustering terms.

The paper is laid out as follows:  In \sec{fact}, we overview how the cross section is computed using SCET and give a summary of the results for each part of the cross section.  In \sec{unc}, we discuss how the perturbative uncertainties are estimated through scale variation and how its various components are combined to estimate the total perturbative uncertainty in the 0-jet cross section, the 0-jet fraction, and the inclusive 1-jet cross section.  In \sec{results}, we present the results of our numerical analysis and our predictions for the LHC for these cross sections.  We conclude in \sec{conclusions}.

\section{\boldmath Factorization with a Jet Algorithm at Small $R$}
\label{sec:fact}

The factorization of the $pp\to H+0$-jet cross section with a jet algorithm has been discussed in Refs.~\cite{Becher:2012qa, Tackmann:2012bt}. For the case of the $p_T$ veto on jets in Higgs production via gluon fusion, the factorized cross section is given by
\begin{align} \label{eq:fact}
&\sigma_0(\pTcut)
\nn \\ & \quad
= \sigma_B H_{gg} (m_t, m_H, \mu) \int\!\df Y
B_g(m_H, \pTcut, R, x_a, \mu, \nu)
\nn \\ & \quad\qquad \times
B_g(m_H, \pTcut, R, x_b, \mu, \nu)\, S_{gg}(\pTcut, R, \mu, \nu)
\nn \\ & \qquad
+ \sigma_0^\Rnons(\pTcut, R) + \sigma_0^\nons(\pTcut, R, \muns)
\,,\end{align}
where
\begin{equation}
x_{a,b} = \frac{m_H}{\Ecm}\,e^{\pm Y}
\,,\quad
\sigma_B = \frac{\sqrt{2} G_F\, m_H^2}{576 \pi \Ecm^2}
\,.\end{equation}
The first term in \eq{fact} provides the leading contribution to the cross section at small $\pTcut$, and contains all the singular logarithmic terms $\alpha_s^i\ln^j(\pTcut/m_H)$. It is factorized into hard, beam, and soft functions, which are discussed below. For instance, the leading double logarithm in \eq{sig0doublelog} is split up as
\begin{equation} \label{eq:logsplit}
\ln^2\!\frac{\pTcut}{m_H}
= \ln^2\!\frac{m_H}{\mu}
+ 2\ln\frac{\pTcut}{\mu}\ln\frac{\nu}{m_H}
+ \ln\frac{\pTcut}{\mu} \ln\frac{\mu\, \pTcut}{\nu^2}
\,,\end{equation}
where the three terms on the right-hand side are contained in the hard, beam, and soft functions, respectively.
In \eqs{fact}{logsplit} $\mu$ is the usual renormalization/factorization scale in virtuality, while $\nu$ denotes the corresponding scale in rapidity~\cite{Chiu:2011qc, Chiu:2012ir}. Hence, we can already see that both invariant mass and rapidity running will be needed to resum the $\ln(\pTcut/m_H)$ terms with renormalization group methods.

The resummation at NNLL$'$ requires determining the functions $H_{gg}$, $B_g$, and $S_{gg}$ to $\ord{\as^2}$, as well as identifying their anomalous dimensions to $\ord{\as^2}$, and their cusp anomalous dimensions to $\ord{\as^3}$.  We present our new two-loop results for the soft and beam functions in \subsecs{factsoft}{factbeam} below, leaving the details of the calculations to a separate publication.

The second term in \eq{fact}, $\sigma_0^\Rnons(\pTcut, R)$, contains $\ord{R^2}$ contributions whose all-orders soft-collinear factorization is challenging and not known at present. In the $R^2\ll 1$ regime, these corrections can formally be treated as subleading power corrections. Numerically, they are indeed very small for the values $R\simeq 0.4$--$0.5$ which are of interest. (As explained in Ref.~\cite{Tackmann:2012bt}, counting $R\sim 1$, they would significantly complicate the soft-collinear factorization already at leading order in the power counting.)
As shown in Ref.~\cite{Banfi:2012jm}, their contribution to the NNLL series is obtained by multiplying them with the same evolution factor as the singular terms, and we will follow this same approach here.  Their contribution to the resummed cross section is discussed in \subsec{factRnons}.

The last term in \eq{fact}, $\sigma_0^\nons(\pTcut, R, \muns)$, contains $\ord{\pTcut/m_H}$ nonsingular corrections, which vanish for $\pTcut\to0$ but become important at large $\pTcut$. These terms are added to the NNLL$'$ result and are required to reproduce the complete NNLO cross section and achieve the full NNLL$'$+NNLO accuracy. Our extraction and analysis of these terms is discussed in \subsec{factnons}.

\subsection{Hard Function}
\label{subsec:facthard}

The hard function, $H_{gg}$, in \eq{fact} is determined by matching QCD onto the gluon fusion operator $O_{ggH}$ in SCET. As discussed in detail in Ref.~\cite{Berger:2010xi}, this matching can be performed as a two-step matching~\cite{Idilbi:2005er, Idilbi:2005ni, Ahrens:2008nc, Mantry:2009qz} or a one-step matching. Since parametrically $m_H/m_t \simeq 1$, we employ the one-step matching, which also makes it easy to include the $m_t$ dependence of the $ggH$ form factor in the matching coefficient $C_{ggH}(m_t, m_H, \mu)$.

The hard matching coefficient satisfies the RG equation
\begin{equation} \label{eq:hardRG}
\frac{\df}{\df\ln\mu} \ln\bigl[C_{ggH} (m_t, m_H, \mu)\bigr] = \gamma_H^g (m_H, \mu)
\,,\end{equation}
where the anomalous dimension has the structure
\begin{equation}
\gamma_H^g (m_H, \mu) = \Gcusp [\as(\mu)] \ln \frac{-m_H^2 - \img 0}{\mu^2} + \gamma_H^g [\as(\mu)]
\,.\end{equation}
The solution of \eq{hardRG} yields the RGE of the matching coefficient from an initial scale $\mu_H$ to some final scale $\mu$,
\begin{align}
&C_{ggH} (m_t, m_H, \mu)
\\\nn & \qquad
= C_{ggH} (m_t, m_H,\mu_H)
\exp\biggl[ \int_{\mu_H}^{\mu}\! \frac{\df\mu'}{\mu'}\, \gamma_H^g (m_H, \mu') \biggr]
\,.\end{align}
The hard function is then given by the absolute value squared of the RG evolved coefficient,
\begin{equation}
H_{gg}(m_t, m_H, \mu) = \bigl\lvert C_{ggH} (m_t, m_H, \mu) \bigr\rvert^2
\,.\end{equation}
For the resummation at NNLL$'$, we require the NNLO result for $C_{ggH}$, the two-loop result for the non-cusp hard anomalous dimension $\gamma_H^g$, and the three-loop result for the gluon cusp anomalous dimension $\Gcusp$~\cite{Vogt:2004mw}. These results as well as the explicit NNLL expression for the evolution factor can be found in App. B of Ref.~\cite{Berger:2010xi}.

To all orders in perturbation theory the matching coefficient contains logarithms of the ratio $(-m_H^2 - \img 0)/\mu_H^2$.  Choosing a real value for the starting scale $\mu_H \sim m_H$ leaves large Sudakov double logarithms $\ln^2(-1-\img 0) = -\pi^2$, leading to a poorly convergent perturbative expansion of the hard function at this scale. Since these terms are associated with the logarithms in the matching coefficient, they can be summed through its RGE by using an imaginary starting scale $\mu_H \simeq -\img m_H$~\cite{Parisi:1979xd, Sterman:1986aj, Magnea:1990zb}. In this way, the double logarithms are fully minimized, leading to a much better perturbative convergence~\cite{Ahrens:2008qu, Ahrens:2008nc}. To illustrate this numerically, for $m_H = 125\GeV$ we find
\begin{align}
\frac{H_{gg}(\mu_H = m_H)}{H_{gg}^{(0)}(\mu_H = m_H)} &= 1 + 0.815 + 0.356 + \dotsb
\,,\nn\\
\frac{H_{gg}(\mu_H = -\img m_H)}{H_{gg}^{(0)}(\mu_H = -\img m_H)} &= 1 + 0.274 + 0.042 + \dotsb
\,,\end{align}
where $H_{gg}^{(0)}$ is the lowest-order result in each case, and the second and third numbers on the right-hand side give the NLO and NNLO corrections, respectively. The substantial improvement in convergence also implies reduced perturbative uncertainties. We therefore use the imaginary hard scale as the default choice in our numerical results.

\subsection{Soft Function}
\label{subsec:factsoft}

The soft function describes the soft radiation across the entire event.  It is defined as a forward scattering matrix element of soft Wilson lines along the two incoming beam directions, with the jet-veto measurement on the final state,
\begin{equation}
S_{gg}(\pTcut, R, \mu, \nu)
= \langle 0 \lvert Y_{n_b} \, Y_{n_a}^{\dagger} \, \cM^\jet(\pTcut, R) \, Y_{n_a} \, Y_{n_b}^\dagger \rvert 0 \rangle
\,.\end{equation}
Here, the measurement function $\cM^\jet(\pTcut, R)$ acts on the soft final state by clustering it into jets of radius $R$ and requiring that all these jets have $p_T < \pTcut$.  This local veto on individual jets can be divided into a global veto and a local correction from the jet algorithm clustering, consequently dividing the soft function into a global term and a jet algorithm correction%
\footnote{Technically, this division into global and clustering contributions is affected by the fact that non-Abelian exponentiation occurs for the soft function, and only specifies how the genuinely new terms at each perturbative order are divided. Since the first nontrivial clustering correction only arises at $\ord{\as^2}$, \eq{DeltaSjet} holds for the soft function through NNLO.  The exponentiation of lower-order results will mix global and clustering contributions at higher orders in the soft function.},%
\begin{equation} \label{eq:DeltaSjet}
S_{gg} (\pTcut, R, \mu, \nu) = S^G_{gg} (\pTcut, \mu, \nu) + \Delta S^\jet_{gg} (\pTcut, R, \mu, \nu)
\,.\end{equation}
This isolates the jet algorithm effects into $\Delta S^\jet_{gg}$, which makes them easier to compute and analyze their resummation properties. Note that these jet algorithm corrections are defined relative to the chosen global veto, while the full soft function on the left-hand side is uniquely defined by specifying the jet-veto measurement, $\cM^\jet(\pTcut, R)$.  At $\ord{\as^2}$, where the clustering corrections are first nonzero, the two-particle phase space constraints of the anti-$\kt$ algorithm are identical to other $\kt$-type jet algorithms, which include $\kt$ and Cambridge-Aachen \cite{Catani:1991hj,Catani:1993hr,Ellis:1993tq,Dokshitzer:1997in}.  This is also true for the jet algorithm effects in the beam function, and thus our calculation does not distinguish between these jet algorithms at the order to which we work.

The soft and beam functions separately contain rapidity divergences. When they are combined in the cross section, the rapidity divergences cancel, leaving large ``rapidity logarithms'' $\ln(\pTcut/m_H)$ at fixed order. We employ the rapidity renormalization group~\cite{Chiu:2011qc, Chiu:2012ir}, which allows one to apply standard effective theory and RG methods to regulate and renormalize the rapidity divergences and perform the resummation of the associated rapidity logarithms. It introduces an arbitrary rapidity renormalization scale $\nu$, whose role in the rapidity RGE is the same as that of the usual renormalization scale $\mu$ in the standard virtuality RGE.

In our case, the soft function is multiplicatively renormalized in both $\mu$ and $\nu$,
\begin{align} \label{eq:softRG}
\frac{\df}{\df\ln\mu} \ln S_{gg}(\pTcut, R, \mu, \nu) &= \gamma_S^g(\mu, \nu)
\,, \nn \\
\frac{\df}{\df\ln\nu} \ln S_{gg}(\pTcut, R, \mu, \nu) &= \gamma_\nu^g(\pTcut, R, \mu)
\,.\end{align}
The anomalous dimensions have the general structure~\cite{Tackmann:2012bt}
\begin{align} \label{eq:softanomdim}
\gamma_S^g(\mu, \nu)
&= 4\Gcusp [\as(\mu)] \ln \frac{\mu}{\nu} + \gamma_S^g[\as(\mu)]
\,, \nn \\
\gamma_\nu^g(\pTcut, R, \mu) &= -4\eta_\Gamma^g(\pTcut,\mu) + \gamma_\nu^g[\as(\pTcut), R]
\,,\end{align}
where
\begin{equation}
\eta^g_{\Gamma}(\mu_0, \mu)
= \int_{\mu_0}^\mu\! \frac{\df\mu'}{\mu'}\, \Gcusp[\as(\mu')]
= \Gcusp \ln\frac{\mu}{\mu_0} + \dotsb
\end{equation}
sums an all-orders set of terms in the anomalous dimension that are determined by the RG consistency.
(They are required to ensure the exact path independence of the evolution in the two-dimensional $\mu$-$\nu$ space~\cite{Chiu:2012ir}.)
The RGE of the soft function is obtained by solving \eq{softRG}. Evolving first in rapidity and then in virtuality, we have
\begin{align}
& S_{gg}(\pTcut, R, \mu, \nu)
\nn\\ & \quad
= S_{gg}(\pTcut, R, \mu_S, \nu_S)
\exp \biggl[\ln\frac{\nu}{\nu_S}\, \gamma_\nu^g(\pTcut, R, \mu_S) \biggr]
\nn\\ & \qquad \times
\exp \biggl[ \int_{\mu_S}^\mu\! \frac{\df\mu'}{\mu'}\, \gamma_S^g(\mu', \nu) \biggr]
\,.\end{align}

We have calculated the complete soft function to $\ord{\as^2}$, which to our knowledge is the first two-loop calculation employing the rapidity renormalization. Our result for the perturbative soft function through $\ord{\as^2}$ is
\begin{align} \label{eq:softFO}
& S_{gg} (\pTcut, R, \mu_S, \nu_S) =
\nn \\ & \quad
1 + \frac{\as(\mu_S)}{4\pi} \Bigl[ 2\Gamma_0^g L_S^\mu \bigl( L_S^\mu - 2 L_S^\nu) - \frac{\pi^2}{3} C_A \Bigr]
\nn \\ & \qquad
+ \frac{\as^2(\mu_S)}{(4\pi)^2}\, \biggl\{
\frac{1}{2} \Bigl[ 2\Gamma_0^g L_S^\mu \bigl( L_S^\mu - 2 L_S^\nu) - \frac{\pi^2}{3} C_A \Bigr]^2
\nn \\ & \qquad
+ 2\beta_0 L_S^\mu \Bigl[2\Gamma_0^g L_S^\mu\Bigl(\frac{1}{3} L_S^\mu - L_S^\nu\Bigr) - \frac{\pi^2}{3} C_A \Bigr]
\nn \\ & \qquad
+ 2\Gamma_1^g L_S^\mu (L_S^\mu - 2 L_S^\nu)
\nn \\ & \qquad
+ \gamma_{S\,1}^g L_S^\mu + \gamma_{\nu\, 1}^g(R)\, L_S^\nu + s_2(R) \biggr\}
\,,\end{align}
where we abbreviated
\begin{equation}
L_S^\mu \equiv \ln \frac{\mu_S}{\pTcut} \,, \qquad L_S^\nu \equiv \ln \frac{\nu_S}{\pTcut}
\,.\end{equation}
Hence, the natural soft scales for which the large logarithms in the soft function are minimized are $\mu_S\sim \pTcut$ and $\nu_S\sim\pTcut$.

In \eq{softFO} and in the following, the $\beta$ function and anomalous dimensions are expanded as
\begin{align}
\beta(\alpha_s) &= -2\as \sum_{n=0}^\infty \beta_n \Bigl(\frac{\as}{4\pi}\Bigr)^{n+1}
\,,\nn\\
\gamma(\alpha_s) &= \sum_{n=0}^\infty \gamma_n \Bigl(\frac{\as}{4\pi}\Bigr)^{n+1}
\,,\end{align}
where the coefficients needed in \eq{softFO} are
\begin{align}
\beta_0 &= \frac{11}{3}\,C_A -\frac{4}{3}\,T_F\,n_f
\,,\nn\\
\beta_1 &= \frac{34}{3}\,C_A^2  - \Bigl(\frac{20}{3}\,C_A\, + 4 C_F\Bigr)\, T_F\,n_f
\,,\nn\\
\Gamma_0^g &= 4C_A
\,,\nn\\
\Gamma_1^g &= 4C_A \Bigl[\Bigl( \frac{67}{9} -\frac{\pi^2}{3} \Bigr)\,C_A  - \frac{20}{9}\,T_F\, n_f \Bigr]
\,,\end{align}
and $C_A = 3$, $C_F = 4/3$, $T_F = 1/2$, and $n_f = 5$ is the number of light quark flavors.  The coefficients $\beta_2$ and $\Gamma_2^g$ are also used in the NNLL resummation.

At one loop, the non-cusp soft and rapidity anomalous dimensions vanish,
\begin{equation}
\gamma_{S\,0}^g = 0
\,,\qquad\
\gamma_{\nu\,0}^g(R) = 0
\,.\end{equation}
The dependence on the jet algorithm starts to enter at two loops through the two-loop $\nu$ anomalous dimension, $\gamma_{\nu\, 1}^g(R)$, which determines the coefficient of the single logarithm of $\ln(\nu/\pTcut)$, as well as the nonlogarithmic two-loop soft constant, $s_2(R)$. For the two-loop coefficients of the non-cusp anomalous dimensions we find
\begin{align} \label{eq:gammaS1munu}
\gamma_{S\, 1}^g
&= 8 C_A \biggl[ \Bigl(\frac{52}{9} - 4 (1+\pi^2) \ln 2 + 11 \zeta_3\Bigr) C_A
\nn\\ & \quad
+ \Bigl(\frac{2}{9} + \frac{7 \pi^2}{12} - \frac{20}{3} \ln 2 \Bigr) \beta_0 \biggr]
\nn\\
&= 16 C_A^2\, (-3.83)
\,, \nn \\
\gamma_{\nu\, 1}^g(R)
&= -16 C_A \biggl[\Bigl(\frac{17}{9} - (1 + \pi^2)\ln2 + \zeta_3 \Bigl) C_A
\nn\\ & \quad
+ \Bigl(\frac{4}{9} +\frac{\pi^2}{12} - \frac{5}{3}\ln2 \Bigl) \beta_0 \biggr] + C_2(R)
\nn\\
&= 16 C_A^2\, (4.16) + C_2(R)
\,.\end{align}
Here, $C_2(R)$ is the clustering correction due to the jet algorithm, and was computed earlier in Ref.~\cite{Tackmann:2012bt}. It is given by
\begin{align} \label{eq:C2value}
C_2(R) &=
 2 C_A \Bigl[\Bigl(1 - \frac{8\pi^2}{3}\Bigr) C_A + \Bigl(\frac{23}{3} - 8\ln2\Bigr)\beta_0\Bigl] \ln R^2
\nn\\ & \quad
+ 15.62 C_A^2 - 9.17 C_A \beta_0 + C_2^\Rnons(R)
\nn\\[1ex]
&= 16 C_A^2\, \bigl( - 2.49 \ln R^2 - 0.49 \bigr) + \ord{R^2}
\,,\end{align}
where $C_2^\Rnons(R) \sim \ord{R^2}$ contains all subleading power corrections in $R^2$.
Note that we define the clustering effects in $C_2(R)$ relative to the global $E_T$ veto. A different choice, such as the $p_T$ of the Higgs used in Ref.~\cite{Banfi:2012yh}, would give a different $R$-independent constant in $C_2(R)$. Nevertheless, the full result for $\gamma^g_{\nu 1}(R)$ is independent of this choice and our final NNLL cross section agrees with that of Ref.~\cite{Banfi:2012yh}.

For the two-loop soft function constant $s_2(R)$, which is not determined from RGE constraints, we find
\begin{align} \label{eq:s2R}
s_2 (R) &=
C_A \biggl[ \Bigl(\frac{19}{3} - 10 \ln 2 + 8 \zeta_3 \Bigr) C_A
\nn\\ & \qquad
+ \Bigl(-\frac{163}{9} + \frac{58}{3}\ln2 + 8 \ln^2 2 \Bigr) \beta_0 \biggr] \ln R^2
\nn\\ & \quad
-18.68 C_A^2 - 3.25 C_A \beta_0 + s_2^\Rnons(R)
\nn\\[1ex]
&= 16 C_A^2\, \bigl(0.43 \ln R^2 - 1.69 \bigr) + \ord{R^2}
\,,\end{align}
where $s_2^\Rnons(R) \sim R^2$. This result for $s_2(R)$ is new and also constitutes the first calculation of the $\pTcut$ independent clustering terms in the soft function.

\begin{figure}[t!]
\begin{center}
\includegraphics[width=\columnwidth]{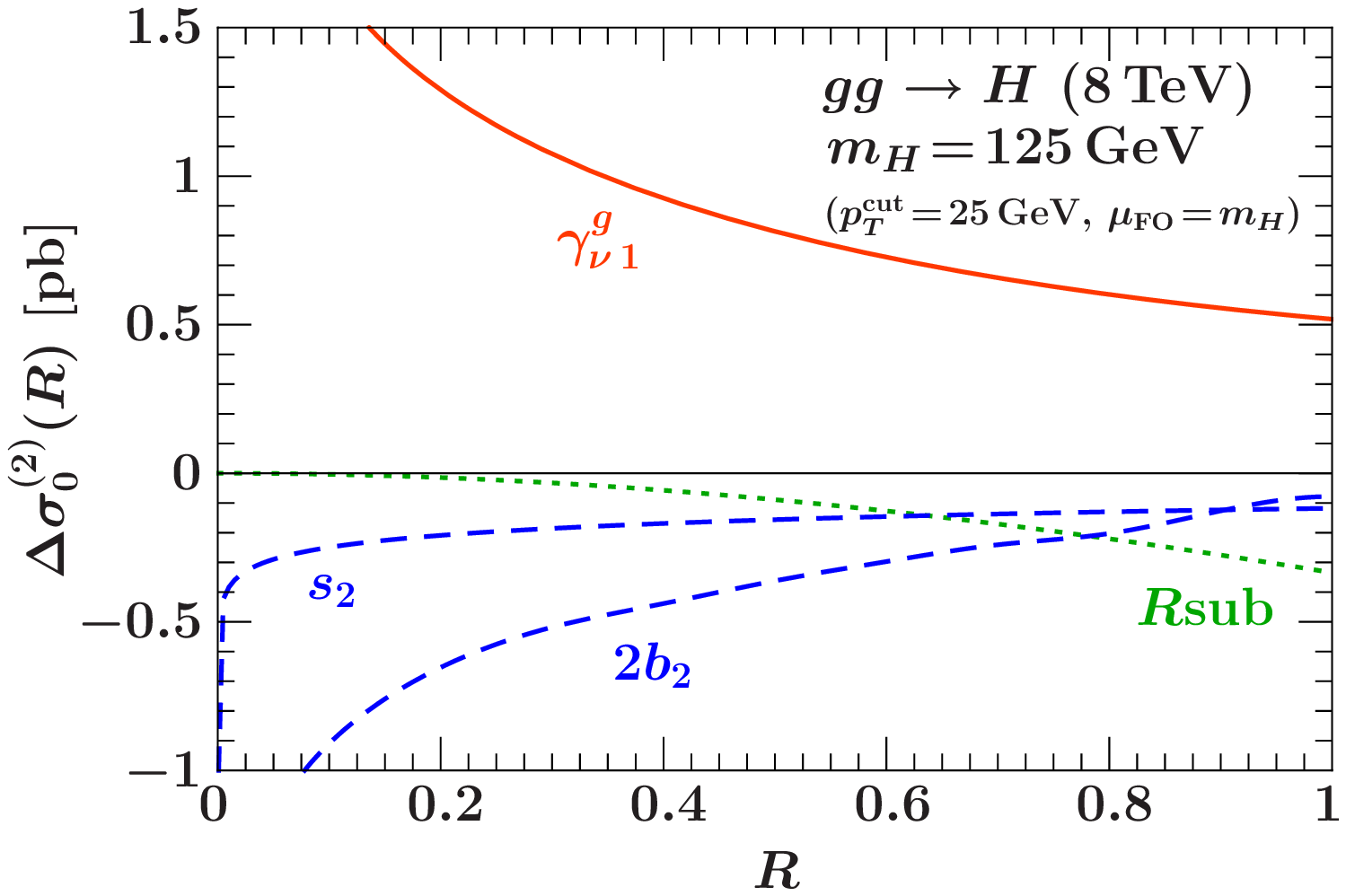}%
\end{center}
\vspace{-3ex}
\caption{Jet-algorithm dependent $\ord{\as^2}$ contributions to the fixed NNLO cross section from different sources, for $\mu_\FO = m_H$ and $\pTcut = 25 \GeV$.  The $\nu$ anomalous dimension coefficient $\gamma_{\nu\,1}^g$ is given in \eq{gammaS1munu}, the $\ord{\as^2}$ soft function constant terms in \eq{s2R}, the beam function constant terms in \eq{Igj2const} and the following paragraph, and the clustering effects on uncorrelated emissions in \eq{sigma0RnonsFO}.}
\label{fig:2looppieces}
\end{figure}

The terms not proportional to $\ln R^2$ in $C_2(R)$ and $s_2(R)$ involve complicated phase-space integrals, which are computed numerically.
The contributions of $\gamma_{\nu\,1}^g(R)$ and $s_2(R)$ to the fixed NNLO cross section including their full $R$ dependence are shown in \fig{2looppieces}.

As mentioned above, the jet algorithm corrections in the soft function start at $\ord{\as^2}$. They have the all-order structure
\begin{align}
&\Delta S^{\jet}_{gg}(\pTcut, R, \mu_S, \nu_S)
\nn\\ & \qquad
= \sum_{n\geq 2} \frac{\as^n(\mu_S)}{(4\pi)^n} \Bigl[ C_n(R) \ln\frac{\nu_S}{\pTcut} + \Delta s_n (R) \Bigr]
\,,\end{align}
where $C_n (R)$ and $\Delta s_n(R)$ contain up to $n - 1$ powers of $\ln R^2$. The $C_n(R)$ in the soft function are the same as in \eq{sigmaclus} for the cross section. The beam functions contain an equivalent set of terms $\sim \alpha_s^n C_n(R) \ln(m_H/\nu_B)$. In the fixed-order cross section (i.e. for $\nu_B = \nu_S = \nu$) they combine with the soft function terms to give the total clustering correction $\sim \alpha_s^n C_n(R) \ln(m_H/\pTcut)$ in \eq{sigmaclus}. For $R^2 \sim \pTcut/m_H$, the leading $\ln^{n-1} R^2$ terms in $C_n(R)$ formally count as NLL in the exponent of the cross section. Similarly, the leading $\ln^{n-1} R^2$ terms in $\Delta s_n(R)$, as well as the $\ln^{n-2} R^2$ terms in $C_n(R)$, formally count as NNLL. The anomalous dimension $\gamma_\nu^g(R)$ includes the $C_n(R)$, so its perturbative series explicitly contains the $\ln R^2$ terms, which means that the NLL and higher logarithmic series from $\ln R^2$ clustering corrections are not resummed here. A formalism for this resummation is not currently known. Since these clustering corrections are numerically large at $\ord{\as^2}$, we perform an estimate of the potential size of the higher-order clustering effects as part of our uncertainty analysis.

\subsection{Beam Function}
\label{subsec:factbeam}

The beam function is defined as the forward proton matrix element of collinear gluon fields. It provides a combined description of collinear initial-state radiation from the incoming gluons together with their extraction from the colliding protons via the nonperturbative parton distribution functions (PDFs)\cite{Stewart:2009yx}.

Like the soft function, the beam function is multiplicatively renormalized in both $\mu$ and $\nu$,
\begin{align} \label{eq:beamRG}
\frac{\df}{\df\ln\mu} \ln B_g(m_H, \pTcut, R, x, \mu, \nu)
&= \gamma_B^g(m_H, \mu, \nu)
\,, \\\nn
\frac{\df}{\df\ln\nu} \ln B_g(m_H, \pTcut, R, x, \mu, \nu)
&= -\frac{1}{2} \gamma_\nu^g(\pTcut, R, \mu)
\,.\end{align}
The anomalous dimensions can be determined from those of the hard and soft functions using the consistency of the factorization theorem. The $\nu$ anomalous dimension, $\gamma_\nu^g$, is the same as in \eq{softanomdim}. The $\mu$ anomalous dimension is given by
\begin{align}
\gamma_B^g(m_H,\mu, \nu)
&= 2 \Gcusp [\as(\mu)] \ln\frac{\nu}{m_H} + \gamma_B^g[\as(\mu)]
\,,\nn\\
\gamma_B^g(\as) &= -\gamma_H^g(\as) - \frac{1}{2} \gamma_S^g(\as)
\,,\end{align}
with the resulting one-loop and two-loop coefficients
\begin{align}
\gamma_{B\,0}^g &= 2\beta_0
\,,\nn\\
\gamma_{B\,1}^g &= 2 \beta_1
+ 8C_A \biggl[ \Bigl(-\frac{5}{4} + 2(1+\pi^2) \ln2 - 6 \zeta_3 \Bigr) C_A
\nn\\ & \quad
+ \Bigl(\frac{5}{24} - \frac{\pi^2}{3} + \frac{10}{3}\ln2 \Bigr) \beta_0 \biggr]
\,.\end{align}

The RGE of the beam function follows from solving \eq{beamRG}, and is analogous to that of the soft function,
\begin{align}
&B_g(m_H, \pTcut, R, x, \mu, \nu)
\nn\\ & \qquad
= B_g(m_H, \pTcut, R, x, \mu_B, \nu_B)
\nn\\ & \quad\qquad\times
\exp \biggl[\frac{1}{2} \ln\frac{\nu_B}{\nu}\, \gamma_\nu^g(\pTcut, R, \mu_B) \biggr]
\nn\\ & \quad\qquad\times
\exp \biggl[\int_{\mu_B}^\mu\! \frac{\df\mu'}{\mu'}\, \gamma_B^g(m_H, \mu', \nu) \biggr]
\,.\end{align}
Note that in contrast to the PDF evolution, the evolution of the beam function does not change its value of $x$. This is a general feature of beam functions and is due to the fact that their evolution describes the initial-state radiation from an incoming parton that is not confined to the proton anymore, while the PDF evolution is frozen out at the beam scale $\mu_B$~\cite{Stewart:2009yx, Stewart:2010qs}.

At the beam scale, the gluon beam function can be computed as a convolution between perturbative matching kernels, $\cI_{gj} (m_H, \pTcut, z, \mu_B, \nu_B)$, and the standard quark and gluon PDFs, $f_j (x, \mu_B)$,
\begin{align}
& B_g(m_H, \pTcut, R, x, \mu_B, \nu_B)
\\\nn & \qquad
= \sum_j \int_x^1\! \frac{\df z}{z}\, \cI_{gj} (m_H, \pTcut, R, z, \mu_B, \nu_B)\, f_j \Bigl(\frac{x}{z}, \mu_B\Bigr)
\,.\end{align}
We expand the matching kernels $\cI_{gj}$ to $\ord{\as^2}$ as (suppressing the arguments for brevity)
\begin{align}
\cI_{gj}
= \delta_{gj} \delta(1-z) + \frac{\as(\mu_B)}{4\pi}\, \cI_{gj}^{(1)} + \frac{\as^2(\mu_B)}{(4\pi)^2}\, \cI_{gj}^{(2)} + \ord{\as^3}
\,.\end{align}
The $\ord{\as}$ coefficients are common to several observables, and we agree with the calculation of $\cI_{gg}^{(1)}$ using the rapidity regulator in Ref.~\cite{Chiu:2012ir}. We find,
\begin{align}
\cI_{gg}^{(1)}(m_H, \pTcut, z, \mu_B, \nu_B)
&= 4C_A L_B^\mu 
\nn \\ & \quad\times
\bigl[2L_B^\nu \delta(1-z) - P_{gg}(z) \bigr]
\,, \nn \\
\cI_{gq}^{(1)}(m_H, \pTcut, z, \mu_B, \nu_B)
&= 2C_F \bigl[-2 L_B^\mu P_{gq}(z) + I_{gq}^{(1)}(z) \bigr]
\,, \nn \\
I_{gq}^{(1)}(z) &= z
\,,\end{align}
where we abbreviated
\begin{equation}
L_B^{\mu} \equiv \ln \frac{\mu_B}{\pTcut}
\,, \qquad
L_B^{\nu} \equiv \ln \frac{\nu_B}{m_H}
\,.\end{equation}
The natural scales for the beam function are thus $\mu_B\sim\pTcut$ and $\nu_B\sim m_H$. Our results for $\cI_{gj}^{(1)}$ agree with Ref.~\cite{Becher:2012qa}, after taking into account the different rapidity regularization.

The $\ord{\as^2}$ kernel for the $gg$ contribution is given by
\begin{align}
&\cI_{gg}^{(2)}(m_H, \pTcut, R, z, \mu_B, \nu_B)
\nn\\ & \quad
= 32 C_A^2 (L_B^\mu)^2 L_B^\nu \bigl[L_B^\nu \delta(1-z) - P_{gg} (z) \bigr]
\nn \\ & \qquad
+ 4 C_A \beta_0 (L_B^\mu)^2 \bigl[2L_B^\nu \delta(1-z) - P_{gg} (z) \bigr]
\nn \\ & \qquad
+ 8 (L_B^\mu)^2 \bigl[C_A^2 (P_{gg} \otimes P_{gg})(z)
\nn \\ & \qquad\quad
+ 2 C_F T_F n_f (P_{gq} \otimes P_{qg}) (z) \bigr]
\nn \\ & \qquad
- 8 L_B^\mu \bigl[P_{gg}^{(1)}(z) + 2C_F T_F n_f (I_{gq}^{(1)} \otimes P_{qg}) (z) \bigr]
\nn \\ & \qquad
+ \Bigl[L_B^\mu \bigl(2\Gamma_1^g L_B^\nu + \gamma_{B\,1}^g \bigr) - \frac{1}{2} \gamma_{\nu\,1}^g(R) L_B^\nu \Bigr] \delta(1-z)
\nn \\ & \qquad
+ I_{gg}^{(2)} (z, R)
\,.\end{align}
The $\ord{\as^2}$ kernel for the $gq$ contribution is given by
\begin{align}
&\cI_{gq}^{(2)}(m_H, \pTcut, R, z, \mu_B, \nu_B)
\nn\\ & \quad
= 16C_A C_F L_B^\mu L_B^\nu \bigl[ -2 L_B^\mu P_{gq} (z) + I_{gq}^{(1)}(z) \bigr]
\nn \\ & \qquad
+ 8 C_F \beta_0 L_B^\mu \bigl[- L_B^\mu P_{gq}(z) + I_{gq}^{(1)} (z) \bigr]
\nn \\ & \qquad
+ 8 C_F (L_B^\mu)^2 \bigl[C_A (P_{gg} \otimes P_{gq})(z)
\\ & \qquad\quad
+ C_F (P_{gq} \otimes P_{qq})(z) \bigr]
\nn \\\nn & \qquad
-8 L_B^\mu \bigl[P_{gq}^{(1)} (z) + C_F^2 (I_{gq}^{(1)} \otimes P_{qq})(z) \bigr]
+ I_{gq}^{(2)} (z, R)
\,.\end{align}
The convolutions $(g \otimes h) (z)$ are defined as
\begin{equation}
(g\otimes h)(z) \equiv \int_z^1\! \frac{\df\xi}{\xi}\, g\Bigl(\frac{z}{\xi}\Bigr) h(\xi)
\,.\end{equation}
The various splitting functions $P_{ij}(z)$ and convolutions between them are given in App. B2 of Ref.~\cite{Berger:2010xi}.
The additional convolutions we need are
\begin{align}
(I_{gq}^{(1)} \otimes P_{qg})(z)
&= 1 + z - 2 z^2 + 2 z \ln z
\,, \\
(I_{gq}^{(1)} \otimes P_{qq})(z)
&= 1 + \frac{z}{2} - z \ln z + 2 z \ln(1-z)
\,. \nn \end{align}

The terms involving logarithms of $\mu$ and $\nu$ in the $\cI_{gj}$ kernels are fully determined by renormalization group (RG) constraints. The nonlogarithmic terms $I_{gi}^{(2)}(z, R)$ require the full two-loop calculation of the beam functions.
Note that the full two-loop $qq$ contribution to the beam function for the transverse momentum of the vector boson has been computed recently in Ref.~\cite{Gehrmann:2012ze}. At two loops, the $p_T^\jet$ beam function needed here is different and requires a separate calculation. Like the soft function, it receives both global and jet clustering contributions. In particular, we can calculate directly the leading clustering corrections proportional to $\ln R^2$, and determine the contribution from the remaining terms numerically, giving
\begin{align} \label{eq:Igj2const}
I_{gg}^{(2)}(z, R)
&=
\frac{C_A}{2} \biggl[\Bigl(1 - \frac{8\pi^2}{3}\Bigr) C_A
+ \Bigl(\frac{23}{3} - 8 \ln2 \Bigr) \beta_0 \biggr]
\nn\\ & \quad
\times P_{gg} (z) \ln R^2
+ I_{gg}^{(2,c)}(z) + I_{gg}^{(2,\Rnons)}(z, R)
\,, \nn \\
I_{gq}^{(2)}(z, R)
&= 2 C_F^2 \Bigl(3 - \frac{\pi^2}{3} - 3\ln 2 \Bigr) P_{gq}(z) \ln R^2
\nn\\ & \quad
+ I_{gq}^{(2,c)}(z) + I_{gq}^{(2,\Rnons)}(z, R) \,.
\end{align}
Here, $I_{gg}^{(2,c)}(z)$ denotes the constant $R$ independent terms, while $I_{gg}^{(2,\Rnons)}(z, R)$ are the $\ord{R^2}$ suppressed contributions.  Their explicit form is not known at present. We extract their total contribution after convolution with the PDFs numerically from the fixed-order cross section as explained in \subsec{factnons} below. This is sufficient for practical purposes, since their effect is found to be numerically small compared to the $\ln R^2$ terms for $R\sim 0.4$--$0.5$. The total contribution (from both beam functions) of the full $I_{gg}^{(2)}(z,R)$ and $I_{gq}^{(2)}(z,R)$ to the fixed NNLO cross section is shown by the blue dashed line in \fig{2looppieces} that is labeled as $2b_2$.

\subsection{\boldmath $\ord{R^2}$ Corrections From Uncorrelated Emissions}
\label{subsec:factRnons}

Starting at $\ord{\as^2}$, the clustering effects from the jet algorithm includes contributions that scale as powers of $R^2$ in the small $R$ limit. Clustering effects from correlated emissions in the soft or collinear sectors are included in the subleading $\ord{R^2}$ corrections in the soft and beam functions. On the other hand, the clustering of uncorrelated emissions from the soft and collinear beam sectors inhibits the factorization of the jet-veto measurement into independent soft and collinear measurements at $\ord{R^2}$. The all-order factorization of the cross section at this level is therefore not known at present.%
\footnote{The statement in Ref.~\cite{Becher:2012qa} that soft-collinear mixing is absent at leading power for $R\sim 1$ relies on a power counting for collinear rapidities ($y_c$) and soft rapidities ($y_s$) where $y_c \gg y_s \sim \ord{1}$ such that $y_c - y_s \gg R \sim 1$. For typical values of $p_T = 25\GeV$ and $Q = 125\GeV$ there is a legitimate power expansion in $\lambda = p_T/Q = 0.2 \ll 1$. But this gives $y_c \simeq \ln(1/\lambda) = 1.6$, which does not clearly satisfy $y_c\gg y_s \sim 1$.  Indeed, physically, emissions at fixed $p_T$ tend to be uniform in rapidity rather than having a rapidity gap between soft and collinear regions. (The analogous statement using light-cone variables is $e^{-R} \gg  e^{y_s - y_c} = \sqrt{(k_s^-/k_s^+)(p_c^+/p_c^-)} \sim \ord{1}\times\lambda$. For $R = 1$, this corresponds to counting $0.37\gg \ord{1}\times\lambda =\ord{1}\times 0.2$.) As discussed in detail in Ref.~\cite{Tackmann:2012bt}, the contribution from clustering a soft and a collinear emission is $\sim\!R^2$, so the only way to expand it to zero is $R^2 \ll 1$.   \\
\indent
The fact that soft and collinear modes in SCET-II are only distinguished by their rapidity does not automatically imply that their rapidities are parametrically widely separated as $y_s \ll y_c$, since in practice amplitudes from each of these modes are integrated over all rapidities and we must worry about contributions from overlapping regions. If there is a double counting for infrared singularities from the overlap region then this is removed by 0-bin subtractions~\cite{Manohar:2006nz}, but in general these subtractions do not suffice to remove finite contributions from the overlap region. Thus a proof of factorization at ${\cal O}(R^2)$, including also soft-collinear mixing contributions, will require additional arguments to all orders in $\alpha_s$, and remains an interesting open question.
}

The full contribution from clustering of uncorrelated emissions to the fixed NNLO cross section is~\cite{Banfi:2012yh}
\begin{equation} \label{eq:sigma0RnonsFO}
\sigma_0^{(2)}(\pTcut)
\supset \sigma_\LO \Bigl(\frac{\as C_A}{\pi}\Bigl)^2 \ln\frac{m_H}{\pTcut} \Bigl( -\frac{\pi^2}{3} R^2 + \frac{R^4}{4} \Bigr)
\,.\end{equation}
It is shown by the green dotted line in \fig{2looppieces} for $\pTcut = 25\GeV$. As one can see, at the $R$ values of interest it is numerically very small compared to the corresponding $\ln R^2$ enhanced clustering corrections contained in $\gamma_{\nu 1}^g(R)$, and can thus safely be treated as a power correction.

As argued in Refs.~\cite{Banfi:2012yh, Banfi:2012jm}, the above $\ord{\alpha_s^2}$ coefficient determines the complete NNLL series coming from this contribution [i.e. no new coefficients appear at $\ord{\as^3 L^2}$ or higher]. Therefore, we can include this correction in the resummed cross section at NNLL by multiplying it with the total evolution factor as follows,
\begin{align} \label{eq:sigma0Rnons}
\sigma_0^\Rnons(\pTcut, R)
&= \frac{\as^2(\sqrt{\mu_B \mu_S})}{\pi^2}C_A^2 \ln\frac{m_H}{\pTcut} \Bigl( -\frac{\pi^2}{3} R^2 + \frac{R^4}{4} \Bigr)
\nn\\ & \quad\times
[F^{(0)} U_0](\mu_H, \mu_B, \mu_S, \nu_B, \nu_S)
\,.\end{align}
Here, $F^{(0)}$ denotes the leading fixed-order contributions from the hard, beam, and soft functions, and $U_0$ is their combined NNLL evolution factor [given explicitly in \eq{U0} below]. Since these corrections come from soft or collinear emissions we choose to evaluate the argument of the $\alpha_s^2$ in the prefactor at the geometric mean of the beam and soft scales.

In Ref.~\cite{Becher:2012qa} this coefficient is absorbed into the two-loop rapidity anomalous dimension, which amounts to writing this contribution as $A \exp(\as^2)$, instead of $A(1+\as^2)$ as in Ref.~\cite{Banfi:2012jm}. Since this contribution first appears at $\ord{\as^2}$, either form gives the same NNLL contribution and the difference is higher order, meaning the results of Refs.~\cite{Banfi:2012yh, Banfi:2012jm} do not determine which is the correct all-order structure beyond NNLL.

\subsection{Nonsingular Contributions}
\label{subsec:factnons}

\begin{figure*}[t!]
\begin{center}
\includegraphics[width=\columnwidth]{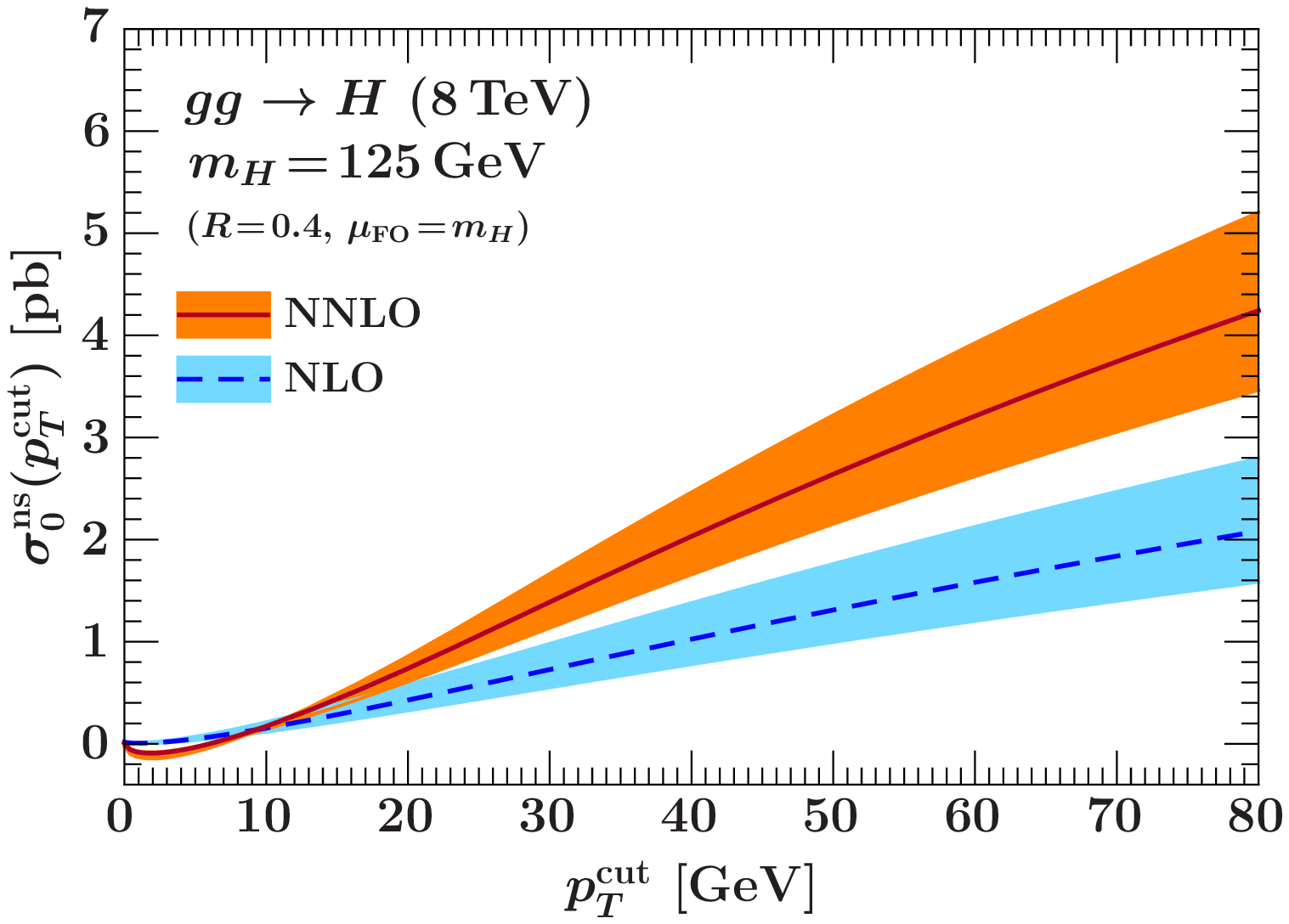}%
\hfill%
\includegraphics[width=\columnwidth]{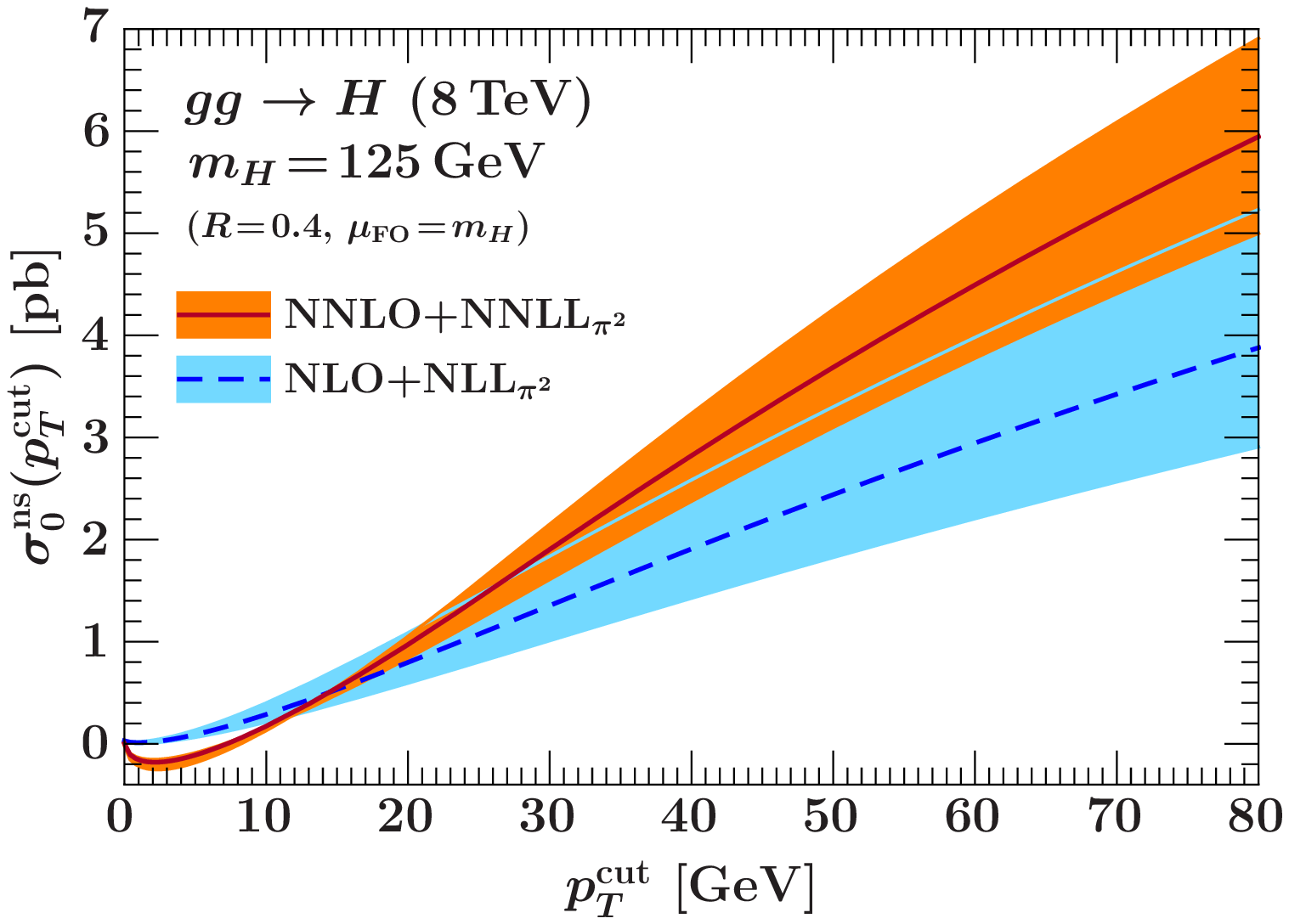}%
\end{center}
\vspace{-3ex}
\caption{The nonsingular cross section at $\muns = m_H$ at NLO (blue, dashed) and NNLO (orange, solid) for $R = 0.4$.  We compare the pure fixed-order nonsingular terms (on the left) with the nonsingular terms that include $\pi^2$ summation (on the right). The latter shows a substantially improved perturbative convergence from NLO to NNLO.}
\label{fig:nonsingularplot}
\end{figure*}

In fixed-order perturbation theory, the cross section at $\mu_f = \mu_r = \mu_\FO$ has the all-order structure
\begin{align}
\sigma_0^\FO(\pTcut, \mu_\FO) &= \sigma_0^{\sing}(\pTcut, \mu_\FO) + \sigma_0^{\nons}(\pTcut, \mu_\FO)
\,,\\\nn
\sigma_0^{\sing}(\pTcut, \mu_\FO) &= \sum_m\sum_{n \leq 2m} \!\!\! c_{mn}(\mu_\FO)\, \as^m(\mu_\FO) \ln^n\frac{\pTcut}{m_H}
\,.\end{align}
Here, the singular cross section, $\sigma_0^\sing$, contains all terms that are nonzero for $\pTcut\to 0$ and which are contained in the resummed cross section. The nonsingular cross section, $\sigma_0^\nons$, scales as $\ord{\pTcut/m_H}$ and vanishes for $\pTcut\to 0$. To reproduce the full fixed-order cross section we have to include the nonsingular terms, in particular when going to large $\pTcut$ where they become important.

An important feature of the NNLL$'$ (NLL$'$) resummed result is that by construction its fixed-order expansion to NNLO (NLO) in terms of $\alpha_s(\mu_\FO)$ can be obtained by simply setting all scales equal to $\mu_\FO$. And this also precisely reproduces the fixed-order singular contributions. Hence, we can determine the nonsingular corrections by subtracting the latter from the full fixed-order cross section,
\begin{align}
\sigma_0^{\nons}(\pTcut, R, \mu_\FO)
&= \sigma_0^\FO(\pTcut, R, \mu_\FO)
\\\nn & \quad
- \sigma_0^{\resum'}(\pTcut, R, \mu_i = \nu_i = \mu_\FO)
\,.\end{align}

At NLO, this procedure is straightforward since the one-loop hard, beam, and soft functions required at NLL$'$ are completely known, while $\sigma_0^\NLO(\pTcut, \mu_\FO)$ is easily obtained numerically e.g. from MCFM.

At NNLO, we obtain the full fixed-order cross section by subtracting the NLO $gg\to H+j$ cross section for a leading jet with $p_T^\jet > \pTcut$, obtained using MCFM~\cite{Campbell:2002tg, Catani:2007vq}, from the total NNLO cross section~\cite{Harlander:2002wh, Anastasiou:2002yz, Ravindran:2003um}. For the resummed NNLL$'$ cross section we include all available contributions through $\ord{\as^2}$ summarized in the previous subsections, including the $\sigma_0^\Rnons$ terms in \eq{sigma0Rnons}. The only missing pieces at two loops are the unknown $I_{gj}^{(2,c)}(z)$ and $I_{gj}^{(2,\Rnons)}(z,R)$ terms in the beam function, which when integrated against the PDFs give a $\pTcut$ independent contribution determined by a constant $b_2^{(c+\Rnons)}(R)$. Hence, we have
\begin{align} \label{eq:NNLOnonsfit}
&\sigma_0^{\nons,\NNLO}(\pTcut\!,R, \mu_\FO) \!+\! \sigma_\LO(\mu_\FO) \frac{\alpha_s^2(\mu_\FO)}{(4\pi)^2} 2b_2^{(c+\Rnons)}(R)
\nn\\ & \qquad
= [\sigma_{\geq 0}^\NNLO(\mu_\FO) - \sigma_{\geq 1}^\NLO(\pTcut, R, \mu_\FO)]
\nn\\ & \qquad\quad
- \sigma_0^{\NNLL'}(\pTcut, R, \mu_i = \nu_i = \mu_\FO)
\,.\end{align}
Here, the right-hand side is obtained numerically and then fit with a set of functions suitable to describe the $\pTcut$ dependence of $\sigma_0^\nons(\pTcut, R, \mu_\FO)$. Since the latter vanishes for $\pTcut\to0$, this fit also allows us to determine the numerical value of $2b_2^{(c+\Rnons)}(R)$ from the intercept at $\pTcut=0$. Note that since there are large numerical cancellations between the full and singular results on the right-hand side, the remaining nonsingular data has large statistical fluctuations for $\pTcut\to 0$. Ensuring a stable fit result therefore required the use of very high statistics from MCFM as well as a careful validation of the fitting procedure.

Note also that the scale $\mu_B$ at which the $b_2^{(c+\Rnons)}$ contribution is evaluated in the beam function is relevant at NNLL$'$ (i.e. it contributes to the subset of N$^3$LL effects that are supposed to be included at NNLL$'$). In the numerical determination above the PDFs are evaluated at a fixed $\mu_B = \mu_\FO$. To account for this we rescale it by the PDF dependence of the LO cross section, as indicated in \eq{NNLOnonsfit}. Since we perform the nonsingular fit at different values of $\mu_\FO$, we are able to check that this captures the PDF scale dependence to very good approximation.

At large $\pTcut$, the distinction between singular and nonsingular contributions becomes meaningless since both are of similar size and there are nontrivial cancellations between them (as can be seen in \fig{singVSnons} below). When using the imaginary scale setting in the hard function, it modifies the cross section at all values of $\pTcut$. Therefore, it is important to implement an analogous improvement for the nonsingular contributions, since otherwise these cancellations would be spoiled. The final expression for the nonsingular cross section entering in \eq{fact} is given by
\begin{align}
&\sigma_0^\nons(\pTcut, R, \muns)
\nn\\ & \qquad
= \biggl\{ \sigma_0^{\nons(1)}(\pTcut, R, \muns) \biggl[ 1 - \frac{\as(\muns)}{2\pi} C_A \pi^2 \biggr]
\nn \\ & \qquad\quad
+ \sigma_0^{\nons(2)}(\pTcut, R, \muns) \biggr\}\, U_H (-\img \muns, \muns)
 \,.\end{align}
Here, $\sigma_0^{\nons(i)}(\pTcut, R, \muns)$ are the $\ord{\alpha_s^i}$ nonsingular terms obtained numerically for given values of $R$ and $\muns$, and $U_H (-\img \muns, \muns)$ is the evolution factor of the hard function. The latter is used to apply the analogous resummation of $\pi^2$ terms to the nonsingular cross section as was induced by the hard function in  the singular terms.

The NLO and NNLO nonsingular contributions for $R = 0.4$ and $\muns = m_H$ are shown in \fig{nonsingularplot} for both real (left panel) and imaginary (right panel) scale setting. We observe that the latter substantially improves the perturbative convergence also in the nonsingular terms at all values of $\pTcut$. This is not unexpected from the point of view of the power expansion in SCET. For $\pTcut \ll m_H$ and at subleading order in the SCET power counting, the nonsingular terms would arise from a combination of leading and subleading hard, beam, and soft functions, and many of the hard functions in these contributions can be expected to require an imaginary hard scale.

\section{Resummation and Perturbative Uncertainties}
\label{sec:unc}

A critical aspect of precision cross section predictions is the theoretical control of perturbative uncertainties. Ultimately, the formal perturbative accuracy in the predictions is only meaningful together with a robust understanding and estimate of theoretical uncertainties.

The categorization of the data into jet bins is used in the experimental analyses to optimize the control of backgrounds and experimental systematic effects. In the end, the information from all measured categories flows together, thereby maximizing the use of the available data. In this context, vetoing jets in the $0$-jet cross section amounts to dividing the total inclusive cross section, $\sigma_\tot \equiv \sigma_{\geq 0}$, into an exclusive $0$-jet bin equivalent to $\sigma_0(\pTcut)$ and the remaining inclusive $1$-jet bin,
\begin{equation}
\sigma_{\geq 0} = \sigma_0(\pTcut) + \sigma_{\geq 1}(\pTcut)
\,.\end{equation}
Therefore, a complete theoretical description of this binning procedure is needed. This requires a framework, which, in addition to the resummation of $\sigma_0(\pTcut)$ at small $\pTcut$, provides a valid description of the cross section at all values of $\pTcut$ as well as the correlations between the perturbative uncertainties in the jet bins and the total cross section.

As we discuss in detail in this section, the framework we use for resummation and fixed-order matching, based on SCET and profile functions, is well-suited for this task. It provides us with direct theoretical handles to reliably assess the perturbative uncertainties and allows us to predict the required correlations by utilizing common underlying theory parameters in the scales $\mu_H$, $\mu_B$, $\mu_S$, $\nu_B$, and $\nu_S$.  These are varied to obtain the uncertainty estimates.

In \subsec{unc} we give an overview of perturbative uncertainties for jet bins, and establish the necessary notation.  As the jet-veto cut is increased our resummed results smoothly reproduce the fixed-order cross section and its standard uncertainties by using profile functions, which are discussed in \subsec{RGEunc}. In \subsec{profileunc} we explain how variations of the hard, soft, and beam scales in the effective theory determine the fixed-order and jet-binning uncertainties. Finally, in \subsec{clusteringuncertainties} we discuss our estimate for the additional uncertainty from clustering effects at higher orders in perturbation theory. Note that we will not discuss additional parametric uncertainties from input parameters such as PDFs or $\as(m_Z)$. These have to be estimated separately and included with the usual uncertainty propagation.

\subsection{Perturbative Uncertainties in Jet Binning}
\label{subsec:unc}

A convenient way to describe the uncertainties involved in the jet binning is in terms of fully correlated and fully anticorrelated components~\cite{Stewart:2011cf, Gangal:2013nxa}, which amounts to parametrizing the covariance matrix for $\{\sigma_0, \sigma_{\geq 1}\}$ as
\begin{equation} \label{eq:Cgeneral}
C(\{\sigma_0, \sigma_{\geq 1}\}) =\!
\begin{pmatrix}
(\Delta^{\rm y}_0)^2 &  \Delta^{\rm y}_0\,\Delta^{\rm y}_{\geq 1}  \\
\Delta^{\rm y}_0\,\Delta^{\rm y}_{\geq 1} & (\Delta^{\rm y}_{\geq 1})^2
\end{pmatrix}
\!+
\begin{pmatrix}
 \Delta_\cut^2 &  - \Delta_\cut^2 \\
-\Delta_\cut^2 & \Delta_\cut^2
\end{pmatrix}
\!.\end{equation}
The first correlated component, denoted with a superscript ``y'', can be interpreted as an overall yield uncertainty shared among all bins. The second anticorrelated component can be interpreted as a migration uncertainty between the two bins, which is introduced by the binning cut and drops out in their sum. The total uncertainty for each bin is given by
\begin{align} \label{eq:DeltaN}
\Delta_{\geq 0} &= \Delta^{\rm y}_{0} + \Delta^{\rm y}_{\geq 1} \equiv \Delta^{\rm y}_{\geq 0}
\,,\nn\\
\Delta_0^2 &= (\Delta_0^{\rm y})^2 + \Delta_\cut^2
\,,\nn\\
\Delta_{\geq 1}^2 &= (\Delta_{\geq 1}^{\rm y})^2 + \Delta_\cut^2
\,.\end{align}

Equation~\eqref{eq:Cgeneral} is a completely generic parametrization of a $2\times2$ symmetric matrix.  This choice of parameters is convenient because of the above physical interpretation. An additional advantage is that the uncertainties are described in terms of two independent components, which are fully correlated or anticorrelated between the different observables, so that the experimental implementation is straightforward (e.g. in a profile likelihood fit, the yield and migration uncertainties can each be implemented by an independent nuisance parameter).

To estimate each uncertainty component in our resummation framework we make the following identifications:%
\begin{equation}
\Delta^{\rm y}_i \equiv \Delta_{\mu i}
\,,\qquad
\Delta_\cut \equiv \Delta_\resum
\,.\end{equation}
Here, $\Delta_{\mu i}$ corresponds to the uncertainties in the cross section that reproduce the fixed-order uncertainty in the total cross section and probe the nonlogarithmic contributions at finite $\pTcut$. This makes it natural to identify these with the yield uncertainties. The resummation uncertainty, $\Delta_\resum$, corresponds to the intrinsic uncertainty in the resummed logarithmic series. The logarithms $\ln(\pTcut/m_H)$ are directly caused by the binning cut and at small $\pTcut$ are the dominant veto-dependent effect, which cancels between $\sigma_0$ and $\sigma_{\geq 1}$. Hence, higher-order logarithms are the primary source of uncertainty in the division of the cross section into bins and we can therefore identify $\Delta_\resum$ with the migration uncertainty. Furthermore, $\Delta_\resum$ vanishes at large $\pTcut$ where the resummation of logarithms becomes unimportant. This is consistent with the fact that in this limit migration effects become irrelevant since $\sigma_{\geq 1}$ becomes numerically much smaller than $\sigma_0(\pTcut)$. Our procedure to estimate $\Delta_{\mu i}$ and $\Delta_\resum$ through scale variations in the resummed cross section is discussed in the following sections.

With these identifications, the full covariance matrix for $\{\sigma_{\geq 0}, \sigma_0, \sigma_{\ge1} \}$ is given by
\begin{equation}
C\bigl(\{\sigma_{\geq 0}, \sigma_0, \sigma_{\ge 1}\}\bigr) = C_{\mu} + C_\resum
\,,\end{equation}
where
\begin{align} \label{eq:Cmures}
C_{\mu} &= \begin{pmatrix}
\Delta_\tot^2 & \Delta_\tot \Delta_{\mu0} & \Delta_\tot \Delta_{\mu\ge1} \\
\Delta_\tot \Delta_{\mu0} & \Delta_{\mu0}^2 & \Delta_{\mu0} \Delta_{\mu\ge1} \\
\Delta_\tot \Delta_{\mu\ge1} & \Delta_{\mu0} \Delta_{\mu\ge1} & \Delta_{\mu\ge1}^2
\end{pmatrix}
, \nn \\
C_\resum &= \begin{pmatrix}
0 & 0 & 0 \\
0 & \Delta_\resum^2 & -\Delta_\resum^2 \\
0 & -\Delta_\resum^2 & \Delta_\resum^2
\end{pmatrix}
,\end{align}
and we can easily read off the uncertainties in the different cross sections
\begin{align}
\Delta_\tot \equiv \Delta_{\mu\geq 0} &= \Delta_{\mu0} + \Delta_{\mu\ge1}
\,, \nn \\
\Delta_0^2 (\pTcut) &= \Delta_\resum^2 + \Delta_{\mu0}^2
\,, \nn \\
\Delta_{\ge1}^2 (\pTcut)
&= \Delta_\resum^2 + (\Delta_{\rm tot} - \Delta_{\mu0})^2
\,.\end{align}
The uncertainties in other observables follow by standard uncertainty propagation. For example, for the 0-jet efficiency, $\e_0 (\pTcut) \equiv \sigma_0 (\pTcut) / \sigma_{\geq 0}$, we have
\begin{align}  \label{eq:e0percent}
\frac{\Delta_{\e_0}^2 (\pTcut)}{\e_0^2 (\pTcut)} &= \frac{\Delta_0^2 (\pTcut)}{\sigma_0^2 (\pTcut)} + \frac{\Delta_\tot^2}{\sigma^2_\tot} - 2\frac{\Delta_\tot \Delta_{\mu0}}{\sigma_{\geq 0} \sigma_0(\pTcut)}
\,.\end{align}
Through the last term the correlation between $\Delta_\tot$ and $\Delta_{\mu0}$ reduces the relative uncertainty in the 0-jet efficiency, which will be noticeable in our numerical analysis. In particular, in the limit of large $\pTcut$ where $\e_0 \to 1$ the uncertainty $\Delta_{\epsilon_0}$ will go to zero as it should.

\subsubsection{Fixed Order}

In a pure fixed-order prediction, there is no way to fully disentangle the two uncertainty components. Using a common fixed-order scale variation for all observables amounts to setting $\Delta_\cut = 0$ and setting $\Delta_i^{\rm y} \equiv \Delta_i^\FO$. However, as demonstrated in detail in Refs.~\cite{Stewart:2011cf, Gangal:2013nxa}, at small values of $\pTcut$, as soon as the logarithmic corrections become sizable, migration effects are important and cannot be neglected. Doing so can lead to a significant underestimate of uncertainties. A more reliable fixed-order estimate is obtained by explicitly taking into account $\Delta_\cut$ by using instead
\begin{equation}  \label{eq:ST}
\Delta_0^{\rm y} = \Delta_{\geq 0}^\FO \equiv \Delta_\tot
\,,\qquad
\Delta_\cut = \Delta_{\geq 1}^\FO
\,,\end{equation}
where $\Delta_{\geq i}^\FO$ are the fixed-order uncertainties in the inclusive cross sections. (As explained in Ref.~\cite{Stewart:2011cf}, this choice is motivated by the fact that the perturbative series in $\sigma_{\geq 1}$ starts as $\as \ln^2(\pTcut/m_H)$ and its fixed-order scale variation therefore directly estimates the size of the $\pTcut$ logarithms. An alternative prescription proposed in Ref.~\cite{Banfi:2012yh} yields very similar results for $\Delta_0(\pTcut)$.)

With the choice in \eq{ST} the uncertainties in the pure fixed-order prediction are described by
\begin{align}
C_{\rm ST} \bigl(\{\sigma_{\geq 0}, \sigma_0, \sigma_{\ge 1}\}\bigr)
\!=\! \begin{pmatrix}
\Delta_\tot^2 & \Delta_\tot^2 & 0  \\
\Delta_\tot^2 & \Delta_\tot^2 \!+\! (\Delta_{\ge1}^\FO)^2 & -(\Delta_{\ge1}^\FO)^2 \\
0 & -(\Delta_{\ge1}^\FO)^2 & (\Delta_{\ge1}^\FO)^2
\end{pmatrix}
\!\!.\end{align}
These are the default fixed-order Higgs jet-binning uncertainties used by the experiments, and also what we will use when comparing our results to fixed order in \sec{results}.

\subsection{Resummation and Matching to Fixed Order with Profile Scales}
\label{subsec:RGEunc}

\begin{figure}[t!]
\includegraphics[width=0.9\columnwidth]{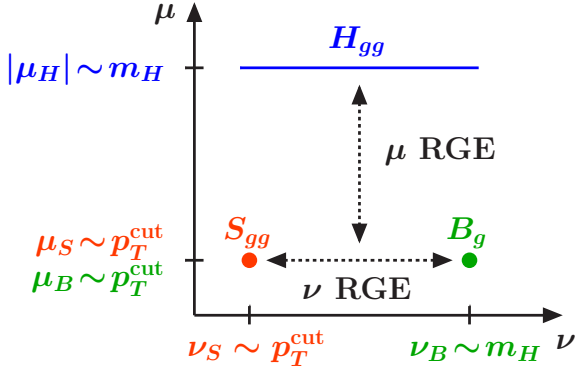}%
\hfill
\vspace{-3ex}
\caption{Combined renormalization group evolution in virtuality and rapidity. The hard, beam, and soft functions are evolved in the virtuality scale $\mu$, where the characteristic scales are $\mu_H \sim m_H$ and $\mu_B \sim \mu_S \sim \pTcut$.  Additionally, rapidity logarithms are summed by evolving the beam and soft functions in the rapidity scale $\nu$, with characteristic scales $\nu_B \sim m_H$ and $\nu_S \sim \pTcut$.}
\label{fig:running}
\end{figure}

\begin{figure*}[t!]
\begin{center}
\includegraphics[width=\columnwidth]{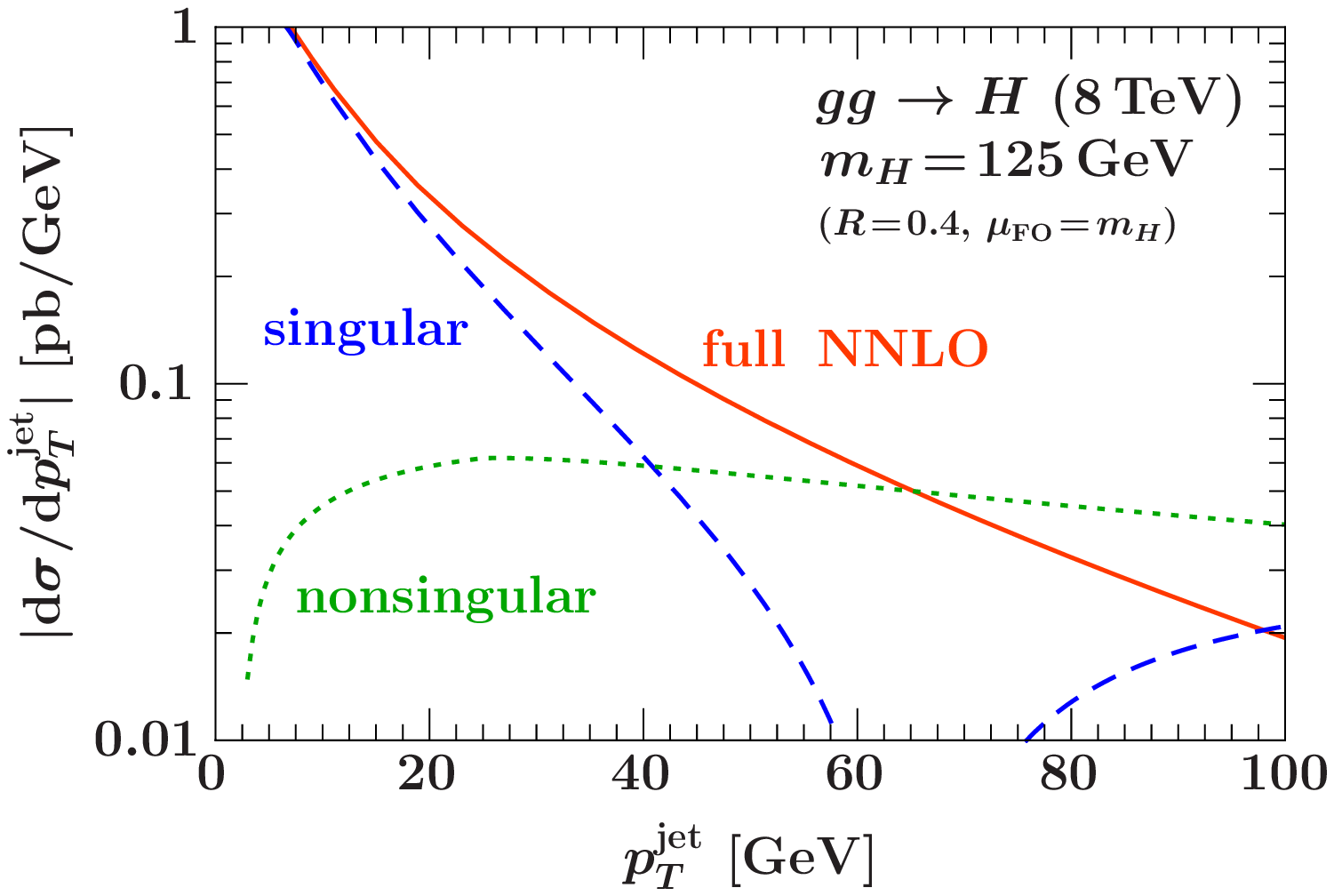}%
\hfill%
\includegraphics[width=0.97\columnwidth]{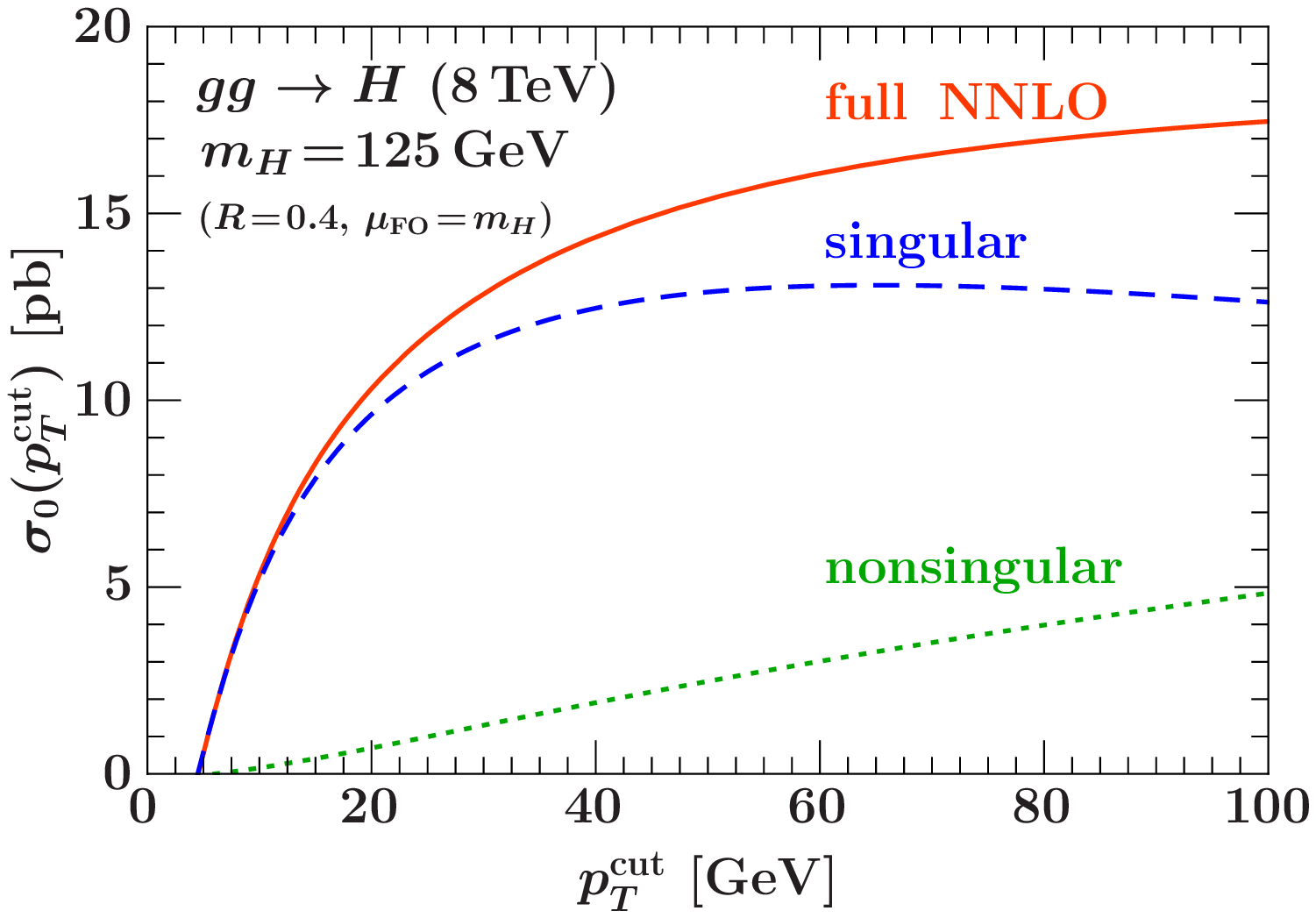}%
\end{center}
\vspace{-3ex}
\caption{Singular and nonsingular contributions to the fixed NNLO cross section (using $R = 0.4$ and $\mu_{\rm FO} = m_H$). Left: The magnitude of the contributions differential in $p_T^\jet$. Right: The corresponding contributions to the integrated cross section as a function of $\pTcut$. The resummation, transition, and fixed-order regions are clearly visible as the relative importance of the singular and nonsingular terms changes with $p_T^\jet$ and $\pTcut$.}
\label{fig:singVSnons}
\end{figure*}

In the effective field theory framework, the resummation is performed by RGE running. First, we evaluate each of the hard, beam, and soft functions appearing in the factorized cross section at their natural virtuality scales $\mu_i$ and rapidity scales $\nu_i$. Next, we evolve them all to arbitrary, common scales: $\mu$ for invariant mass and $\nu$ for rapidity.  This resums the logarithms of the invariant mass ratios $\mu_i/\mu_j$ and rapidity ratios $\nu_i/\nu_j$. As we saw in \sec{fact}, the beam and soft functions evolve in both virtuality and rapidity space, while the hard function only evolves in virtuality. The evolution together with the natural scales is illustrated in \fig{running}. Finally, the evolved functions are combined together  in the cross section at the common scales $(\nu,\mu)$, which is a point in the plane shown in this figure.

The resummed cross section is explicitly independent of the arbitrary scales $\mu$ and $\nu$ at each order in resummed perturbation theory, which means we are free to pick any convenient values. Taking $\mu = \mu_B$ and $\nu = \nu_S$, and combining all the ingredients detailed in \sec{fact}, the cross section in \eq{fact} takes the form
\begin{align}
&\sigma_0(\pTcut)
\nn \\ & \quad
\!\! = \sigma_B H_{gg} (m_t, m_H, \mu_H) \! \int\!\!\df Y
B_g(m_H, \pTcut, R, x_a, \mu_B, \nu_B)
\nn \\ & \quad\qquad \times
B_g(m_H, \pTcut, R, x_b, \mu_B, \nu_B)\, S_{gg}(\pTcut, R, \mu_S, \nu_S)
\nn \\ & \quad\qquad \times
U_0(\pTcut, R; \mu_H, \mu_B, \mu_S, \nu_B, \nu_S)
\nn \\[3pt] & \qquad
+ \sigma_0^\Rnons(\pTcut, R) + \sigma_0^\nons(\pTcut, R, \muns)
\,,\end{align}
where the combined renormalization group evolution factor $U_0$ is given by
\begin{align} \label{eq:U0}
&U_0 (\pTcut, R; \mu_H, \mu_B, \mu_S, \nu_B, \nu_S)
\nn\\ & \qquad
= \biggl\lvert\exp\biggl[ \int_{\mu_H}^{\mu_B}\! \frac{\df\mu'}{\mu'}\, \gamma_H^g (m_H, \mu') \biggr] \biggr\rvert^2
\nn \\ & \qquad\quad\times
\exp \biggl[ \int_{\mu_S}^{\mu_B}\! \frac{\df\mu'}{\mu'}\, \gamma_S^g(\mu', \nu_S) \biggr]
\nn\\ & \quad\qquad\times
\exp \biggl[\ln\frac{\nu_B}{\nu_S}\, \gamma_\nu^g(\pTcut, R, \mu_B) \biggr]
\,.\end{align}

Next, we discuss how to choose numerical values for the scales $\mu_H, \mu_B, \mu_S, \nu_B$, and $\nu_S$ as a function of $\pTcut$, which are referred to as profile scales~\cite{Ligeti:2008ac, Abbate:2010xh}. For this purpose we can distinguish three different regimes according to the relative importance of the singular and nonsingular cross section contributions. In \fig{singVSnons}, the singular and nonsingular terms are plotted against the total fixed-order cross section at $\ord{\as^2}$.

In the resummation region at low values of $\pTcut$, the singular contributions dominate and must be resummed, while the nonsingular contributions are perturbative power corrections. To resum the logarithms, the scales should parametrically follow their canonical values dictated by the RGE,
\begin{align} \label{eq:canonicalscales}
\mu_H &\sim -\img m_H \,, & \mu_B &\sim \mu_S \sim \pTcut
\,,\nn \\
\qquad \nu_B &\sim m_H \,, & \nu_S &\sim \pTcut
\,.\end{align}
At large $\pTcut \gtrsim m_H/2$, the singular and nonsingular contributions are equally important, and fixed-order perturbation theory should be used. In this fixed-order region it is essential that the resummation is turned off to ensure that the correct fixed-order cross section is obtained. The reason is that there are important cancellations between singular and nonsingular terms, which are spoiled if the resummation is kept on too long. In this region, all virtuality scales must approach a common fixed-order scale and the rapidity scales must be equal,
\begin{equation}
\lvert\mu_H\rvert = \mu_B = \mu_S = \muns = \mu_\FO
\,,\qquad
\nu_B = \nu_S
\,.\end{equation}

Finally, in the transition between the resummation and fixed-order regions typically both the logarithmic resummation as well as the fixed-order corrections are important. To obtain a proper description of this transition region, which in our case also includes the experimental range of interest, we have to use profiles that incorporate the constraints imposed by the resummation toward small $\pTcut$ and the fixed-order matching toward large $\pTcut$, together with a smooth interpolation between these two regimes. There is a growing body of literature on the construction of appropriate profiles in a variety of contexts~\cite{Ligeti:2008ac, Abbate:2010xh, Berger:2010xi, Bauer:2011uc, Jain:2012uq, Liu:2012sz, Alioli:2012fc, Jouttenus:2013hs, Gritschacher:2013pha, Liu:2013hba, Kang:2013nha, Chang:2013iba}.

For the central profiles we take
\begin{align} \label{eq:munu}
\mu_H &= -\img \mu_\FO
\,,\qquad
\muns = \mu_\FO
\,,\nn\\
\nu_B &= \mu_\FO
\,,\nn\\
\mu_B &= \mu_S = \nu_S = \mu_{\rm FO}\, f_{\rm run} (\pTcut / m_H)
\,.\end{align}
That is, we take fixed values for $\mu_H$, $\muns$, and $\nu_B$, while $\mu_B$, $\mu_S$, and $\nu_S$ are constructed in terms of the common profile function
\begin{align} \label{eq:profile}
f_{\rm run} (x) =
\begin{cases}
x_0 \bigl[ 1 + (x / x_0)^2/4 \bigr] & x \le 2x_0 \,, \\
x & 2x_0 \le x \le x_1 \,, \\
x + \dfrac{(2 - x_2 - x_3) (x - x_1)^2}{2(x_2 - x_1) (x_3 - x_1)} & x_1 \le x \le x_2 \,, \vspace{1ex} \\
1 - \dfrac{(2 - x_1 - x_2) (x - x_3)^2}{2(x_3 - x_1) (x_3 - x_2)} & x_2 \le x \le x_3 \,, \\
1 & x_3 \le x \,.
\end{cases}
\end{align}
The first regime, $x \le 2x_0$, is the nonperturbative region and the scales $\mu_{B,S}$ and $\nu_S$ asymptote as $x\to0$ to a fixed scale $x_0\mu_\FO \gtrsim \Lambda_{\rm QCD}$.  This ensures that factors of $\alpha_s(\mu_i)$ that enter from solving perturbatively defined anomalous dimension equations, never become nonperturbative.  The second regime has the canonical scaling for resummation. The third and fourth have quadratic scaling (of positive and negative second derivative, respectively) and simply provide a smooth transition to the final (constant) region where all scales are equal and resummation is turned off.  This profile function and its first derivative are both continuous.

\begin{figure}[t!]
\begin{center}
\includegraphics[width=\columnwidth]{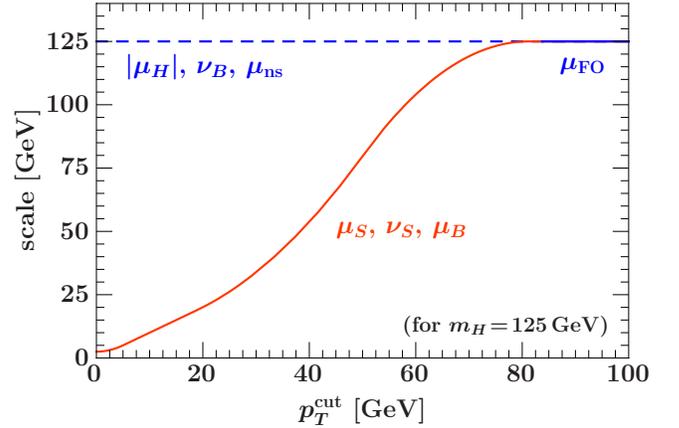}%
\end{center}
\vspace{-3ex}
\caption{The central profile scale for the low scales $\mu_B, \mu_S, \nu_S$ as a function of $\pTcut$, together with the central value for the high scales $\lvert \mu_H \rvert, \nu_B$.}
\label{fig:profilecentral}
\end{figure}

For the overall scale parameter we have $\mu_\FO\sim m_H$ and for our central result we will use $\mu_\FO=m_H$
in \eq{munu}.  In \eq{profile} the parameters $x_i$ mark the boundary between the different regimes, and their values are chosen by considering the importance of the singular versus nonsingular contributions plotted in \fig{singVSnons}.  The singular and nonsingular contributions become comparable near $\pTcut=40\,{\rm GeV}$ so the profile must transition towards the fixed-order result beyond this value. For our central profiles we choose
\begin{equation}
x_0 = 2.5 \GeV / \mu_\FO \,, \quad \{x_1, x_2, x_3\} = \{0.15 ,\, 0.4 ,\, 0.65 \}
\,.\end{equation}
For $\mu_\FO=m_H=125\,{\rm GeV}$ the $\{x_1,x_2,x_3\}$ values correspond to $\{19,50,81\}\,{\rm GeV}$. 
The resulting central profile scales are shown in \fig{profilecentral}, so we see that the transition occurs roughly between $30$--$65\,{\rm GeV}$. In the next subsection, we discuss in detail the profile scale variations that we use to evaluate perturbative uncertainties.

Note that in the transition from small to large $\pTcut$, we are essentially forced to keep the hard scale at its imaginary value $\mu_H = -\img m_H$. In principle, one could contemplate rotating it to the real axis as a function of $\pTcut$ to turn off the resulting resummation of large $\pi^2$ terms in the hard virtual corrections. However, this would inevitably lead to an unphysical result of a decreasing cross section with increasing $\pTcut$. What this means is that the significantly improved perturbative stability observed in the small $p_T$ region also directly translates into an improved convergence in the fixed-order cross section at large $\pTcut$, simply because a large part of the total cross section comes from the small $p_T$ region. Furthermore, as we have seen in \fig{nonsingularplot}, the imaginary scale also translates into an improved convergence of the nonsingular contributions themselves. The total cross section for $\mu_H = -\img m_H$ increases by about 7\% compared to the NNLO cross section evaluated at $\mu_\FO = m_H/2$. This increase is quite consistent with the expected increase in the total cross section at N$^3$LO from the recent estimate in Ref.~\cite{Ball:2013bra}.

\subsection{Yield and Resummation Uncertainties Via Profile Scale Variations}
\label{subsec:profileunc}

To evaluate the perturbative uncertainties in our predictions we vary the profile scales about the central profiles defined in the previous section.  We consider several types of variation in turn, and discuss how they are used to determine the yield and resummation uncertainties that appear in the matrices $C_\mu$ and $C_\resum$ in \eq{Cmures}.

The first type of variation is a collective variation of all of the scales up or down by a factor of $2$.  This is accomplished by taking $\mu_\FO=2m_H$ or $\mu_\FO=m_H/2$ in \eq{munu}. At large $\pTcut$, where all scales become equal to $\mu_\FO$, this variation becomes equivalent to the usual scale variation in the fixed-order cross section.  Indeed, in the limit of very large $\pTcut$ it reproduces the fixed-order scale variation of the total cross section.\footnote{For $\pTcut > x_3 m_H$ and real $\mu_H = \mu_\FO$ we exactly reproduce the fixed-order cross section scale variation for equal factorization and renormalization scales. If these two scales are varied independently they give essentially the same final result since the renormalization scale variation dominates by far.}  When varying $\mu_\FO$, all scale ratios are kept fixed, so this does not change any of the arguments inside the logarithms $\ln(\mu_H/\mu_{B,S})$ and $\ln(\nu_B/\nu_S)$ that sum up the large $\ln(m_H/\pTcut)$ terms. Hence, this variation is clearly identified as contributing to the yield uncertainties.

\begin{figure*}[t!]
\begin{center}
\includegraphics[width=\columnwidth]{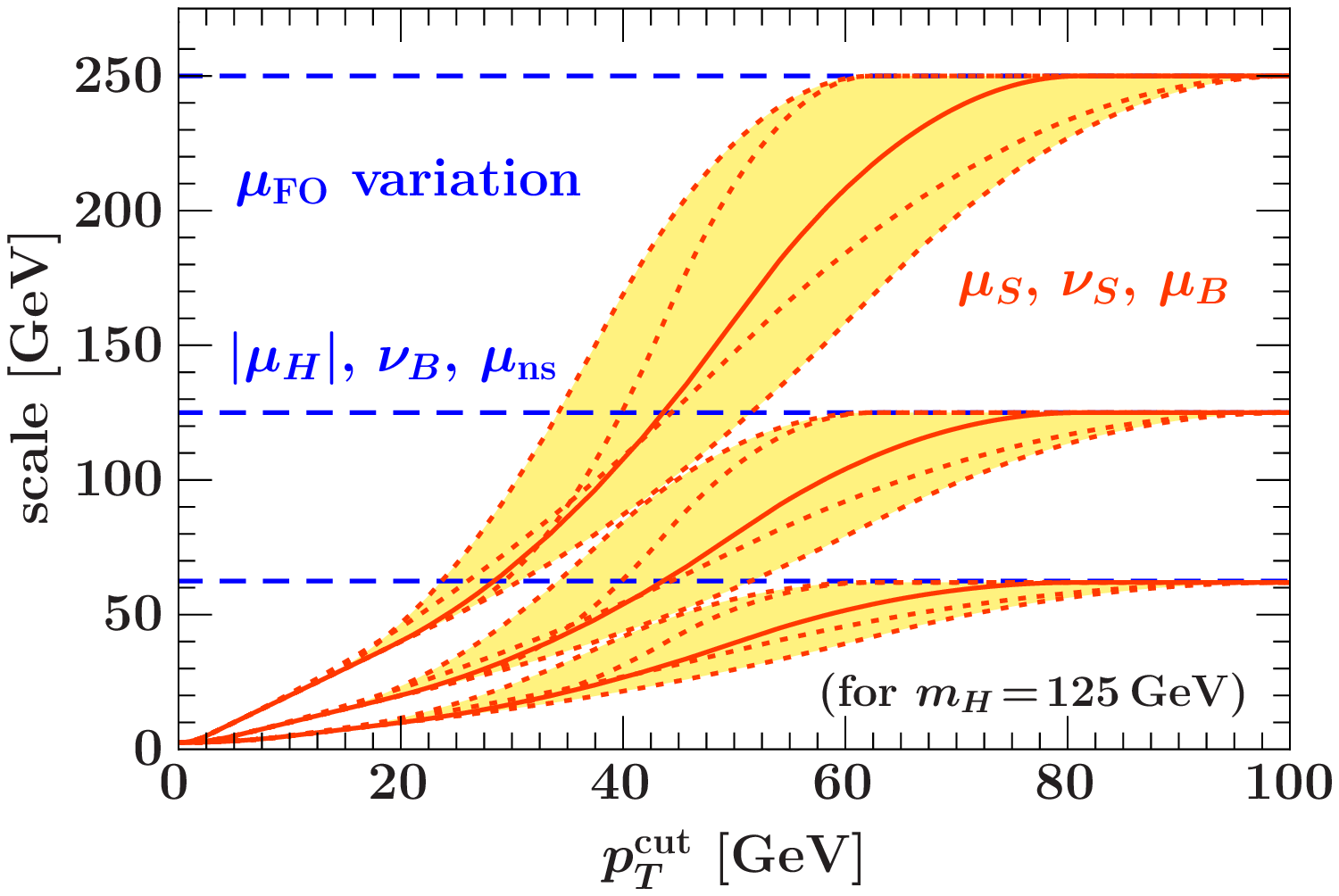}%
\hfill%
\includegraphics[width=\columnwidth]{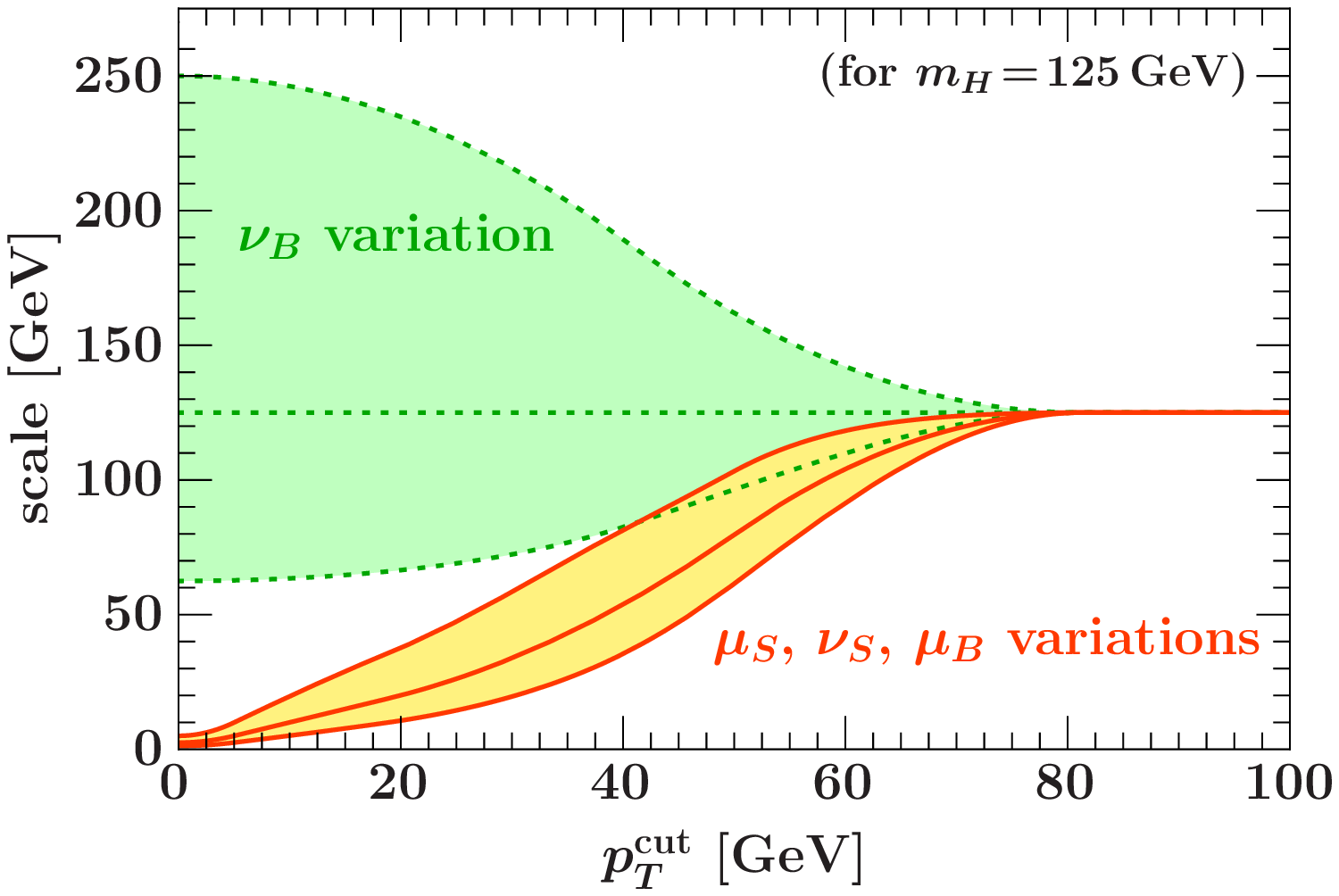}%
\end{center}
\vspace{-3ex}
\caption{The variations of the central profiles as described in the text.  On the left, the variations are shown that contribute to the yield uncertainty, where all scales are collectively multiplied by a factor 2 or 1/2, for all four profile shapes.  The central profile shape is shown with thick lines, while the other profile shapes are shown with dotted lines, and we shade between the shapes.  On the right, the variations of $\mu_B$, $\mu_S$, and $\nu_S$ (solid lines, yellow shading) and $\nu_B$ (dotted lines, green shading) are shown which contribute to the resummation uncertainty.  Combinations of variations of these scales make up the set of variations that we perform to asses the uncertainties in our prediction.}
\label{fig:profilevariations}
\end{figure*}

A second type of variation is to the profile shape.  The values $\{x_1, x_2, x_3\}$ determine the boundaries between the different scaling regions of the low-scale profiles as a function of $\pTcut$.  We account for the ambiguity in this shape by using four different choices for $\{x_1, x_2, x_3\}$ to provide a variation away from the central scale choice $\{x_1, x_2, x_3\} =\{0.15, 0.4, 0.65\}$:
\begin{align}
\{x_1, x_2, x_3\} \; : \; &\{0.1, 0.3, 0.5\} \,, \{0.2, 0.5, 0.8\} \,, \nn \\
& \{0.04, 0.4, 0.8\}\,, \{0.2, 0.35, 0.5\} \,.
\end{align}
These changes to the profile have an impact on the uncertainty from varying $\mu_\FO$ since they determine the transition between the region where the resummation is active and where the fixed-order prediction is used and hence the extent of the fixed-order region.  They also vary the logarithms $\ln(\mu_H/\mu_{B,S})$, and hence have some impact on uncertainties that would be associated to resummation. In practice, with $\mu_\FO = m_H$ the effect of varying the $x_i$ in the central profile is smaller than the other resummation uncertainties (discussed below), whereas when varying $\mu_\FO$ up and down there is a noticeable impact on the yield uncertainties.  Therefore we will group this variation with the yield uncertainty, and use each of the five profiles specified by $\{x_1,x_2,x_3\}$ together with each of the three values of $\mu_\FO$.  This set of profile variations is plotted in the left panel of \fig{profilevariations}. We still note that the range of cross section values obtained from changing $\mu_\FO$ with a fixed profile is significantly larger than the range from changing the profile via $x_{1,2,3}$ for a fixed $\mu_\FO$, and hence the $\mu_\FO$ variation is the more important variation by far.

The total yield uncertainty for the $0$-jet cross section is thus defined as the maximum absolute deviation from the central scale over all 14 variations,
\begin{equation}
\Delta_{\mu 0} (\pTcut) = \max_{v_i \in V_\mu} \bigl\lvert \sigma_0^{v_i} (\pTcut) - \sigma_0^{\rm central} (\pTcut) \bigr\rvert
\,.\end{equation}
where $V_\mu$ is the set of variations.    To determine the total uncertainty in the fixed-order cross section we make use of the fact that $\lim_{\pTcut\to \infty} \Delta_{\mu 0}(\pTcut) = \Delta_{\rm tot}$, and in practice we extract $\Delta_{\rm tot}$ for $\pTcut = 600\,{\rm GeV}$.  Together this determines the two parameters occurring in the yield covariance matrix $C_\mu$.

Resummation uncertainties are estimated through variations of the beam and soft scales, while keeping $\mu_\FO = m_H$ at its central value. The variations of the beam and soft scales are performed with a multiplicative variation factor $f_{\rm vary} (\pTcut)$.  For a generic beam or soft scale $\mu_i$ or $\nu_i$, the up and down variations are performed via the variations
\begin{align}
\mu_i^{\rm up} (\pTcut) &= \mu_i^{\rm central} (\pTcut) \times f_{\rm vary}(\pTcut/m_H)\,,
\nn\\
\mu_i^{\rm down} (\pTcut) &= \mu_i^{\rm central} (\pTcut) \,/ \, f_{\rm vary}(\pTcut/m_H)\,,
\nn\\
 \nu_i^{\rm up} (\pTcut)  &= \nu_i^{\rm central} (\pTcut) \times f_{\rm vary}(\pTcut/m_H)\,,
 \nn \\
 \nu_i^{\rm down} (\pTcut) &= \nu_i^{\rm central} (\pTcut) \,/ \, f_{\rm vary}(\pTcut/m_H)\,.
\end{align}
The variation factor is defined by
\begin{align}
f_{\rm vary} (x) =
\begin{cases}
2(1 - x^2 / x_3^2) & 0 \le x \le x_3 / 2 \,, \\
1 + 2(1 - x/x_3)^2 & x_3 / 2 \le x \le x_3 \,, \\
1 & x_3 \le x \,.
\end{cases}
\end{align}
It is designed to smoothly turn off these variations, since they must turn off when the resummation is turned off at high $\pTcut$ values.  These variations for $\mu_B, \mu_S, \nu_B$, and $\nu_S$ are plotted in the right panel of \fig{profilevariations}.

The resummation uncertainty is a combination of a set of up, down, and central values for the $\mu_B, \mu_S, \nu_B$, and $\nu_S$ scales.  The dependence on each of these scales cancels between RG evolution and the fixed-order contributions at the order we are working, while the remaining residual dependence probes the higher-order contributions in resummed perturbation theory.

The purpose of an individual scale variation is to vary the argument of the logarithms it appears in by a factor in order to probe the potential size of higher-order logarithms of that scale.  For our profiles the variation factor above is $1/2$ or $2$ for $\pTcut\to 0$ and goes towards $1$ for $\pTcut \to x_3 m_H$ where the resummation is turned off.   Certain combinations of scale variations are undesirable as they double the variations of the logarithms, for example $\{\nu_B^{\rm up},\, \nu_S^{\rm down} \}$ gives a factor of $4$ variation for the logarithm of $\nu_B / \nu_S$.  To avoid varying the scales in logarithms outside of the desired factor of $2$ range, we consider all the ratios of beam and soft scales that appear in the factorization,
\begin{align}
 \frac{\mu_S}{\mu_B} \sim \frac{\mu_S}{\nu_S}
\sim 1 \,, 
 \qquad 
 \frac{\nu_B}{\nu_S} \sim \frac{m_H}{\pTcut} \,.
\end{align}
All of these scalings are respected by the central profiles. We then constrain the variations about the central profiles to not violate any of these scaling relations by more than a factor of $2$ (as would happen for instance by varying $\mu_B$ up and $\mu_S$ down). We make one additional constraint on the variations by considering the evolution factor $U_0$ in \eq{U0}.  The summation of rapidity logarithms contains the factor
\begin{equation}
\exp \biggl[ \ln \Bigl( \frac{\nu_B}{\nu_S} \Bigr) \gamma_{\nu}^g(\pTcut, R, \mu_B) \biggr] \,.
\end{equation}
This is a unique combination as it features a large logarithm of $\nu_B/\nu_S$ multiplying a rapidity anomalous dimension that depends on $\mu_B$. A simultaneous variation of $\mu_B$ down with either $\nu_S$ down or $\nu_B$ up gives sensitivity to small scales $\alpha_s(\mu_B)$, and the effect is effectively doubled by the $\ln(\nu_S/\nu_B)$ variation, leading us to eliminate these two combinations from the set of scale variations we consider.

With these restrictions, there are 35 remaining (of an original possible 80) profile scale variations of $\mu_B, \mu_S, \nu_B,$ and $\nu_S$  away from their central profile which probe the resummation uncertainty.  We note that without separately varying $\mu_B$ and $\mu_S$, and without explicit variations of the $\nu_B$ and $\nu_S$ scales there would be only a single up/down variation and a significant reduction in the resummation uncertainty.  Exploring a much larger space for the scale variations is crucial to reliably estimate the uncertainty from the summation of logarithms. Note that at small $R$ the large $\ln R^2$ effects appear through the rapidity RGE, so it is important to vary the rapidity scales to probe the effect of these terms on the $\pTcut$ resummation. For the final resummation uncertainty we use
\begin{equation}
\Delta_\resum (\pTcut) = \max_{v_i \in V_\resum} \bigl\lvert \sigma_0^{v_i} (\pTcut) - \sigma_0^{\rm central} (\pTcut) \bigr\rvert
\,,\end{equation}
where $V_\resum$ is the above set of 35 resummation scale variations. This uncertainty determines the covariance matrix $C_\resum$, and together with $C_\mu$ gives the full covariance matrix.

\subsection{Uncertainties from Clustering Effects}
\label{subsec:clusteringuncertainties}

The purpose of the profile scale variations is to estimate the effect of uncalculated higher-order terms in the cross section. This includes the higher-order corrections in the perturbative series of the various anomalous dimensions, which would be needed for the resummation at N$^3$LL.  While this is effective for the logarithms of $\pTcut/m_H$, which are being resummed, the clustering effects generate an all-orders series of logarithms of $\pTcut/m_H$ and logarithms of $R^2$. In particular, as explained at the end of \subsec{factsoft}, the $\ln R^2$ terms appear as an unresummed series of large logarithms in the rapidity anomalous dimension. The effect of these terms on the resummed cross section is not necessarily well estimated from scale variation of the lowest order term alone.

\begin{table*}[t!]
  \centering
  \begin{tabular}{c | c c c c c c | c c}
  \hline \hline
  order & matching ($H_{gg}$, $B_g$, $S_{gg}$) & nonsingular & $\gamma^g_{H,B,S}$ & $\gamma^g_\nu$ & $\Gamma^g_\mathrm{cusp}$ & $\beta$ & PDF & $\as(m_Z)$ \\ \hline
  NLL$_{p_T}$ & LO & - & $1$-loop & $1$-loop & $2$-loop & $2$-loop & LO & $0.13939$ \\
  NLL$'_{p_T}+$NLO & NLO & NLO & $1$-loop & $1$-loop & $2$-loop & $2$-loop & NLO & $0.12018$ \\
  NNLL$'_{p_T}+$NNLO & NNLO & NNLO & $2$-loop & $2$-loop & $3$-loop & $3$-loop & NNLO & $0.11707$ \\
  \hline\hline
  \end{tabular}
  \caption{Perturbative ingredients entering at each order in resummed perturbation theory.}
\label{table:counting}
\end{table*}

The new clustering effects (those not determined from soft function exponentiation) arising at $\ord{\as^n}$ depend on a coefficient $C_n(R)$, whose small $R$ limit has the form in \eq{CnRdef}.  The term with the most factors of $\ln R^2$ at ${\cal O}(\alpha_s^n)$ gives a contribution to the cross section of the form
\begin{align} \label{eq:clusHO}
\ln\frac{\sigma_0^{(n)}(\pTcut)}{\sigma_\LO} &\supset C_{n,n-1} \biggl[ \frac{\as(\pTcut) C_A}{\pi} \ln R^2 \biggr]^{n-1}
\nn \\ & \quad
\times \biggl[ \frac{\as (\pTcut) C_A}{\pi} \ln \frac{m_H}{\pTcut} \biggr]
\,,\end{align}
where only the lowest $\ord{\as^2}$ clustering coefficient $C_{2,1} = -2.49$ is known [see \eq{C2value}].  Note that $\ln R^2$ dependent terms with more powers of $\ln(m_H/\pTcut)$ are determined by exponentiation through the rapidity RGE [i.e. the terms in \eq{clusHO} arise as higher-order corrections in $\gamma_\nu^g(R)$].

Until a calculation of any of the higher-order clustering coefficients exists, the best we can do is to estimate their effect on the cross section.  To derive an uncertainty estimate from higher-order clustering effects, we use the ansatz $C_{3,2} = \pm C_{2,1}$ and add the corresponding $\ord{\as^3}$ term to the rapidity anomalous dimension.
We have chosen the above way of factoring out color factors and defining the higher-order clustering coefficients, such that $C_{2,1}$ is roughly an $\ord{1}$ number and the higher-order corrections scale with a power of
\begin{equation}
\frac{\as(\pTcut) C_A}{\pi} \ln R^2
\,.\end{equation}
In this way, taking $C_{3,2} = \pm C_{2,1}$ leads to a reasonable estimate of the potential size of the higher-order clustering corrections. For example, for $R = 0.4$, $\pTcut = 25 \GeV$, this factor is $-0.25$, so taking $C_{3,2} = \pm C_{2,1}$ the $\ord{\as^3}$ clustering term would give a 25\% correction to the $\ord{\as^2}$ clustering term. This leads to a clustering uncertainty which is not negligible but fortunately does not dominate the uncertainty.  Numerical results for different parameters of phenomenological interest are given in the next section.

\begin{table*}[t!]
\begin{tabular}{c|cccc}
\hline\hline
& $\sigma_{\geq 0}$ [pb] & $\sigma_0 (\pTcut)$ [pb] & $\sigma_{\geq 1}(\pTcut)$ [pb] & $\e_0 (\pTcut)$
\\\hline
NLL$'_{p_T}$+NLO
\\
$\pTcut = 25 \GeV$ &
$20.46 \pm 3.37\,\, (16.5\%)$ & $11.19 \pm 1.98\,\, (17.7\%)$ & $9.27 \pm 2.76\,\, (29.7\%)$ & $0.547 \pm 0.086\,\, (15.8\%)$
\\
$\pTcut = 30 \GeV$ &
$20.46 \pm 3.37\,\, (16.5\%)$ & $12.70 \pm 2.07\,\, (16.3\%)$ & $7.76 \pm 2.67\,\, (34.5\%)$ & $0.621 \pm 0.090\,\, (14.5\%)$
\\[0.5ex] \hline
NNLL$'_{p_T}$+NNLO ($R = 0.4$)
\\
$\pTcut = 25 \GeV$ &
$21.68 \pm 1.49\,\, (6.9\%)\,\,$ & $12.67 \pm 1.22\,\, (9.6\%)\,\,$ & $9.01 \pm 1.06\,\, (11.8\%)$ & $0.584 \pm 0.040\,\, (6.8\%)\,\,$
\\
$\pTcut = 30 \GeV$ &
$21.68 \pm 1.49\,\, (6.9\%)\,\,$ & $14.09 \pm 0.96\,\, (6.8\%)\,\,$ & $7.60 \pm 0.93\,\, (12.3\%)$ & $0.650 \pm 0.028\,\, (4.4\%)\,\,$
\\[0.5ex] \hline
NNLL$'_{p_T}$+NNLO ($R = 0.5$)
\\
$\pTcut = 25 \GeV$ &
$21.68 \pm 1.49\,\, (6.9\%)\,\,$ & $12.40 \pm 1.12\,\, (9.0\%)\,\,$ & $9.28 \pm 1.03\,\, (11.1\%)$ & $0.572 \pm 0.036\,\, (6.2\%)\,\,$
\\
$\pTcut = 30 \GeV$ &
$21.68 \pm 1.49\,\, (6.9\%)\,\,$ & $13.85 \pm 0.87\,\, (6.3\%)\,\,$ & $7.83 \pm 0.94\,\, (12.0\%)$ & $0.639 \pm 0.026\,\, (4.1\%)\,\,$
\\[0.5ex] \hline
NNLL$'_{p_T}$+NNLO ($R = 0.7$)
\\
$\pTcut = 25 \GeV$ &
$21.68 \pm 1.49\,\, (6.9\%)\,\,$ & $11.97 \pm 1.05\,\, (8.8\%)\,\,$ & $9.71 \pm 0.97\,\, (10.0\%)$ & $0.552 \pm 0.032\,\, (5.7\%)\,\,$
\\
$\pTcut = 30 \GeV$ &
$21.68 \pm 1.49\,\, (6.9\%)\,\,$ & $13.48 \pm 0.83\,\, (6.2\%)\,\,$ & $8.20 \pm 0.92\,\, (11.2\%)$ & $0.622 \pm 0.024\,\, (3.8\%)\,\,$
\\[0.5ex]
\hline\hline
\end{tabular}
\caption{Predictions for various cross sections with complex scale setting $\mu_H = -\img \mu_\FO$ and $\mu_\FO = m_H$ as the central scale choice, and with the total combined perturbative uncertainties.  For convenience we also show the equivalent percent uncertainty in brackets after each result.}
\label{table:numbers}
\end{table*}

\section{Predictions for the LHC}
\label{sec:results}

In this section we present our predictions for the exclusive 0-jet cross section $\sigma_0$, the inclusive 1-jet cross section $\sigma_{\ge 1}$, and the exclusive 0-jet fraction $\epsilon_0$.  In analyzing our results we will consider varying: the perturbative order  (NLL$_{p_T}$, NLL$'_{p_T}+$NLO, and NNLL$'_{p_T}+$NNLO), the choice of jet radius $R$, and the choice of $\pTcut$.  The Higgs mass dependence may also be examined, but we will fix $m_H = 125 \GeV$.  The order of the hard, beam, and soft functions, nonsingular corrections, and anomalous dimensions entering at each order in the resummed cross section are given in \tab{counting}. We use the MSTW 2008 PDFs~\cite{Martin:2009iq} with their $\as(m_Z)$ at the relevant order as shown in \tab{counting}.%
\footnote{At NLL, the $\as$ running order required by the LO PDFs and the resummation differ. In this case, we use the pragmatic solution of including the required $2$-loop beta function coefficients in the RGE evolution kernels, but use the $1$-loop running required by the LO PDFs to obtain the numerical value of $\as$ at a given scale. This mismatch does not happen at the higher orders.}

We start with a summary of our main results. In \tab{numbers} we give our predictions for each of $\sigma_{_\geq 0}$, $\sigma_0$, $\sigma_{\ge 1}$, and $\epsilon_0$ using $\pTcut \in \{25, 30\} \GeV$ and $R \in \{0.4, 0.5, 0.7\}$.  The uncertainties are determined by the  covariance matrix in \eq{Cmures}. The basic parameters in the matrix are the resummation uncertainty $\Delta_\resum$ and the fixed-order uncertainties $\Delta_\tot$, $\Delta_{\mu0}$, and $\Delta_{\mu\ge1} = \Delta_\tot - \Delta_{\mu0}$. The values of these uncertainties for two examples are
\begin{align}  \label{eq:primUnc}
& & \pTcut &= 25 \GeV & \pTcut &= 30 \GeV \nn \\[-3pt]
& & R &= 0.4 & R &= 0.5 \nn \\[3pt]
\Delta_\tot & \; : \; &  & 1.49 \,{\rm pb} & & 1.49\,{\rm pb} \nn \\
\Delta_\resum & \; : \; &  & 0.86\,{\rm pb} & & 0.52\,{\rm pb}  \\
\Delta_{\mu0} & \; : \; &  & 0.87\,{\rm pb} & & 0.70\,{\rm pb} \nn \\
\Delta_{\mu\ge1} & \; : \; &  & 0.62\,{\rm pb} & & 0.79\,{\rm pb} \nn 
\end{align}
which can be compared to total uncertainties quoted in \tab{numbers}.   In \eq{primUnc} the reduction in resummation uncertainties at larger $R$ and $\pTcut$ is to be expected, and is mainly driven by the increase in $\pTcut$.  This is also the main reason for the reduced uncertainties with increasing $\pTcut$ in $\sigma_0$ and $\epsilon_0$  at NNLL$'_{p_T}$+NNLO, seen in \tab{numbers}.

We will discuss additional aspects of \tab{numbers} and associated figures for $\sigma_0$, $\sigma_{\geq 1}$, and $\e_0$ in the following subsections. In \figsthree{sigma0}{sigma1}{eff0} we will show predictions at different orders and compare our most accurate prediction to the NNLO result.  In \eq{clusternum} we will estimate the uncertainty from higher-order clustering terms. Then in \fig{correl} we will plot various correlation coefficients as a function of $\pTcut$, and in \tab{correl} give correlation coefficients for two different values of $R$.  In \app{nopi2}, we will discuss in more detail the impact of the $\pi^2$ summation on our analysis.

\subsection{The 0-Jet Cross Section}
\label{subsec:0jet}

\begin{figure*}[t!]
\begin{center}
\includegraphics[width=\columnwidth]{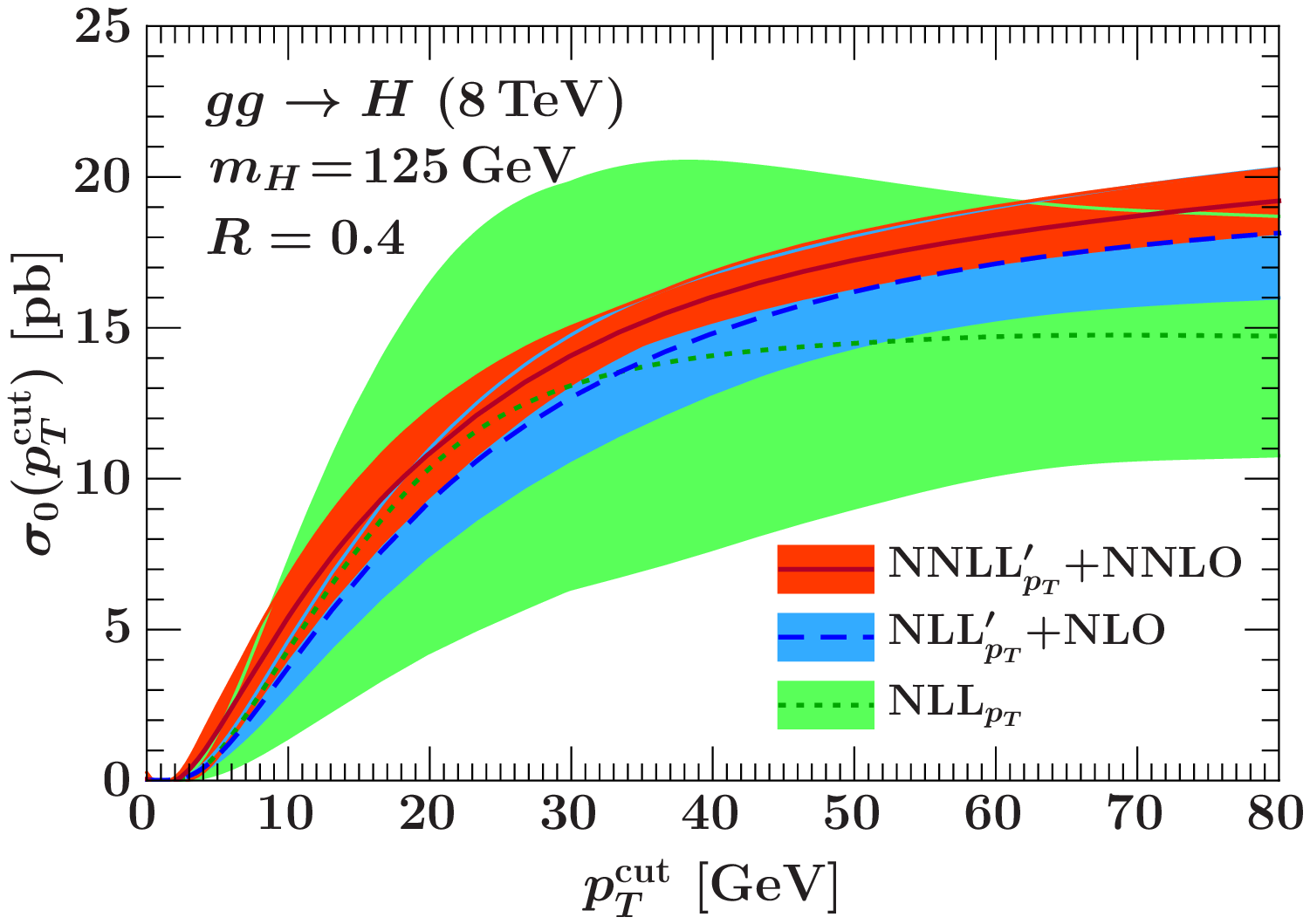}%
\hfill%
\includegraphics[width=\columnwidth]{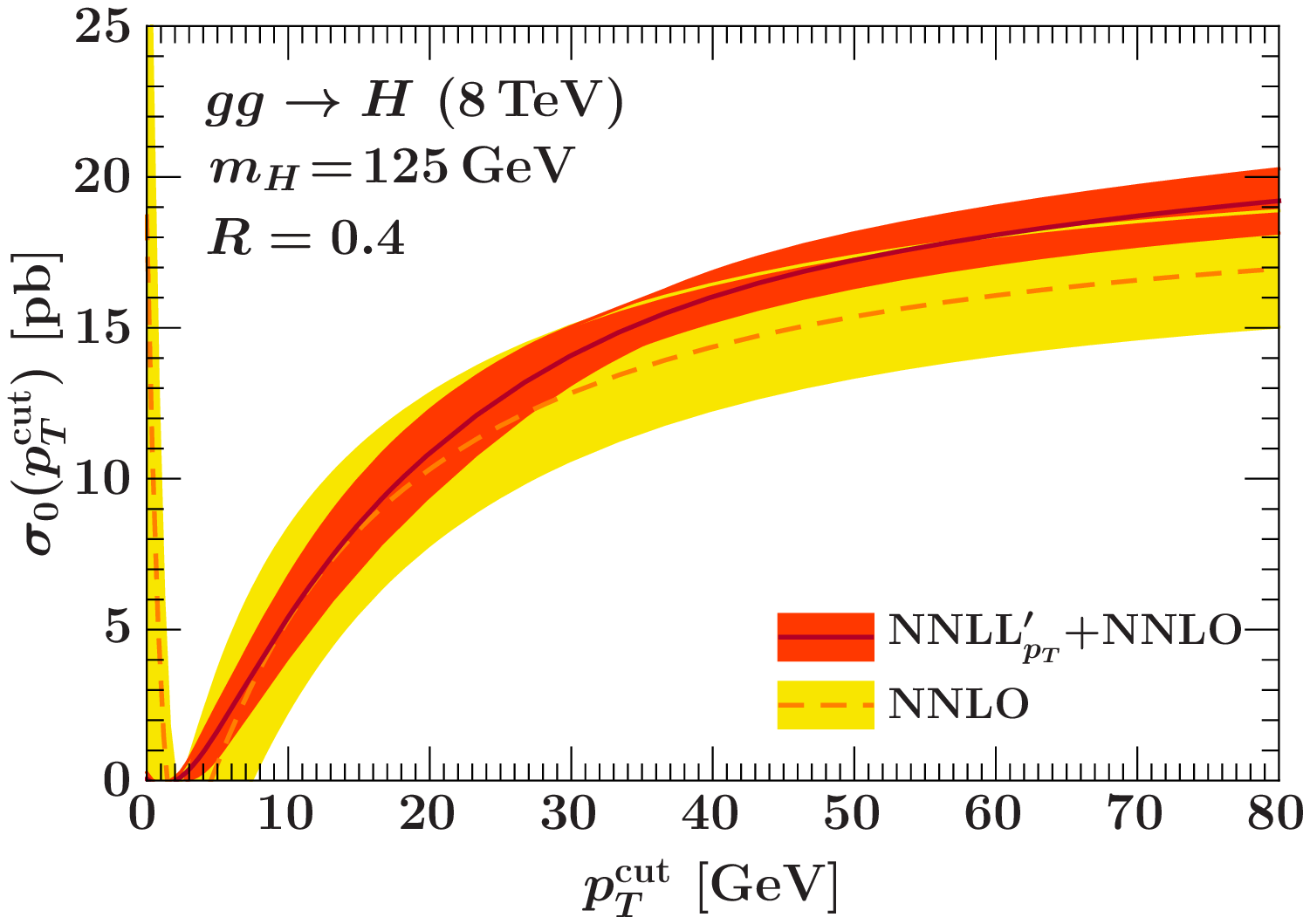}%
\\
\includegraphics[width=\columnwidth]{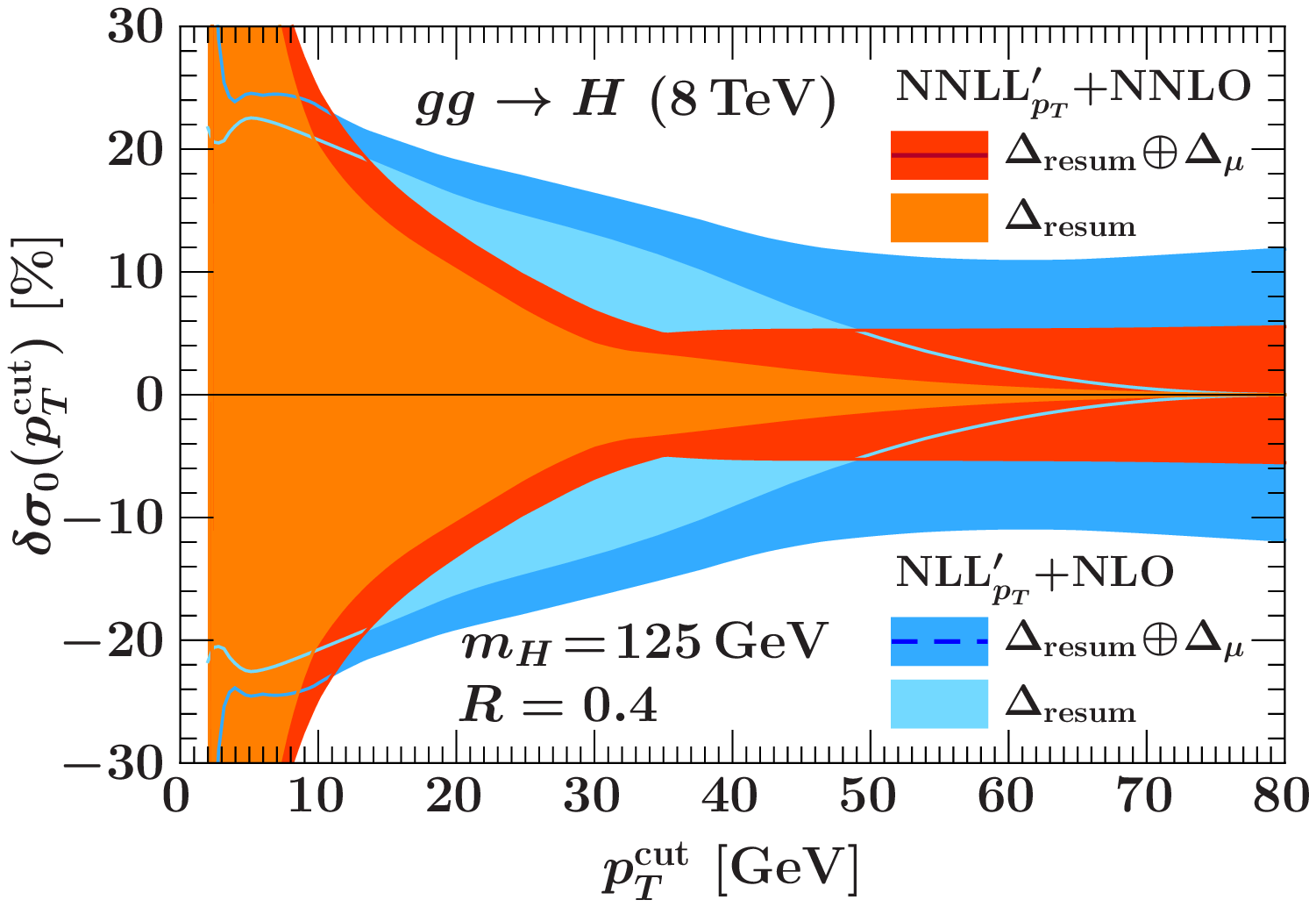}%
\hfill%
\includegraphics[width=\columnwidth]{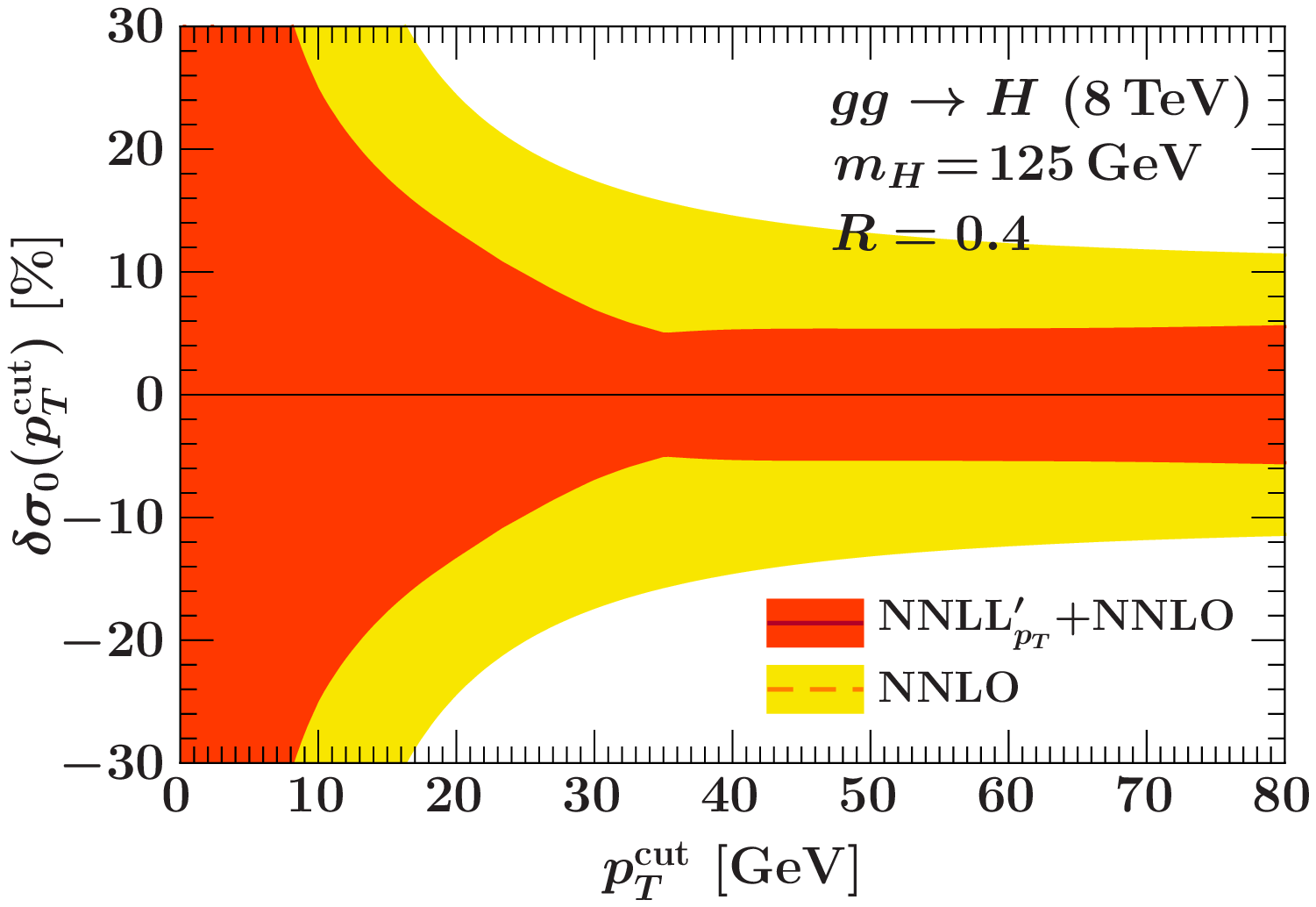}%
\end{center}
\vspace{-3ex}
\caption{The 0-jet cross section for $R = 0.4$ and $m_H=125\GeV$.  On the left we show the NLL$_{p_T}$, NLL$'_{p_T}+$NLO, and NNLL$'_{p_T}+$NNLO predictions. A good convergence and reduction of uncertainties at successively higher orders is observed.  On the right we compare our best prediction at NNLL$'_{p_T}+$NNLO to the fixed NNLO prediction. The lower plots show the relative uncertainty in percent for each prediction. On the lower left the lighter inside bands show the contribution from $\Delta_\resum$ only, while the darker outer bands show the total uncertainty from adding $\Delta_\resum$ and $\Delta_\mu$ in quadrature.}
\label{fig:sigma0}
\end{figure*}

The fundamental quantity measured by experiments that needs to be calculated theoretically is $\sigma_0 (\pTcut, R)$, the fiducial cross section in the $0$-jet bin.  For this reason the predictions discussed here for the 0-jet cross section at NNLL$^\prime_{p_T}$+NNLO are our main results.  The purpose of the resummation is to improve the precision and accuracy of the fixed-order cross section when $\pTcut \ll m_H$, so it is natural to  compare the resummed result to the NNLO cross section. For NNLO we use the central scale $\mu_\FO=m_H$ throughout.  In addition, to verify the validity of our uncertainty analysis it is important to study the convergence of the resummation by studying different orders in the resummed perturbation theory.  We make these comparisons in \fig{sigma0} using $R = 0.4$.  From the top left panel one sees that there is indeed a substantial reduction of uncertainties when increasing the accuracy of the resummation/matching, with higher orders falling inside the uncertainty bands of the lower order results, as desired. From the top right panel one sees that the NNLL$'_{p_T}$+NNLO prediction has noticeably smaller uncertainties than the NNLO prediction.  This is expected for smaller $\pTcut$, but even remains true for larger $\pTcut$ due to the $\pi^2$ summation that is present in the NNLL$'_{p_T}$+NNLO result, but not in the NNLO result.  (The corresponding comparisons for $R = 0.5$ are quite similar, yielding the same conclusions.)

The bottom left panel shows percent uncertainties for the two highest order resummation results, and also breaks them down into the contributions from the resummation uncertainty $\Delta_{\resum}$ and the total uncertainty from combining yield and resummation uncertainties in quadrature. For large $\pTcut$ the yield uncertainties dominate at both NLL$^\prime_{p_T}$+NLO and NNLL$^\prime_{p_T}$+NNLO, since the resummation is not important in this region.  For both of these orders the resummation uncertainty starts to have a relevant impact for $\pTcut \lesssim 40\,{\rm GeV}$.

In the bottom right panel of \fig{sigma0} we show the percent uncertainties relative to the central curve for the NNLL$'_{p_T}$+NNLO and NNLO cross sections.  In this figure the size of the improvement is clear.  For instance, for $R=0.4$ and $\pTcut=25\GeV$ the uncertainty decreases from $20\%$ at NNLO to $9.6\%$ at  NNLL$'_{p_T}+$NNLO. Similar improvements by roughly a factor of $2$ are observed for $\pTcut=30\GeV$ and for $R=0.5$.  Jet binning is a key aspect of the experimental $H\to WW$ and $H\to\tau\tau$ analyses, which will therefore directly benefit from this substantial improvement in the theoretical uncertainties.

The clustering effects provide an additional uncertainty.  Using the procedure discussed in \subsec{clusteringuncertainties}, the relative uncertainty from clustering, $\Delta_0^{\rm clus} (\pTcut) / \sigma_0 (\pTcut)$, is
\begin{align} \label{eq:clusternum}
&\!\!\!\!\!\!\!\!\!\!\!\!\!\!(\Delta_0^{\rm clus} / \sigma_0) (\pTcut) & \pTcut &= 25 \GeV & \pTcut &= 30 \GeV \nn \\
R &= 0.4 \; : \; &  & 3.6\% & & 2.9\% \nn \\
R &= 0.5 \; : \; &  & 2.1\% & & 1.7\%  \\
R &= 0.7 \; : \; &  & 0.5\% & & 0.4\% \nn
\end{align}
Since our method of estimating these uncertainties is likely to be improved in the future by calculations or a better understanding of clustering effects, we have not included them in the plots or in our numbers in \tab{numbers}.  These clustering uncertainties are small compared to the perturbative uncertainties discussed above and shown in \tab{numbers}, but are nonnegligible, so we will quote them as an additional uncertainty on each $0$-jet cross section.  One should interpret these with care since they come from a rough estimate of the higher-order clustering coefficient which could easily be twice as large or one-half as large. As representative final results we quote the following values for $\sigma_0(\pTcut,R)$ with both theoretical uncertainties:
\begin{align} \label{eq:sigma0final}
  \sigma_0(25\,{\rm GeV},0.4 ) &=  12.67  \pm  1.22_\text{pert}  \pm  0.46_\text{clust}  \: {\rm pb}
  \,,\nn\\
   \sigma_0(30\,{\rm GeV},0.5 ) &=  13.85  \pm  0.87_\text{pert}  \pm  0.24_\text{clust}  \: {\rm pb}
 \,.
\end{align}

It is interesting to compare our results and uncertainties for $\sigma_0$ to the NNLL+NNLO results presented earlier in Ref.~\cite{Banfi:2012jm}. Our results build on their results in a few ways.  In particular, our RG approach includes $\pi^2$ resummation, our results are quoted as NNLL$^\prime$ because they go beyond NNLL by including the complete NNLO singular terms in the fixed-order matching (which are the correct boundary conditions for the N$^3$LL resummation), and finally we use a factorization based approach to uncertainties, which also makes predictions for the correlations between the different jet bins.

Comparing $\sigma_0$ at $\pTcut=25\,{\rm GeV}$ and $R=0.4$ our central values agree with those in Ref.~\cite{Banfi:2012jm}, and are well within each other's uncertainties.  Our perturbative uncertainty of $9.6\%$ is a bit smaller than the $13.3\%$ uncertainty for $\sigma_0$ of Ref.~\cite{Banfi:2012jm} which seems reasonable given the above mentioned additions. One important ingredient in this comparison is the inclusion of the $\pi^2$ resummation which improves the convergence of our results and decreases our uncertainty. On the other hand, in Ref.~\cite{Banfi:2012jm} the central scale is chosen to be $\mu_\FO=m_H/2$ which also works in the same direction, decreasing the uncertainty relative to the choice $\mu_\FO=m_H$.  For the total cross section Ref.~\cite{Banfi:2012jm} has a $7.4\%$ uncertainty, whereas we have
$6.9\%$ uncertainty using $\mu_\FO=m_H$ and including $\pi^2$ resummation (see \tab{numbers}).  From \tab{numbersnopi2} in appendix~\app{nopi2} we see that our perturbative uncertainty for $\sigma_0(25 \GeV, 0.4)$ would increase to $12.8\%$ if the $\pi^2$ resummation were turned off (while still taking the central $\mu_\FO=m_H$), and that at this level the uncertainty would become comparable to that of Ref.~\cite{Banfi:2012jm}.  For $\pTcut=30\,{\rm GeV}$ and $R=0.5$ our central values remain perfectly compatible with Ref.~\cite{Banfi:2012jm}, and the uncertainties follow a pattern similar to the case above.

\subsection{The Inclusive 1-Jet Cross Section}
\label{subsec:1jet}

\begin{figure*}[t!]
\begin{center}
\includegraphics[width=\columnwidth]{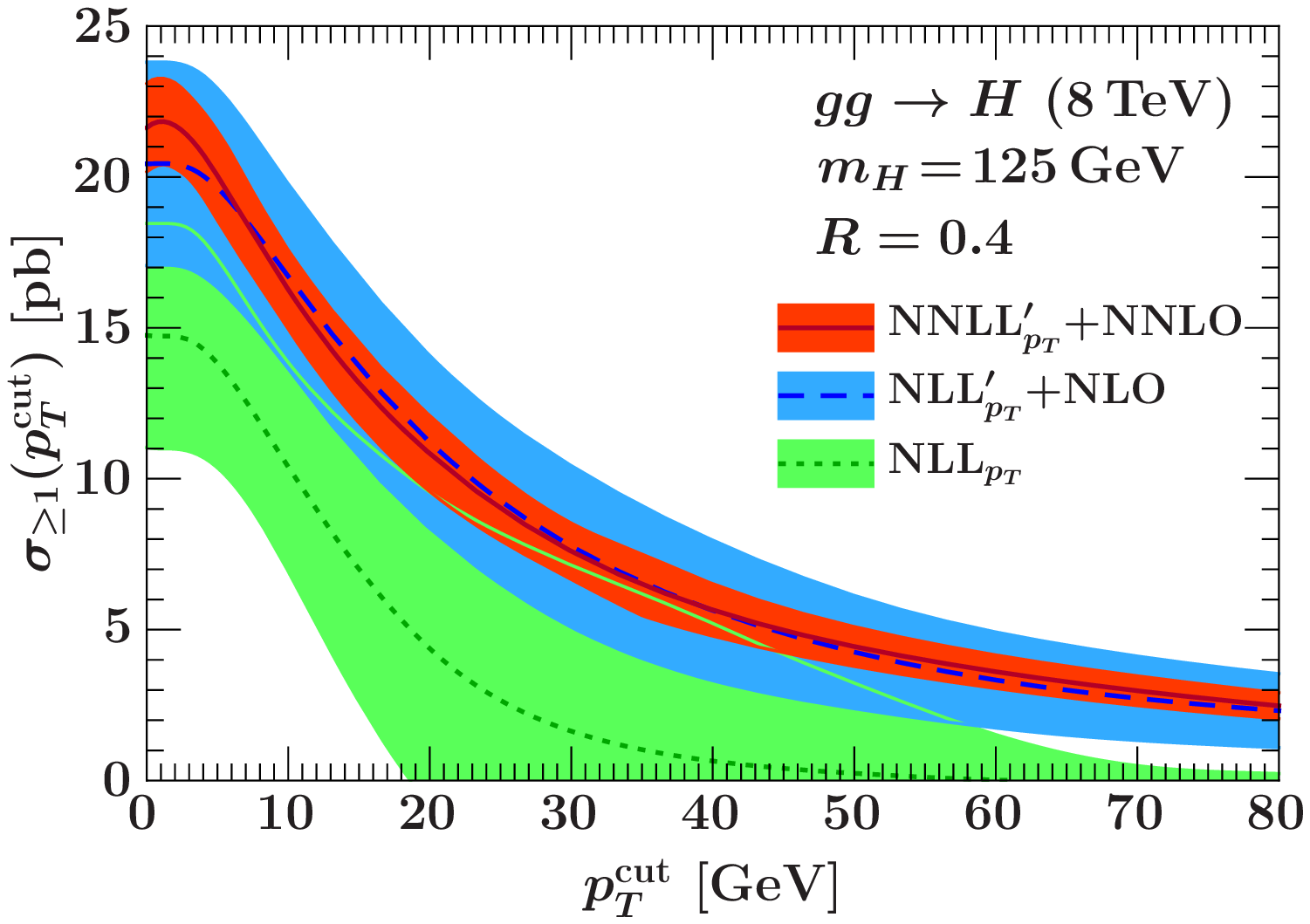}%
\hfill%
\includegraphics[width=\columnwidth]{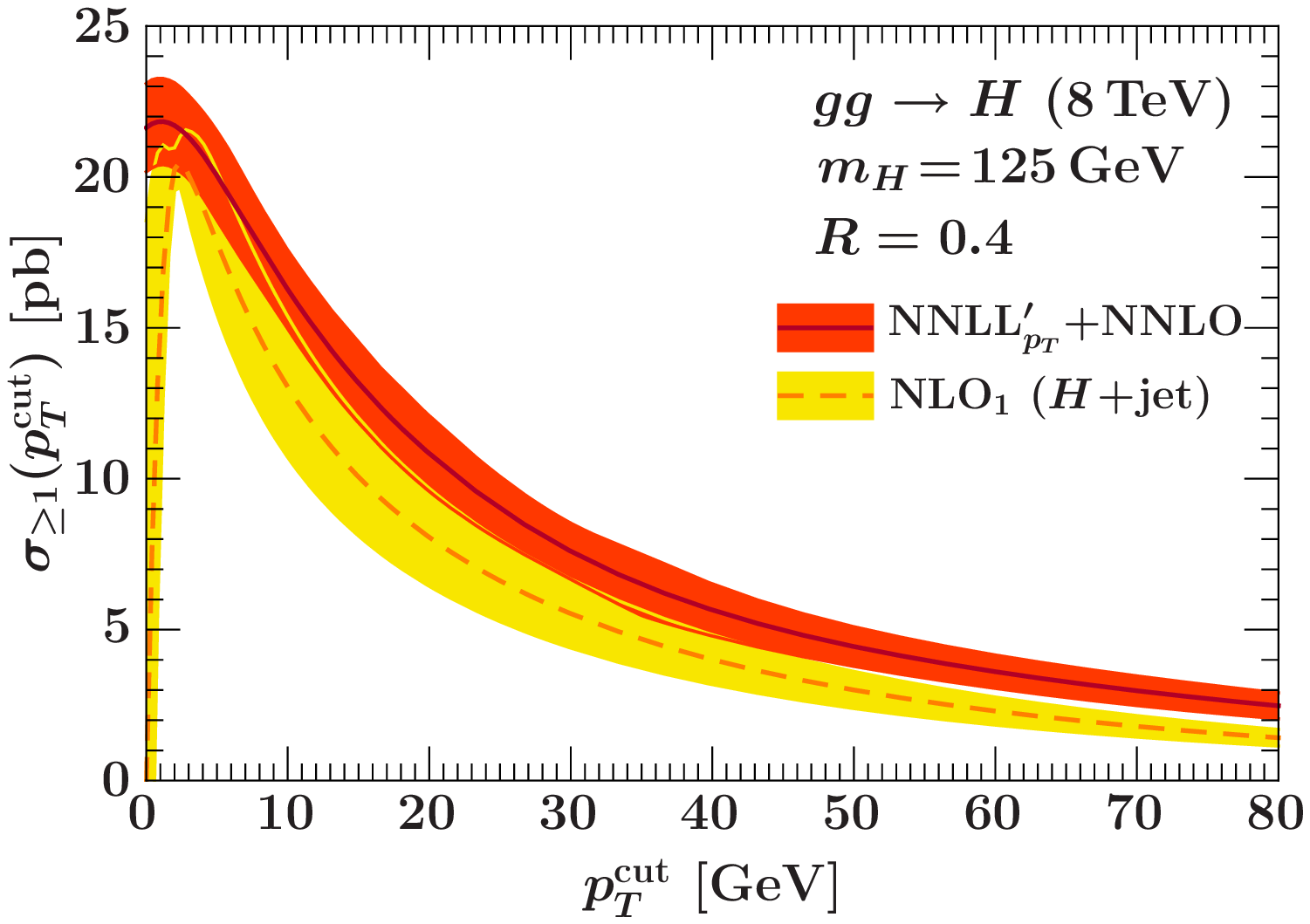}%
\end{center}
\vspace{-3ex}
\caption{The inclusive 1-jet cross section for $R = 0.4$ and $m_H = 125\GeV$.  On the left we show the different orders of our resummed predictions, and on the right we compare our best prediction to that derived from the fixed NNLO cross section.  As in the 0-jet cross section, we observe a good convergence and reduction in uncertainties at successively higher orders of accuracy.}
\label{fig:sigma1}
\end{figure*}

The inclusive 1-jet cross section contains the same jet-veto logarithms as the exclusive 0-jet cross section, 
\begin{equation} \label{eq:sig1pheno}
\sigma_{\ge 1}(\pTcut) = \sigma_{\ge 0} - \sigma_0(\pTcut)
\,.\end{equation}
Here, $\pTcut$ in $\sigma_{\geq 1}(\pTcut)$ is now the lower limit on the $p_T$ of the leading jet in this inclusive cross section. Since our resummation framework consistently includes both $\sigma_0(\pTcut)$ and $\sigma_{\ge 0}$, we can determine a resummed prediction for $\sigma_{\ge 1}(\pTcut)$ from their difference. A nontrivial ingredient in this prediction is determining its perturbative uncertainty via the theory covariance matrix determined in \sec{unc}.

In \fig{sigma1}, we show the convergence of the resummed and matched predictions at different orders, as well as the comparison to the fixed-order cross section.  The total cross section used to obtain $\sigma_{\ge1} (\pTcut)$ is evaluated with an accuracy equal to the fixed-order matching results contained in $\sigma_0(\pTcut)$.  This is required to enforce $\sigma_{\ge1} (\pTcut \to \infty) \to 0$.  For this reason in the left panel of \fig{sigma1} the NLL$_{p_T}$ distribution (whose matching does not even include the full tree-level matrix element for the $H+1$ jet rate) is lower than the higher-order distributions. The NNLL$'_{p_T}$+NNLO distribution is well contained within the  NLL$'_{p_T}$+NLO uncertainty band, with the expected improvement in accuracy. Note that when including the resummation, $\sigma_{\ge 1}(\pTcut)$ approaches the total cross section as $\pTcut\to 0$, whereas it would diverge at fixed order.

In the right panel of \fig{sigma1} we compare the fixed-order result for $\sigma_{\geq 1}(\pTcut)$ with the result obtained from \eq{sig1pheno} using our NNLL$'_{p_T}$+NNLO  $0$-jet distribution. (We label the NNLL$'_{p_T}$+NNLO prediction as such to be consistent with our predictions for other observables, although in terms of the fixed-order contributions it is not beyond the NLO result for $H+\!\ge$1 jet denoted as NLO$_1$ in the figure.)   The resummed prediction for $\sigma_{\geq1}(\pTcut)$ is larger than the NLO$_1$ result due to the summation of $\pi^2$ terms in $\sigma_{\ge0}$ and $\sigma_0(\pTcut)$ in \eq{sig1pheno}. Without this $\pi^2$ summation, the resummed $\sigma_{\ge 1}(\pTcut)$ would give a slightly lower rate than at fixed order.  For $R=0.4$ and $\pTcut=25\,{\rm GeV}$ the fixed-order uncertainty is $20\%$. It is reduced to $11.8\%$ at NNLL$^\prime_{p_T}$+NNLO (see \tab{numbers}).  This reduction is similar to what was observed for $\sigma_0$, as is the mild dependence on $R$. On the other hand, increasing $\pTcut$ to $30\GeV$ does not really change the relative uncertainty for $\sigma_{\ge 1}$, unlike for $\sigma_0$.   Note the importance of the theory correlations here, since we can see from \eq{primUnc} that the yield uncertainty $\Delta_{\mu\ge 1}$ alone behaves in the opposite fashion.

\begin{figure*}[t!]
\begin{center}
\includegraphics[width=\columnwidth]{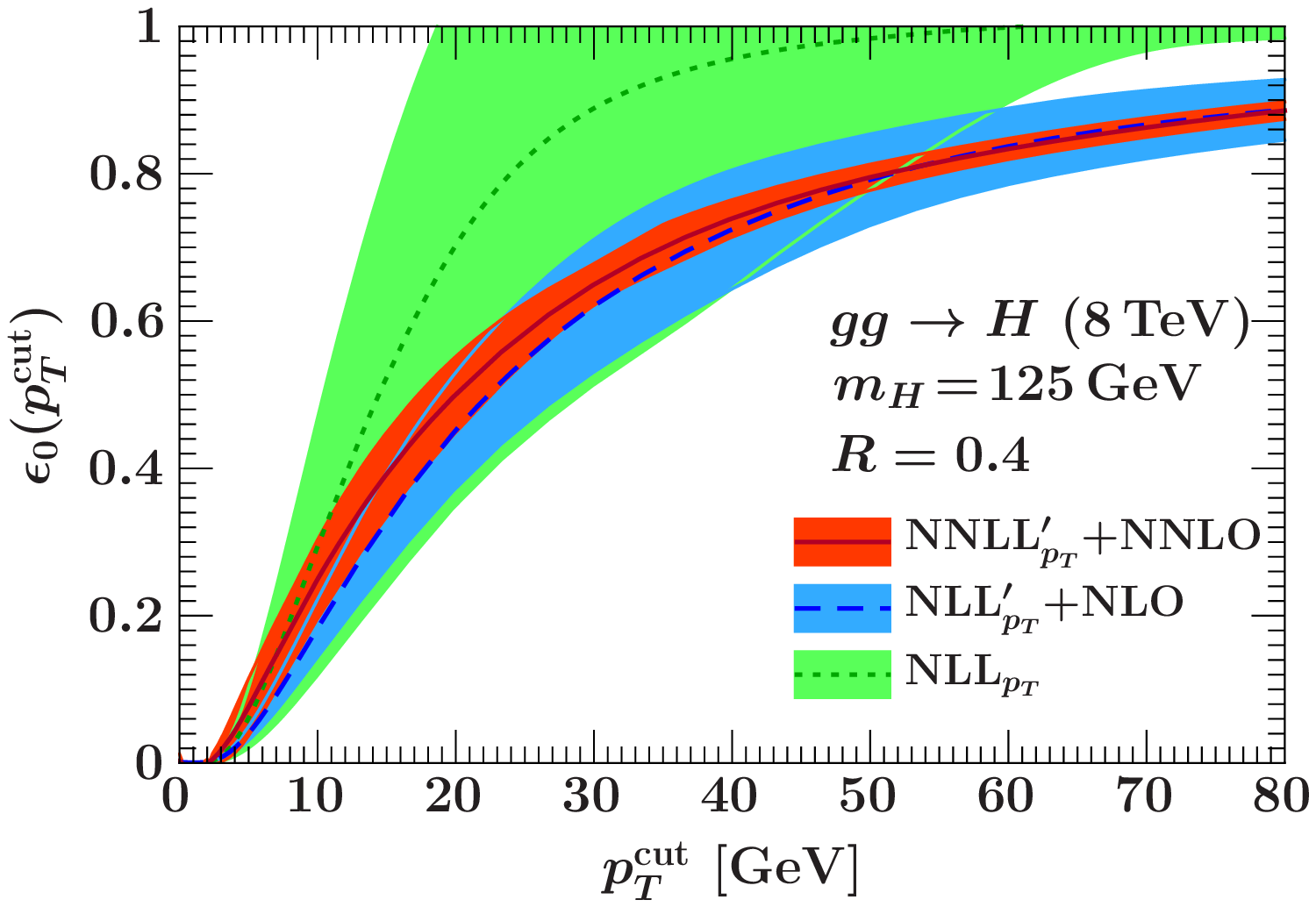}%
\hfill%
\includegraphics[width=\columnwidth]{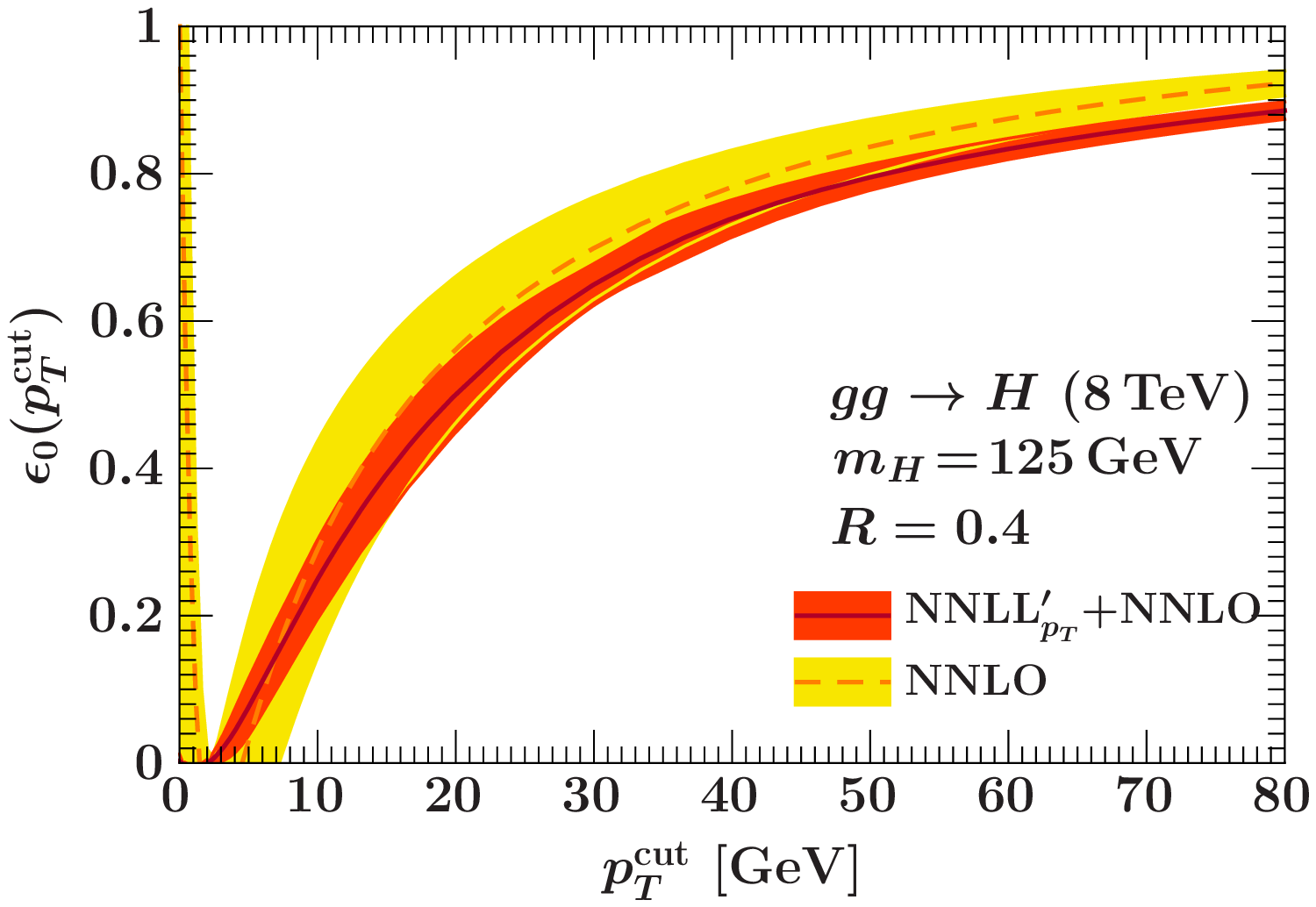}%
\end{center}
\vspace{-3ex}
\caption{The 0-jet efficiency for $R = 0.4$ and $m_H = 125\GeV$.  On the left we show the different orders in our resummed predictions, and on the right we compare our best prediction to that derived from the fixed NNLO cross section.  Because the efficiency is the ratio of the 0-jet and total cross sections, the correlated fixed-order scale uncertainty in each quantity reduces the uncertainty in the 0-jet efficiency, making it relatively more accurate than the cross section itself.}
\label{fig:eff0}
\end{figure*}

Our resummed results for the inclusive 1-jet cross section $\sigma_{\ge 1}(\pTcut, R)$ provide improved predictions compared to the accuracy of its NLO result, but should be used together with the appropriate theory uncertainty correlations determined here. As representative final results for $\sigma_{\ge 1}(\pTcut,R)$, where we also include the uncertainty from clustering estimated as in \eq{sigma0final}, we quote
\begin{align}
  \sigma_{\ge 1}(25\,{\rm GeV},0.4 ) &=  9.01  \pm  1.06_\text{pert}  \pm  0.46_\text{clust}  \: {\rm pb}
  \,,\nn\\
   \sigma_{\ge 1}(30\,{\rm GeV},0.5 ) &=  7.83  \pm  0.94_\text{pert}  \pm  0.24_\text{clust}  \: {\rm pb}
 \,.
\end{align}
Note that the clustering uncertainties have a larger relative size here ($5.1\%$ and $3.0\%$) since $\sigma_{\geq 1}$ is numerically smaller than $\sigma_0$.

Recently, the $gg\to Hg$ contribution to the $H + \!\ge$1-jet cross section has been calculated at NNLO \cite{Boughezal:2013uia}.  This calculation includes all $\ord{\as^3}$ corrections which include logarithms of $\pTcut / m_H$, $\pi^2$ terms, and nonsingular contributions.  Our resummed calculation captures all of the logarithms of $\pTcut / m_H$ except for the single logarithms (which would require N$^3$LL resummation) as well as the $\pi^2$ terms at $\ord{\as^3}$, but does not include any nonsingular contributions.  In contrast, the fixed-order calculation does not include the resummation of the $\pTcut$ logarithms or $\pi^2$ terms beyond $\ord{\as^3}$.  The different theoretical ingredients in these  two calculations makes a comparison between them interesting.  In fact, for phenomenologically relevant parameters the $gg\to Hg$ NNLO calculation finds a $K$-factor relative to NLO that is quantitatively similar to the increase over the NLO cross section that we observe between the two central curves in the right panel of \fig{sigma1}.  As mentioned above, in our case the resummation of the $\pTcut$ logarithms lowers the 1-jet inclusive cross section relative to fixed NLO, but including also the $\pi^2$ summation raises it above.  Although the purely virtual $\pi^2$ terms from the hard function cancel out in \eq{sig1pheno}, there are real-virtual cross terms involving $\pi^2$ factors  in $\sigma_{\ge 0}$ that are not canceled.  This suggests that these $\pi^2$ terms may play an important role in determining the magnitude of the NNLO $K$-factor.  (In contrast, the $\pi^2$ terms that can be determined from imaginary scale setting in the exclusive $H+$1-jet cross section are known to not play a dominant role at NLO~\cite{Jouttenus:2013hs}.)

\subsection{The 0-Jet Efficiency}
\label{subsec:0jeteff}

Another observable that can be predicted using our results is the 0-jet efficiency,
\begin{align}  \label{eq:e0pheno}
  \epsilon_0(\pTcut) &= \frac{\sigma_0(\pTcut)}{\sigma_{\ge 0}}\,.
\end{align}
Once again it is important to account for the correlations in theoretical uncertainties when computing the uncertainty in this observable according to \eq{e0percent}.  In \fig{eff0}, we plot $\epsilon_0(\pTcut)$ and its uncertainty as a function of $\pTcut$ for $R = 0.4$ and we give explicit numbers in \tab{numbers}.  At NLL$^\prime_{p_T}$+NLO the relative uncertainties for $\sigma_0$ and $\epsilon_0$ are similar, but this is no longer the case at NNLL$^\prime_{p_T}$+NNLO.  With the decreased uncertainties that occur at this order,
a more significant amount of the uncertainties in the numerator and denominator of \eq{e0pheno} become positively correlated and cancel.  As a result, our 0-jet efficiency at NNLL$^\prime_{p_T}$+NNLO has smaller relative uncertainties than our 0-jet cross section. This is reflected in both the numbers in \tab{numbers} and in the results shown in \fig{eff0}.

In the left panel of \fig{eff0} we show results for the efficiency at different orders. The results at NNLL$^\prime_{p_T}$+NNLO are within the uncertainty band of the lower order NLL$^\prime_{p_T}$+NLO results, and again display an improved level of precision. In the right panel of \fig{eff0} we see that the comparison of $\epsilon_0(\pTcut)$ between NNLL$^\prime_{p_T}$+NNLO  and pure NNLO follows a similar pattern of improvement to what we have already observed for the 0-jet and inclusive 1-jet cross sections.  

Since the 0-jet efficiency is the more fundamental quantity in the framework of Ref.~\cite{Banfi:2012jm}, it makes sense to extend the comparison made in \subsec{0jet} to this observable, again taking $R=0.4$ and $\pTcut=25\,{\rm GeV}$.  At NNLL+NNLO Ref.~\cite{Banfi:2012jm} has a $11.5\%$ perturbative uncertainty for $\epsilon_0$, which in their framework is assumed to be independent from the uncertainty in the total cross section.  Thus, their uncertainty for $\sigma_0$ is always larger than that for $\epsilon_0$.  This $11.5\%$ uncertainty for their $\epsilon_0$ is close to the $9.6\%$ uncertainty for our $\sigma_0$, but larger than the $6.8\%$ uncertainty for our $\epsilon_0$.  For the analysis of Ref.~\cite{Banfi:2012jm} there is no corresponding cancellation of uncertainties between the numerator and denominator of \eq{e0pheno}, and hence the same cancellation that we observe does not occur.

\subsection{Correlations}
\label{subsec:correlations}

\begin{figure*}[t!]
\begin{center}
\includegraphics[width=\columnwidth]{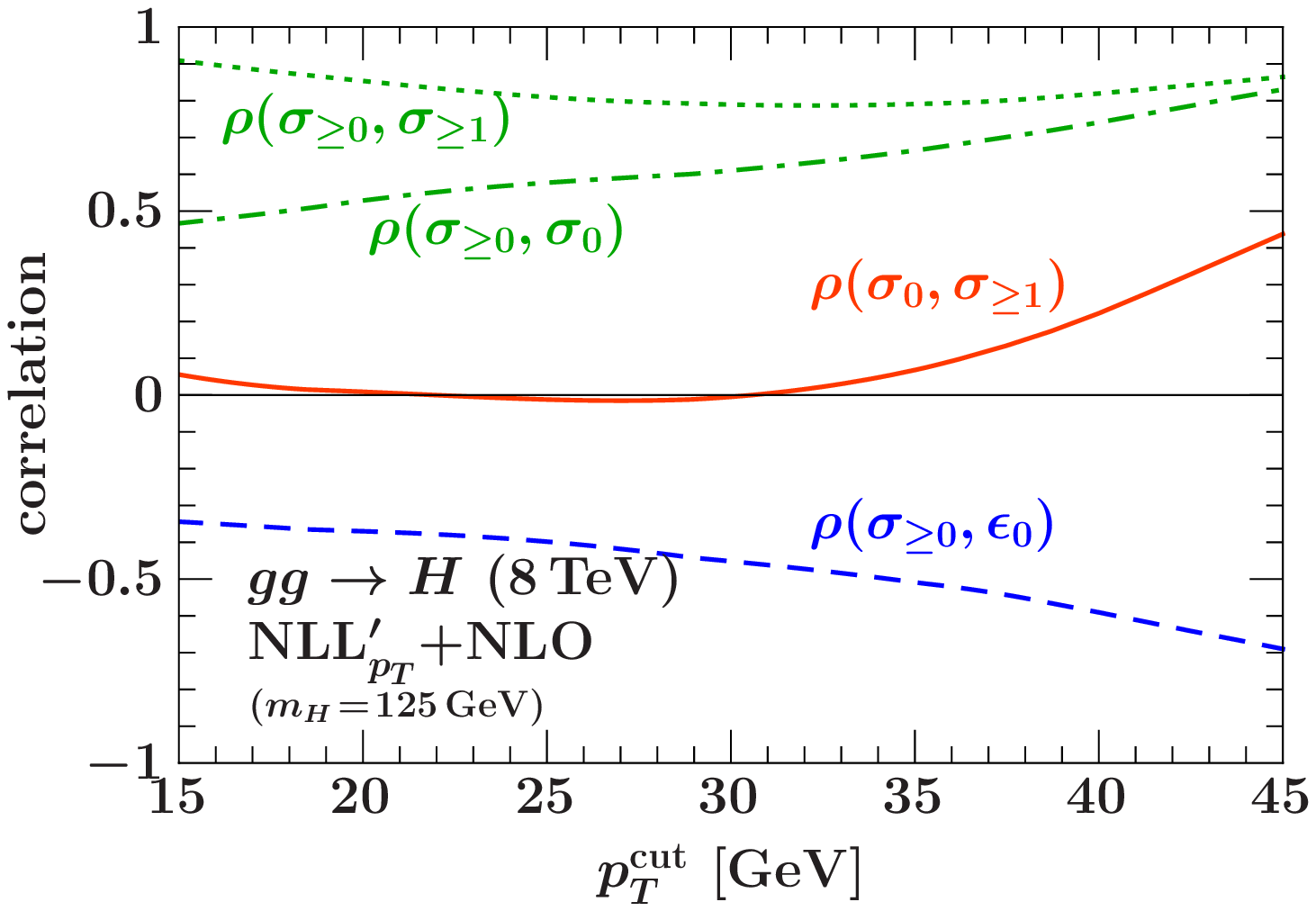}%
\hfill%
\includegraphics[width=\columnwidth]{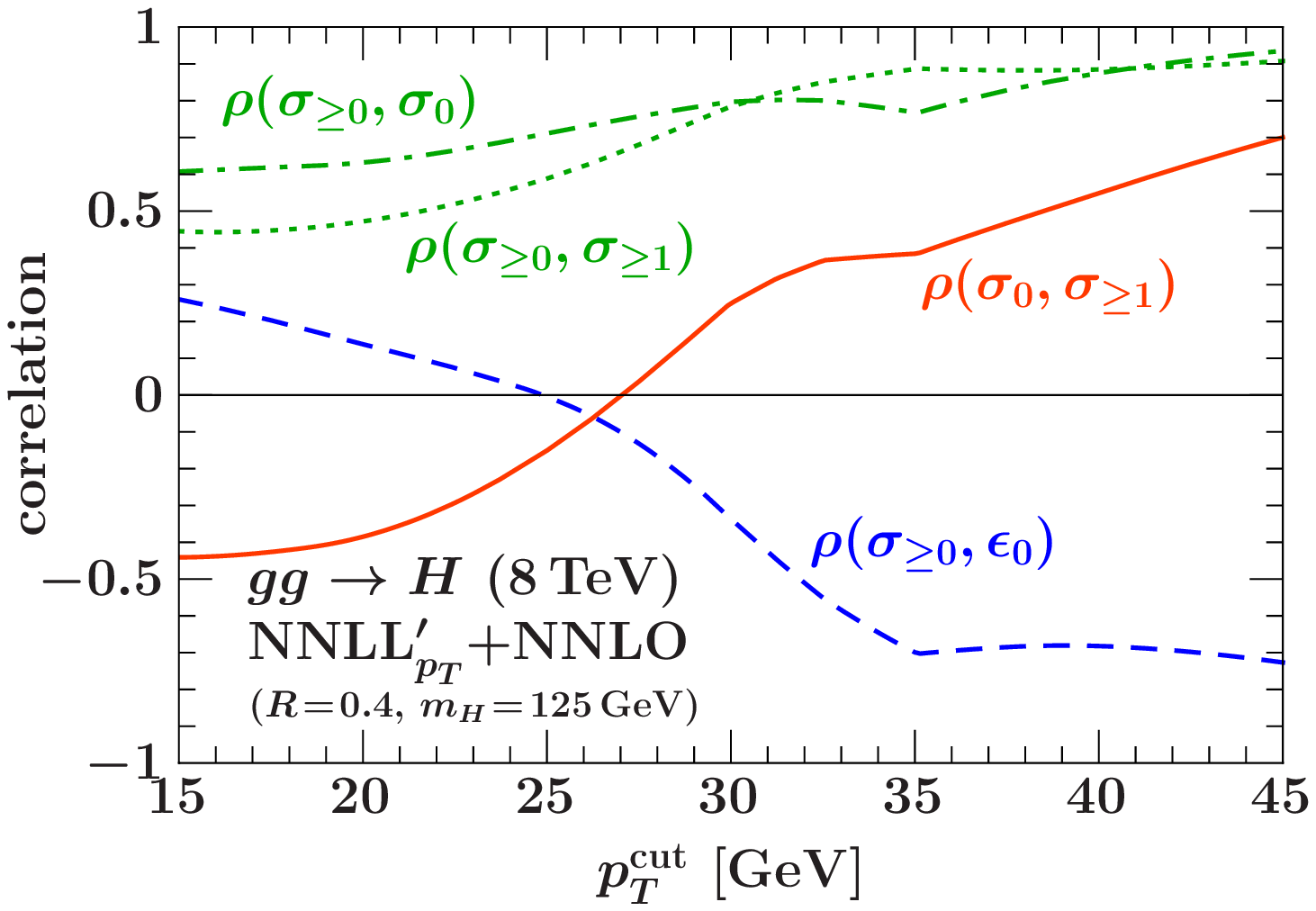}%
\end{center}
\vspace{-3ex}
\caption{Predicted correlation coefficients for the total perturbative uncertainties within the resummed predictions between different observables at NLL$_{p_T}'+$NLO (left) and NNLL$_{p_T}'+$NNLO (right, with $R=0.4$). Since the correlations result from the interplay between the relative sizes of the $C_\mu$ and $C_\resum$ components, the changes between orders is not unexpected.}
\label{fig:correl}
\end{figure*}

\begin{table}[t!]
\begin{tabular}{c|cccc}
\hline\hline
$\pTcut = 30\GeV$ & $\sigma_{\geq 0}$ & $\sigma_0(\pTcut)$ & $\sigma_{\geq 1}(\pTcut)$ & $\e_0(\pTcut)$
\\\hline
$R = 0.4$
\\
$\sigma_{\geq 0}$ &
$1$ & $0.80$ & $0.78$ & $-0.34$
\\
$\sigma_0(\pTcut)$ &
 & $1$ & $0.25$ & $0.30$
\\
$\sigma_{\geq 1}(\pTcut)$ &
 &  & $1$ & $-0.85$
\\
$\e_0(\pTcut)$ &
 &  &  & $1$
\\[0.5ex]\hline
$R = 0.5$
\\
$\sigma_{\geq 0}$ &
$1$ & $0.81$ & $0.84$ & $-0.44$
\\
$\sigma_0(\pTcut)$ &
 & $1$ & $0.35$ & $0.18$
\\
$\sigma_{\geq 1}(\pTcut)$ &
 &  & $1$ & $-0.86$
\\
$\e_0(\pTcut)$ &
 &  &  & $1$
\\[0.5ex]\hline\hline
\end{tabular}
\caption{Correlations in the perturbative uncertainties between different observables at $\pTcut = 30 \GeV$ for $R = 0.4$ and $R = 0.5$.}
\label{table:correl}
\end{table}

When evaluating the perturbative uncertainties via the profile scale variations as discussed in \subsec{profileunc}, the correlations in the total perturbative uncertainties between the different observables are automatically predicted by the resulting total covariance matrix $C_\mu + C_\resum$.  In previous subsections we have highlighted a few cases where these correlations are important for determining uncertainties, and in this section we discuss them in more detail.

As an example, in \tab{correl} we give the correlation coefficients obtained at $\pTcut = 30\GeV$ for both $R = 0.4$ and $R = 0.5$. One observes that they have a fairly mild dependence on $R$. On the other hand, since the correlations arise from the interplay between the relative size of the anticorrelated component $C_\mu$ and correlated component $C_\resum$, they can have a much stronger dependence on $\pTcut$.  Similarly, the correlation matrix can also change by a large amount between perturbative orders because uncertainties are decreased by going to higher order, and therefore the relative importance of $C_\mu$ and $C_\resum$ can change. These two features are illustrated in \fig{correl}. The $\pTcut$ dependence is strongest in the correlations between the inclusive cross section, $\sigma_{\geq 0}$, and the exclusive $0$-jet observables $\sigma_0$ (solid orange lines) and $\epsilon_0$ (blue dashed lines). The reason for this is that the $0$-jet observables receive contributions from $C_\resum$, whose importance relative to $C_\mu$ depends on $\pTcut$, while $\sigma_{\geq 0}$ has no contribution from $C_\resum$. We also see that at NNLL$'_{p_T}+$NNLO the correlation between $\sigma_0$ and $\sigma_{\geq 1}$ decreases toward smaller $\pTcut$ and turns negative below $\lesssim 30\GeV$, because the anticorrelated migration uncertainties from $C_\resum$ start dominating over their common correlated yield uncertainty in $C_\mu$.  This anti-correlation is not so evident in the resummed result at NLL$'_{p_T}+$NLO since $C_\mu$ plays a bigger role at this order. Finally, we observe that in the large $\pTcut$ regime, where the resummation turns off and the $C_\mu$ contributions dominate, the correlations between the 0-jet efficiency and the total cross section in our formalism approaches $-1$, as it must.  For large $\pTcut$ the correlations between any two cross sections tends to $1$, also as they must.

From this discussion it should also be apparent that we do not expect the correlations obtained after resummation to be the same as in the pure fixed-order calculation. Indeed, including the resummation the perturbative uncertainties in the logarithmic series induced by the jet binning are significantly reduced compared to in the fixed-order case. This means the correlation between the uncertainties in $\sigma_0$ and $\sigma_{\geq 1}$ should be more negative at fixed order. This is indeed what happens when using the method of Ref.~\cite{Stewart:2011cf}, for which at pure NNLO we find $\rho(\sigma_0, \sigma_{\geq1})$ rises from $-0.7$ to $-0.2$ over the $\pTcut$ range shown in \fig{correl}. The added advantage of the resummation framework used here is that it automatically provides theory based handles to estimate both the correlated contributions $C_\mu$ and anticorrelated contributions in $C_\resum$ without having to make an assumption about the correlation between any two quantities.  As a final cautionary note, we remark that one should recall that the magnitude of the correlation coefficients does not indicate the relative importance of their entries in determining the final uncertainties since the size of the corresponding diagonal uncertainties is also required.

\section{Conclusions}
\label{sec:conclusions}

In this paper we have presented results for  Higgs production via gluon fusion  with a jet veto.  Jets are identified with a $\kt$-type clustering algorithm (which includes the experimentally used anti-$\kt$ algorithm) with jet radius $R$, and are vetoed via the requirement $p_T^\jet < \pTcut$.  The logarithms of $\pTcut/m_H$ are resummed to NNLL$'$ and the resummation is matched to the full fixed NNLO cross section.  Our analysis is based around the small $R$ limit, where the cross section can be factorized into hard, beam, and soft functions.  To achieve NNLL$'$ order we computed the relevant soft function to ${\cal O}(\alpha_s^2)$ and computed the full $\alpha_s^2\ln R^2$ term for the beam function, determining the remaining $\pTcut$ independent ${\cal O}(\alpha_s^2)$ terms in the beam function numerically.  Our resummation results also include $\pi^2$ summation in the hard corrections through imaginary scale setting. To consistently incorporate the full NNLO result we made use of profile functions that properly handle both the small and large $\pTcut$ regions, and in particular the experimentally relevant transition region in between. We also included a precise numerical determination of the ${\cal O}(\alpha_s^2)$ nonsingular terms. Our results include predictions for the exclusive 0-jet cross section, the 0-jet efficiency, and the inclusive 1-jet cross section.  

A key aspect of our numerical analysis is the robust estimation of perturbative uncertainties.  The uncertainty comes from two independent components: overall yield uncertainties (which are correlated between jet bins) and resummation uncertainties (related to predicting the migration between jet bins as we vary $\pTcut$). Each of these can be estimated through the variation of various invariant mass and rapidity scales in the factorization theorem.  The uncertainty framework discussed in \sec{unc} allows us to construct the complete covariance matrix for the total, exclusive 0-jet, and inclusive 1-jet cross sections.  

In \sec{results}, we presented results for the 0-jet cross section, the inclusive 1-jet cross section, and the 0-jet efficiency.  Our numerical results for several phenomenological points of interest ($\pTcut = 25, 30 \GeV$ and $R = 0.4, 0.5, 0.7$) are given in \tab{numbers}.  The precision of the predictions increases significantly as the resummation and matching is improved, from NLL$_{p_T}$ to NLL$'_{p_T}$+NLO to NNLL$'_{p_T}$+NNLO.  For the most precise predictions, the uncertainties are significantly smaller than the fixed-order NNLO uncertainties, which are currently the nominal benchmark uncertainties for the experimental $H\to WW$ and $H\to \tau\tau$ analyses. Our results add a few additional ingredients on top of the NNLL results in Ref.~\cite{Banfi:2012jm}, in particular: by including $\pi^2$ summation~\cite{Ahrens:2008qu}, by including the complete NNLO singular terms in the fixed-order matching for soft and beam functions at ${\cal O}(\alpha_s^2)$, and because our factorization based framework also makes predictions for both correlated and anticorrelated contributions to the theory uncertainty correlation matrix between different jet bins. We observe a corresponding modest improvement in the size of the uncertainties, where details can be found in \subsec{0jet} and \subsec{0jeteff}.

Our results are part of an ongoing effort to more completely understand jet vetoes for Higgs production and their associated uncertainties. The $H + 0$-jet cross section is an excellent testing ground for the new methods being developed to improve the theoretical predictions. Currently, the fixed-order perturbative uncertainties due to the jet binning in the $H \to WW$ analysis are the dominant systematic uncertainties. Our results can be directly applied to provide improved theory predictions with substantially reduced perturbative uncertainties.

\begin{acknowledgments}
We thank Robert Schabinger for assistance in converting our numerical values of the 2-loop non-cusp anomalous dimensions into analytic expressions.  The authors thank each other's institutions and the Erwin Schr\"odinger Institute program ``Jets and Quantum Fields for LHC and Future Colliders'' for hospitality while portions of this work were completed.

This work was supported in part by the Director, Office of Science, Offices of
Nuclear Physics and High Energy Physics of the U.S. Department of Energy under the
Grant No.  DE-FG02-94ER40818 and the Contract No. DE-AC02-05CH11231,
the DFG Emmy-Noether grant TA 867/1-1, 
and the US National Science Foundation, grant NSF-PHY-0705682, the LHC Theory Initiative.

This research used resources of the National Energy Research Scientific Computing Center,
which is supported by the Office of Science of the U.S. Department of Energy under
Contract No. DE-AC02-05CH11231.
\end{acknowledgments}

$\\$
\noindent\textit{\textbf{Note Added:}}
$\\[3pt]$
While finalizing this paper Ref.~\cite{Becher:2013xia} appeared, which also makes predictions for the $H+0$-jet cross section including contributions beyond the NNLL results of Ref.~\cite{Banfi:2012jm}.  In this note added we compare their theoretical ingredients with ours.

Regarding the derivation of factorization for terms of $\ord{R^2}$ we believe the discussion of rapidity scaling in our footnote 2 still applies to Ref.~\cite{Becher:2013xia}.

A common goal of both our work and Ref.~\cite{Becher:2013xia} is the inclusion of fixed-order corrections from the low-energy matrix elements (corresponding to beam and/or soft functions) that are needed as ingredients in a calculation at N$^3$LL order.  In our analysis we have fully calculated the $\ord{\alpha_s^2}$ soft function, including the $R$-dependent anomalous dimension and finite corrections that depend on $\ln R^2$.  In addition, we have calculated the finite corrections in the beam function that depend on $\ln R^2$. Thus, the dominant $R$ dependence has been fully determined analytically, and the only numerical ingredient is the remaining contribution in the beam function.  In contrast, in Ref.~\cite{Becher:2013xia} an analytic calculation is done for the anomalous dimension terms, but a numerical extraction is done for the combined finite soft $+$ beam contributions including their $R$ dependence.  We make use of the rapidity renormalization group in our analysis, including rapidity scale variations in our uncertainties to estimate the size of higher-order rapidity logarithms, while Ref.~\cite{Becher:2013xia} accounts for these contributions using the ``collinear anomaly'' formalism without variations of the rapidity scales.  A resummation of $\pi^2$ contributions through imaginary scale setting is used in both our work and their work.

Ref.~\cite{Becher:2013xia} refers to the accuracy of their resummation as ``N$^3$LL$_{\rm p}$'', where ``p'' stands for partial, which can be contrasted with our NNLL$'$.   As far as perturbative ingredients that have been either calculated analytically or extracted numerically, both our results include the same theoretical ingredients.  Ref.~\cite{Becher:2013xia} makes an additional ansatz about the anomalous dimensions required for N$^3$LL resummation, since none of the required coefficients are currently known. Their method of estimating and varying the size of these coefficients in some range is another method for estimating uncertainties from unknown higher-order perturbative corrections. It does not however improve the perturbative accuracy of the resummation beyond NNLL$'$ order.

In our analysis we have used profile scales to properly describe the transition between the resummation and fixed-order regimes, which ensures that we have canonical scales in the small $\pTcut$ region and also reproduce the fixed-order cross section in the large $\pTcut$ limit.  In contrast, Ref.~\cite{Becher:2013xia} limit themselves to using canonical scales, which can only be used to properly describe the cross section in the small $\pTcut$ region below the transition region.  As we have seen in our analysis, for phenomenologically relevant values of $\pTcut$, the cross section and its uncertainties are influenced by the transition region.  The connection to the fixed-order cross section also provides an important constraint when predicting correlations (which are not considered in Ref.~\cite{Becher:2013xia}).  Overall, our method of calculating perturbative uncertainties by varying all scales appearing in the RGE is therefore quite different from Ref.~\cite{Becher:2013xia}.  Numerically, the resummed perturbation theory as organized in Ref.~\cite{Becher:2013xia} show a slower convergence (as shown, e.g., in their Figs. 8 and 11) compared to our results shown in \fig{sigma0}.

\appendix

\begin{figure*}[ht!]
\includegraphics[width=\columnwidth]{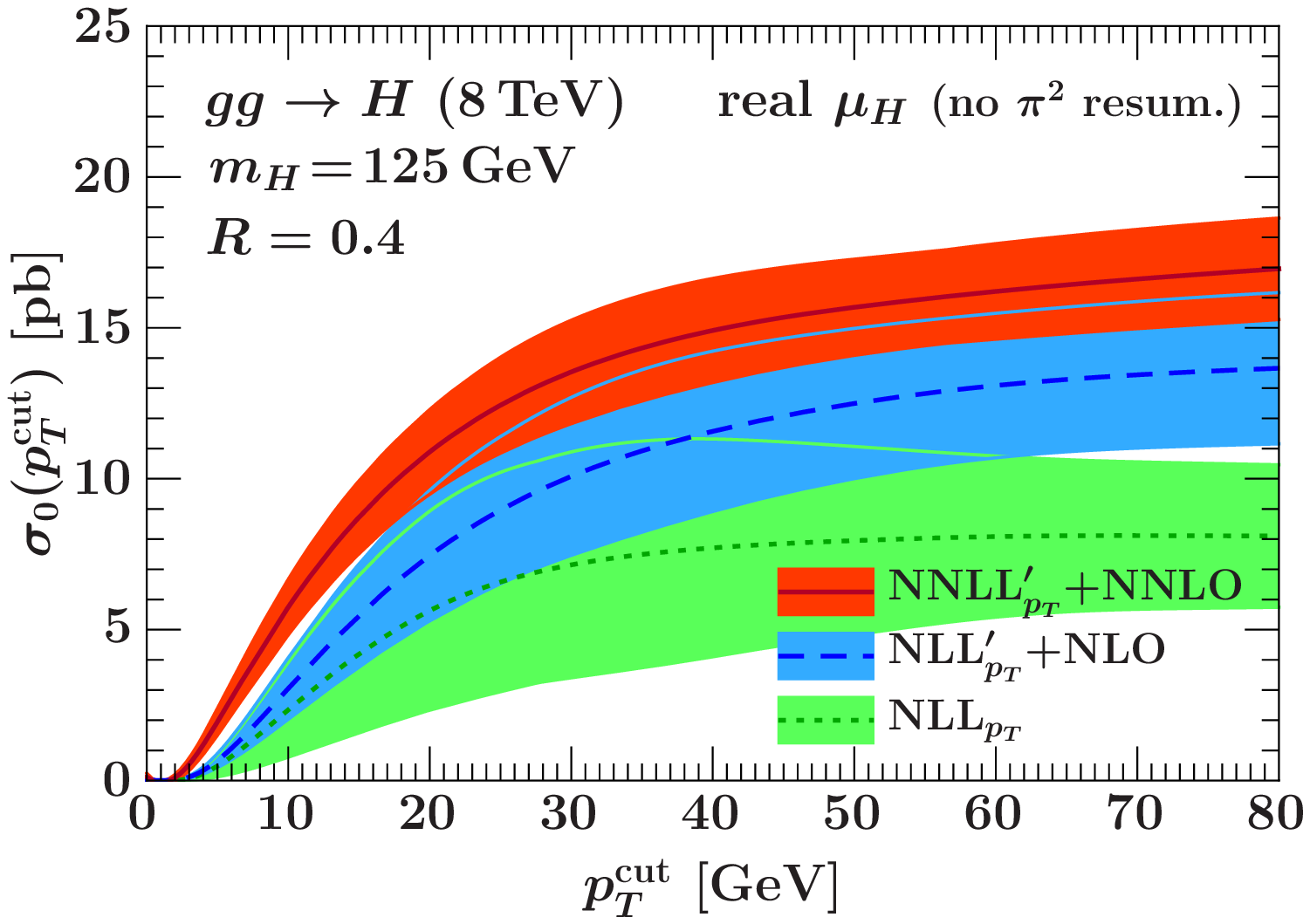}%
\hfill%
\includegraphics[width=1.02\columnwidth]{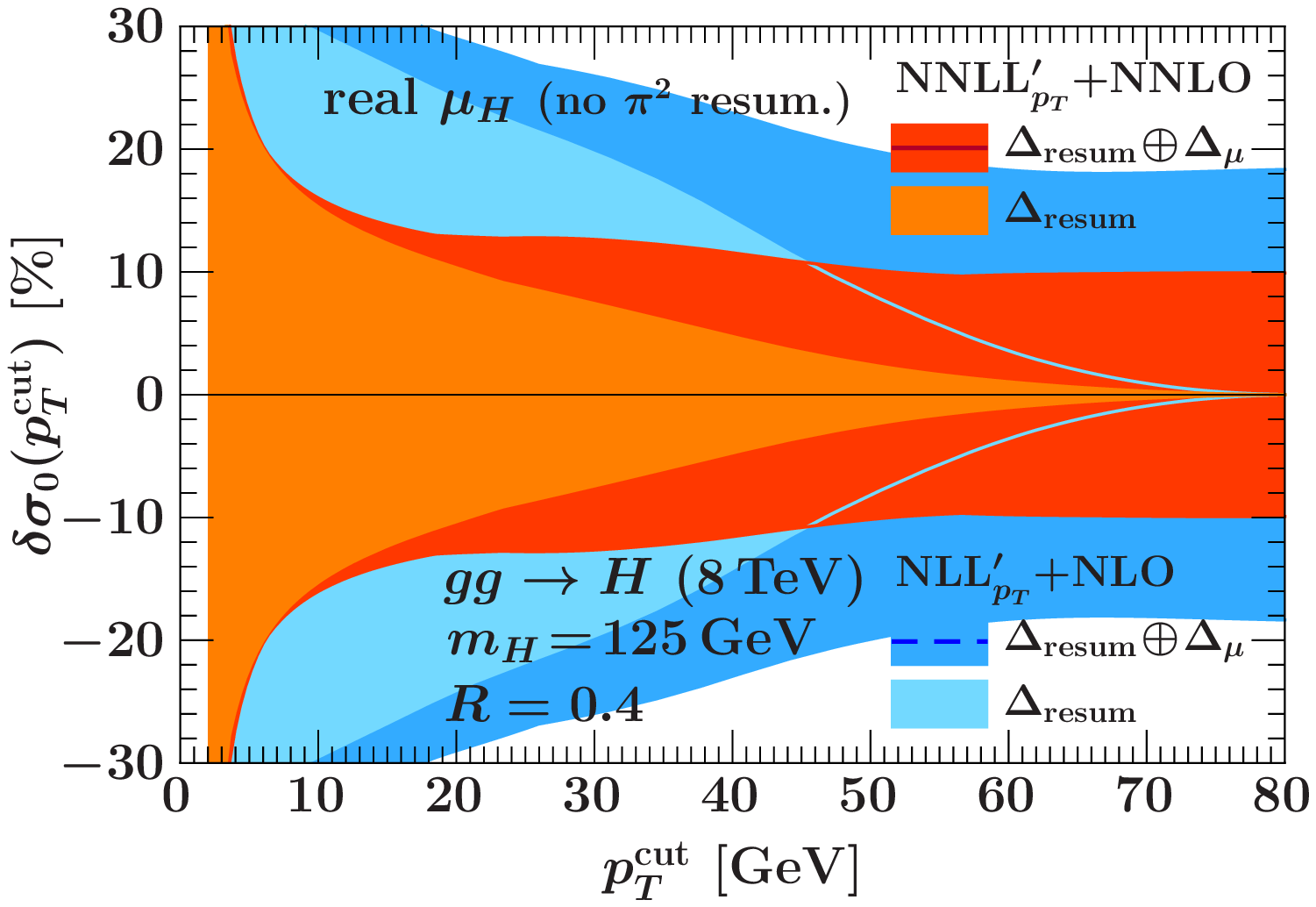}%
\vspace{-3ex}
\caption{The 0-jet cross section for $m_H = 125\GeV$ and $R = 0.4$ using the real scale setting $\mu_H = \mu_\FO$, which excludes the $\pi^2$ resummation. The poor convergence of the hard function results in larger uncertainties and a poorer convergence of the cross section at all values of $\pTcut$.}
\label{fig:sigma0nopi2}
\end{figure*}

\begin{table*}[t!]
\begin{tabular}{c|cccc}
\hline\hline
& $\sigma_{\geq 0}$ [pb] & $\sigma_0 (\pTcut)$ [pb] & $\sigma_{\geq 1}(\pTcut)$ [pb] & $\e_0 (\pTcut)$
\\\hline
NLL$'_{p_T}$+NLO
\\
$\pTcut = 25 \GeV$ &
$14.57 \pm 2.91\,\, (20.0\%)$ & $8.96 \pm 2.44\,\, (27.2\%)$ & $5.61 \pm 2.44\,\, (43.5\%)$ & $0.615 \pm 0.136\,\, (22.1\%)$
\\
$\pTcut = 30 \GeV$ &
$14.57 \pm 2.91\,\, (20.0\%)$ &$10.08 \pm 2.62\,\, (26.0\%)$ & $4.49 \pm 2.32\,\, (51.7\%)$ & $0.692 \pm 0.138\,\, (19.9\%)$
\\[0.5ex] \hline
NNLL$'_{p_T}$+NNLO ($R = 0.4$)
\\
$\pTcut = 25 \GeV$ &
$18.38 \pm 1.91\,\, (10.4\%)$ & $12.44 \pm 1.59\,\, (12.8\%)$ & $5.94 \pm 1.32\,\, (22.2\%)$ & $0.677 \pm 0.059\,\, (8.8\%)\,\,$
\\
$\pTcut = 30 \GeV$ &
$18.38 \pm 1.91\,\, (10.4\%)$ & $13.54 \pm 1.71\,\, (12.6\%)$ & $4.84 \pm 1.13\,\, (23.4\%)$ & $0.737 \pm 0.055\,\, (7.4\%)\,\,$
\\[0.5ex] \hline
NNLL$'_{p_T}$+NNLO ($R = 0.5$)
\\
$\pTcut = 25 \GeV$ &
$18.38 \pm 1.91\,\, (10.4\%)$ & $12.14 \pm 1.50\,\, (12.4\%)$ & $6.24 \pm 1.29\,\, (20.7\%)$ & $0.661 \pm 0.056\,\, (8.4\%)\,\,$
\\
$\pTcut = 30 \GeV$ &
$18.38 \pm 1.91\,\, (10.4\%)$ & $13.29 \pm 1.63\,\, (12.2\%)$ & $5.09 \pm 1.12\,\, (21.9\%)$ & $0.723 \pm 0.052\,\, (7.2\%)\,\,$
\\[0.5ex] \hline
NNLL$'_{p_T}$+NNLO ($R = 0.7$)
\\
$\pTcut = 25 \GeV$ &
$18.38 \pm 1.91\,\, (10.4\%)$ & $11.69 \pm 1.41\,\, (12.1\%)$ & $6.68 \pm 1.33\,\,(19.9\%)$ & $0.636 \pm 0.055\,\, (8.6\%)\,\,$
\\
$\pTcut = 30 \GeV$ &
$18.38 \pm 1.91\,\, (10.4\%)$ & $12.91 \pm 1.54\,\, (11.9\%)$ & $5.47 \pm 1.18\,\, (21.6\%)$ & $0.703 \pm 0.052\,\, (7.5\%)\,\,$
\\[0.5ex]
\hline\hline
\end{tabular}
\caption{Predictions for various cross sections with real scale setting $\mu_H = \mu_\FO$ and $\mu_\FO = m_H$ as central scale.}
\label{table:numbersnopi2}
\end{table*}

\section{\boldmath Results for Real $\mu_H$}
\label{app:nopi2}

For completeness and to demonstrate the benefit of the imaginary scale setting for $\mu_H$, in this Appendix we give predictions for the real scale setting $\mu_H = \mu_\FO$, which excludes the large $\pi^2$ terms from the resummation in the hard function.

In \fig{sigma0nopi2}, we plot the analog of \fig{sigma0} for $\sigma_0 (\pTcut)$ but using real $\mu_H$.  Comparing these two figures, it is clear that including the $\pi^2$ terms in the resummation significantly improves the convergence and precision of the 0-jet predictions at small $\pTcut$.  This improvement also translates into an improved convergence and reduced uncertainties at larger values of $\pTcut$.  In \tab{numbersnopi2}, we give the analogous values without $\pi^2$ summation to those in \tab{numbers}.  For $\pTcut = 25 \GeV$, $R = 0.4$ and $\pTcut = 30 \GeV$, $R = 0.5$, the corresponding components of the uncertainty are
\begin{align}
& & \pTcut &= 25 \GeV & \pTcut &= 30 \GeV \nn \\
& & R &= 0.4 & R &= 0.5 \nn \\
\Delta_\tot & \; : \; &  & 1.91 & & 1.91 \nn \\
\Delta_\resum & \; : \; &  & 1.08 & & 0.95 \\
\Delta_{\mu0} & \; : \; &  & 1.16 & & 1.32 \nn \\
\Delta_{\mu\ge1} & \; : \; &  & 0.75 & & 0.59 \nn \\\nn
\end{align}
Both the resummed and fixed-order uncertainties for the 0-jet cross section are larger when the $\pi^2$ terms are excluded from the resummation, indicating that these large $\pi^2$ terms have an effect on the shape as well as the normalization of the cross section.  This is also reflected in \fig{sigma0nopi2}.

\bibliographystyle{../physrev4}
\bibliography{../jets}

\end{document}